\documentclass[fleqn,usenatbib]{mnras}
\usepackage{newtxtext,newtxmath}
\usepackage[T1]{fontenc}

\DeclareRobustCommand{\VAN}[3]{#2}
\let\VANthebibliography\thebibliography
\def\thebibliography{\DeclareRobustCommand{\VAN}[3]{##3}\VANthebibliography}

\usepackage{graphicx}
\usepackage{amsmath}    
\usepackage{dcolumn}
\usepackage{bm}

\usepackage{multirow}
\usepackage{mathtools}
\usepackage{subfig}
\usepackage{xcolor}

\usepackage{xr}

\usepackage[para]{threeparttable}
\usepackage{verbatim}   
\usepackage{gensymb} 
\usepackage{wasysym} 
\usepackage{hyperref}   
\usepackage{fnpos}
\usepackage{nth}
\def\gcm3{\hbox{g cm$^{-3}$}}       

\def\degr{\hbox{$^\circ$}}

\def \1s{$1\,\sigma$}

\def \t0{T$_0$}

\def\m2s2{\hbox{\,m$^{2}$\,s$^{-2}$}}

\def\degr{\hbox{$^\circ$}}

\title[The study on HAT-P-13, HAT-P-16 and WASP-32 systems]{The study on transiting systems HAT-P-13, HAT-P-16 and WASP-32 through combining ground-based and TESS photometry}
\author[L. Sun et al.]{L. Sun,$^{1,2}$
S.~Gu,$^{1,2,3}$
X.~Wang,$^{1,2,3}$\footnotemark[1]
L.~Bai,$^{1,2,3}$
J.~H.~M.~M.~Schmitt,$^{4}$
V.~Perdelwitz,$^{5,4}$
\newauthor
P.~Ioannidis$^{4}$
\\
$^{1}$Yunnan Observatories, Chinese Academy of Sciences, Kunming 650216, China\\
$^{2}$Key Laboratory for the Structure and Evolution of Celestial Objects, Chinese Academy of Sciences, Kunming 650216, China\\
$^{3}$ School of Astronomy and Space Science, University of Chinese Academy of Sciences, Beijing 101408, China\\
$^{4}$Hamburger Sternwarte, Universität Hamburg, Gojenbergsweg 112, 21029 Hamburg, Germany\\
$^{5}$Department of Physics, Ariel University, Ariel 40700, Israel
}

\date{Accepted XXX. Received YYY; in original form ZZZ}

\pubyear{2022}

\begin{document}
\label{firstpage}
\pagerange{\pageref{firstpage}--\pageref{lastpage}}
\maketitle

\begin{abstract}

High-precision transit photometry supplies ideal opportunities for detecting new exoplanets and characterizing their physical properties, which usually encode valuable information for unveiling the planetary structure, atmosphere and dynamical history. We present revised properties of three transiting systems (i.e., HAT-P-13, HAT-P-16 and WASP-32) through analyzing TESS photometry and ground-based transit observations, which were obtained at the 1m and 2.4m telescopes of Yunnan Observatories, China, and the 1.2m telescope of Hamburg Observatory, Germany, as well as the data in the literature. During modelling the transit light curves, Gaussian process is employed to account for the potential systematic errors. Through comprehensive timing analysis, we find that both HAT-P-13b and HAT-P-16b show significant timing variations (TTVs) that can be explained by apsidal precession. TTVs of WASP-32b may be led by a decaying orbit due to tidal dissipation or apsidal precession. However, the current observations can not rule out the origins of three systems' TTVs from gravitational perturbations of close planetary companions conclusively.

\end{abstract}


\section{Introduction}
\label{sect:intro}

The rapid growth in the number of transiting exoplanets plays a fundamental role in the advance of exoplanetary science, thanks to many dedicated exoplanet survey projects like SuperWASP, HATNet, XO, Kepler, K2, TESS, and so on \citep{Bakos2004,McCullough2005,Pollacco2006,Borucki2011}. High-precision photometry for transit events has become one of the most successful methods for detecting new exoplanets and characterizing the exoplanetary systems. Up to 2022 March 15, transiting exoplanets of 3513 had been discovered and confirmed, which reside in 2645 planetary systems; transiting exoplanets are dominant (i.e., $\sim70$ \%) in the confirmed exoplanet population \footnote{http://exoplanet.eu/}. Furthermore, follow-up observations for known transiting exoplanetary systems could not only refine the parameters of the systems, but also provide excellent opportunities to uncover the structure, atmosphere and dynamical evolution of the planets, which allow us to understand the formation and evolution of planetary systems.

Long-term high-precision photometric monitoring for transit events of known exoplanetary systems is of importance for hunting extra planets and planning planning new observations. In general, a single exoplanet transits its host star with a steady orbital period on a time scale of years, on which tidal and general relativity effects have no significant accumulation. But the transit timing variation \citep[TTV;][]{Agol2005,Holman2005}, which represents the deviation of transit times from a linear ephemeries, will appear in principle when extra bodies exist in the same system. Due to the relatively longer time scales than the orbital periods, on which the gravitational interactions among the exoplanets work to generate significant TTVs, the space missions, such as {\it Kepler}, TESS and upcoming PLATO, and ground-based long-term high-precision photometry are more possible to seize such signals \citep{Mazeh2013,Rowe2014,Holczer2016}. The overall shapes, amplitudes and frequencies of these TTVs induced by  the gravitational interactions, primarily determined by their orbital parameters and masses \citep[see e.g.][]{Agol2005,Holman2005,Nesvorny2008,Lithwick2012,Xie2014}. Therefore, TTV technique is often served as an important tool for characterizing exoplanetary systems: it can supply limits for the hidden exoplanets, hence remedying the absent information on the genuine planetary orbital architectures due to the detection bias inherent to the transit method \citep{Xie2014,Zhu2018,Sun2022}, which benefits the study on synthetic planetary system population \citep[see e.g.][]{Mordasini2009,Mordasini2018,Wu2019}. In addition, TTV technique is able to weigh the exoplanets in muti-planetary systems \citep[see e.g.][]{Nesvorny2012,Lithwick2012}, and therefore their densities, which provide tight constraints on their internal structures, such as the analyses for Kepler-411 system \citep{Sun2019}, Trappist-1 system \citep{Grimm2018} and so on. The discovery of distinct dynamically active systems could also be used to constrain the planetary formation and evolution models, since the present orbital architectures of such systems may hold the imprints from their past orbital migrations \citep[see e.g.][]{Delisle2017,Nesvorny2022}. 

There are some exoplanetary systems exhibiting temporal variations in their orbital periods, however, other origins may cause the change in their orbital periods \citep{RN212,RN209,RN214,Baluev2019,RN787,RN788}, such as tidal dissipation \citep{Barker2010}, apsidal precession \citep{RN265}, spot crossing events \citep{Sun2017}, planetary mass loss driven by atmospheric escape \citep{Fujita2022}, and so on; these origins would likely imitate the TTV signals originated from gravitational interactions with other planetary bodies. However, orbital decay driven by tidal dissipation is another case deserving for further investigation, in addition to the case of gravitational interactions with other bodies.  Such kind of TTV signals could place constraints on the tidal quality factors and experienced dynamical history of the planets \citep{Goldreich1966}. Associated theory suggests that it is likely to exhibit in massive hot Jupiters. 

Here we present TTV analyses on HAT-P-13, HAT-P-16 and WASP-32 planetary systems. Since 2010, we have monitored their transit events using the 1 m and 2.4 m telescopes of Yunnan Observatories (hereafter, YO-1m and YO-2.4m) in China and the 1.2 m Oskar-Lühning telescope (hereafter, OLT-1.2m) at Hamburg Observatory in Germany. As a result, a series of high-precision photometric transit data were obtained. In addition, HAT-P-13, HAT-P-16 and WASP-32 were observed by TESS in sectors 47, 17 and 42, respectively. Furthermore, HAT-P-16 was observed by TESS in Sector 57 (2022 September 30 to 2022 October 29). In this paper, we present the timing analyses of these planetary systems. First, we briefly introduce each target in Section \ref{sec:tar}. Subsequently, the description for the observations and data reduction strategy are present in Section \ref{sec:obs} and \ref{sec:dat}, respectively. In Section \ref{sec:tra} the transit modeling is described. We then discuss the results in Section \ref{sec:dis}. At last, we summarize the study in Section \ref{sec:con}. 

\section{Targets}
\label{sec:tar}

\subsection{HAT-P-13}

The multi-planet system HAT-P-13 was discovered by \citet{Bakos2009}. This system has a inner transiting hot Jupiter HAT-P-13b and a outer extremely massive planet HAT-P-13c, which orbits a metal-rich ($[Fe/H]=+0.41\pm0.08$) host star with periods of 2.9 and 428.5 days, respectively. \citet{Winn2010} found the spin-orbit of this system is well aligned by modeling the Rossiter-McLaughlin (R-M) effect, and evidence for the third companion based on the long term radial acceleration of the host star. \citet{Szabo2010} observed HAT-P-13 system to detect a possible transit of HAT-P-13c by their multi-site campaign, they found HAT-P-13c was most likely not a transiting exoplanet. Moreover, \citet{Szabo2010} obtained two additional transit events and refined the orbital period of HAT-P-13b. \citet{Nascimbeni2011}, \citet{pal2011}, \citet{RN767}, \citet{Payne2011}, \citet{Southworth2012}, \citet{Sada2016}, and \citet{Baluev2019} acquired new transiting light curves of HAT-P-13b based on several ground-based telescopes, refined the physical parameters and analyzed the TTVs. Although the available transit times deviated from a linear ephemeris, the residuals of the transit times were complex rather than expected sinusoid. \citet{Turner2016} observed a transit event of HAT-P-13b in the near-UV, and found the available multi-wavelength photometric transit depths from near-UV to optical wavelengths were not able to well constrain the atmospheric model for HAT-P-13 b due to the large error bars. \citet{Kramm2012} and \citet{Batygin2016} modeled the interior structure of HAT-P-13b using the observationally determined tidal Love number $k_{2}$, then \citet{Buhler2016} and \citet{Hardy2017} imposed constraints on the orbital eccentricity, the tidal Love number $k_{2}$, and the interior structure of HAT-P-13b based on the same set of two secondary-eclipse observations collected by Spitzer Space Telescope.

\subsection{HAT-P-16}

HAT-P-16b was discovered by \citet{Buchhave2010}, which is a massive transiting hot Jupiter (i.e., $M_{p}=4.193\pm0.094 M_{Jup}$; $R_{p}=1.289\pm0.066 R_{Jup}$) orbiting a relatively bright (V=10.8) F-type host star with a period of 2.7 days. \citet{Moutou2011} studied the R-M effect of HAT-P-16 system, and the projected spin-orbit angle is $\lambda=-10\pm16\degr$. \citet{Sada2012}, \citet{Ciceri2013}, \citet{Pearson2014}, \citet{Aladag2021} and \citet{Wang2021} acquired new transiting light curves of HAT-P-16b using several ground-based telescopes, refined the system parameters and found no significant TTV.  \citet{Davoudi2020} re-analyzed several relative high quality transit light curves of HAT-P-16b collected from the Exoplanet Transit Database (ETD) \footnote{http://var2.astro.cz/ETD/}. They found that the newly derived physical parameters of HAT-P-13 were in good agreement with those in NASA exoplanet archive.

\subsection{WASP-32}

WASP-32b is a massive transiting hot Jupiter (i.e., $R_p = 1.190 \pm 0.047 R_{Jup}$; $M_p = 3.59 \pm 0.06 M_{Jup}$ ), discovered by \citet{Maxted2010} in the SuperWASP survey, orbiting a Sun-like lithium-poor star ($V$ = 11.3 mag) every 2.72 days. \citet{Sada2012} obtained one $J$-band transiting light curve of WASP-32 utilizing the Kitt Peak National Observatory's 2.1m telescope and refined the orbital period of the planet. \citet{Brown2012} recalculated the physical parameters of the system and derived a spin-orbit angle of $10\degree.5^{+6.4}_{-6.5} $ using both the R-M effect and Doppler tomography for WASP-32. \citet{Brothwell2014} confirmed the spin-orbit alignment angle of \cite{Brown2012} utilizing the R-M effect. \citet{Sun2015} recalculated the physical parameters of the system and confirmed no apparent transit timing variation signal for WASP-32. \citet{Grauzhanina2017} found that the intensity and the equivalent width of the H$_{\alpha}$ line showed a significant change and gave the evidence for WASP-32b filling the Roche lobe and having a comet-like tail based on spectroscopic observations by utilizing the 6m BTA telescope.  At last, \citet{Valeev2019} found the presence of Na, K and probable HI based on the transmission spectrum of WASP-32b observed by using 10.4 m GTC telescope.

\section{Observations}
\label{sec:obs}

\subsection{YO-1m photometry}

We observed the transit events of HAT-P-13, HAT-P-16 and WASP-32 using the YO-1m telescope \citep[see e.g.][]{RN747,RN338,RN748,Sun2015,Sun2017} between 2010 December and 2018 March (see Table \ref{tab1}). The Johnson-Cousins R filter was used during all observations. The 2K$\times$2K CCD camera with a field of view (FOV) of 7.3$\times$7.3 arcmin$^{2}$ was adopted, except on 2017 March 23 when another 4K$\times$4K CCD camera with a FOV of 15$\times$15 arcmin$^{2}$ was utilized. During the observations, the instrument statuses were good, and the exposure times were set according to the weather conditions. Some transit event observations were incomplete, whose light curves had a relatively larger dispersion and/or offset, mostly because the weather rapidly changed to be bad (see Figure \ref{fig3}). 

\subsection{YO-2.4m photometry}

On 2015 November 7, one transit event of WASP-32b was observed by utilizing Yunnan Faint Object Spectrograph and Camera (YFOSC) attached to the YO-2.4m telescope \citep{RN716,RN715} at Yunnan Observatories, China. As YFOSC is not a multi-channel imaging instrument, we obtained nearly simultaneous four-color transit photometry of WASP-32b in Johnson B and V as well as Johnson-Cousins Rc and Ic passbands through rapidly switching the passband filters. The FOV is 9.7 $\times$ 9.7 arcmin$^2$. The telescope pointing was maintained using the auto-guiding system, but it was broken down due to a temporal instrument issue during our observation and was reactivated rapidly. The weather was clear with a good seeing. The observing log is listed in Table \ref{tab1}. 

\subsection{OLT-1.2m photometry}

A transit event of HAT-P-16 on 2016 December 8 was observed using an Apogee Alta U9000 CCD camera mounted on the OLT-1.2m telescope at Hamburg Observatory, Germany. The FOV is 9.0 $\times$ 9.0 arcmin$^2$. The sky was quite clear and Johnson-Cousins R filter was used in the observation. We used its auto-guiding system during the observation. To improve the photometric precision, the defocusing technique was adopted during the observation. The exposure time was set to 120 s (see Table \ref{tab1}).

\subsection{TESS photometry}

The TESS spacecraft observed HAT-P-13, HAT-P-16 and WASP-32 during its Sector 47 (2021 December 30 to 2022 January 28), Sector 17 (2019 October 7 to 2019 November 2) and Sector 42 (2021 August 20 to 2021 September 16), respectively. All of the data products with a 2-minute cadence mode reduced by the pipeline created by the Science Processing Operations Center \citep[SPOC;][]{RN702}  were used in this analysis. The light curve files were retrieved from the archives at Mikulski Archive for Space Telescopes (MAST)\footnote{https://archive.stsci.edu/mission/tess/} and the light curves were obtained by employing $Lightkurve$ \citep{RN521}. 
\begin{table*}
\caption[]{The observing log of all transit photometry.}
\label{tab1}
\centering
\begin{threeparttable}
\begin{tabular}{c c c c c c c c}
\hline
No. & Date & Telescope & Target & Exposure time & Number of & Duty cycle & Filter\\
& & & &(s) & data points & hour$^{-1}$ & \\
\hline
1 & 2013-01-06 & YO-1m & HAT-P-13 & 90 & 184 & 36 & Rc\\
2 & 2013-12-19 & YO-1m & HAT-P-13 & 180 & 104 & 19 & Rc\\
3 & 2014-01-23 & YO-1m & HAT-P-13 & 120 & 212 & 27 &Rc\\
4 & 2014-01-26 & YO-1m & HAT-P-13 & 120 & 127 & 27 &Rc\\
5 & 2014-03-08 & YO-1m & HAT-P-13 & 120 & 120 & 27 &Rc\\
6 & 2014-12-01 & YO-1m & HAT-P-13 & 150 & 92 & 22 &Rc\\
7 & 2015-11-16 & YO-1m & HAT-P-13 & 150 & 110 & 22 &Rc\\
8 & 2016-12-02 & YO-1m & HAT-P-13 & 120 & 114 & 27 &Rc\\
9 & 2017-03-23 & YO-1m & HAT-P-13 & 60 & 251 & 51 &Rc\\
10 & 2018-03-08 & YO-1m & HAT-P-13 & 90 & 178 & 36 & Rc\\
11 & 2010-12-24 & YO-1m & HAT-P-16 & 30 & 167 & 90 &Rc\\
12 & 2011-12-17 & YO-1m & HAT-P-16 & 100 & 96 & 33 &Rc\\
13 & 2016-12-08 & OLT-1.2m & HAT-P-16 & 120 & 159 &28 & Rc\\
14 & 2015-11-07 & YO-2.4m & WASP-32 & 50 & 79 &26 & B\\
15 & 2015-11-07 & YO-2.4m & WASP-32 & 30 & 78 &26 & V\\
16 & 2015-11-07 & YO-2.4m & WASP-32 & 10 & 78 &26 & Rc\\
17 & 2015-11-07 & YO-2.4m & WASP-32 & 10 & 78 &26 & Ic\\
18 & 2015-11-18 & YO-1m & WASP-32 & 150 & 62 &22 & Rc\\
\hline
\end{tabular}
\end{threeparttable}
\end{table*}

\section{Data reduction} 
\label{sec:dat}

\subsection{Aperture photometry for ground-based observations}\label{subsec:ap}

All of CCD images collected by ground-based telescopes were reduced according to the standard procedure as described in \citet{RN338}, which includes aperture photometry and systematic error correction, and has been integrated into a pipeline. The pipeline was written in Python based on the IRAF package, including image trimming, bias subtraction, flat-field correction and cosmic ray removal. The instrumental magnitudes of the target and reference stars were measured by adopting aperture photometry technique with an optimal aperture, which was to minimize the dispersion of the light curve. The optimal aperture was acquired through comparing the photometric precision for extracted light curves obtained from a series of trial apertures. Then the method proposed by \citet{RN379} was used to convert the observing time into the format of $BJD_{TDB}$. Because the transit signals of exoplanets are normally shallow, we employed coarse de-correlation \citep{RN230} and SYSREM algorithms \citep{RN231} to diagnose potential systematic errors and correct them so as to enhance the signal-to-noise ratio (SNR) of ground-based transit observations. These systematic errors are mainly due to the extinction of Earth atmosphere, the change of point spread function, the apparent drifts of stars on CCD position associated with the imperfect flat-filed correction and so on. In this correction procedure, the least square algorithm was iteratively used to model the patterns of significant systematic errors shared by most reference stars, in which high-quality reference stars with trivial variability was simultaneously identified. We refer interested readers to \citet{RN752} for further details. The efficiency of coarse de-correlation \citep{RN230} and SYSREM \citep{RN231} for systematic errors correction would be limited, however, when only several high-quality reference stars were observed in CCD images, whereas other reference stars were much fainter than the target. This issue was apparently presented in the transit observations of HAT-P-13b obtained by employing the 2K$\times$2K CCD camera on YO-1m, because only one reference star has comparable brightness to the target in the FOV. To handle this issue, we employed the Gaussian Process (GP) regression to account for the remaining systematic noises (see \ref{sec:Juliet} for details) and eliminated them in photometric data through subtracting the optimal results of GP regression.

\section{Transit Modeling}\label{sec:tra}

\subsection{Initial Transit Modelling }\label{sec:Juliet}

Usually, astronomical photometric data are contaminated by noises, which will reduce the SNR of the signals of interest and hamper the accurate measurements of the physical properties of transiting exoplanets. The correlated components in those noises (i.e., also called systematic noises) are typically associated with instrumental and/ or astrophysical sources, such as imperfect flat fielding, variations of the telluric atmosphere, the stellar activity and so on, which cannot generally be eliminated through adopting specific observation technique \citep{2009MNRAS.396.1023S}. Therefore, appropriately accounting for the systematic noises by using some noise reduction techniques \citep[see e.g.][]{RN230,RN231,RN382,RN525}, becomes quite essential in the analysis of the transit observations.

In exoplanet community, a widely adopted way to account for systematic noises is to model them with Gaussian Process \citep[see e.g.][]{RN524,RN525,RN526,RN704,Ahrer2022,2022MNRAS.tmp..222L}. By choosing different forms of the kernel, GPs are able to represent a wide range of signals.  Here, we employed publicly available $Juliet$ \citep{RN527}, which is able to model the systematic errors in time series data with GP technique, to diagnose and correct the remaining long-term trends and systematic errors in both TESS and ground-based photometric data.

According to the Gaia DR3 database \citep{RN515}, all of our targets are in the relatively evacuated fields and there is no other source bright enough to dilute the transit light curves, so the dilution factors of all targets were fixed to 1 during our transit modelling with $Juliet$. Other fitted transit parameters included the orbital period $P$ (fixed to the optimal values in the literature), the mid-time $T_{0}$, the scaled semi-major axis $a/R_{A}$, the eccentricity $e$ (fixed to 0), the argument of periastron $\omega$ (fixed to 90 degrees), the impact parameter $b$, the planet-to-star radius ratio $R_{b}/R_{A}$, the mean out-of-transit flux and the jitter. During modelling the transit light curve, a truncated Gaussian distribution was imposed on the prior of the fitted transit parameter. The centre value and sigma of the Gaussian prior were set to the median and 3 times sigma of its posterior in \citet{Bakos2009} for HAT-P-13, \citet{Buchhave2010} for HAT-P-16 and \citet{Maxted2010} for WASP-32, respectively. 

We individually modeled the GPs of each light curve, which were defined for each instrument and observation. We chose the celerite (approximate) Matern multiplied exponential kernel for the GP regression. The hyperparameters of this GP kernel includes the amplitude of the GP, two length scales corresponding to the Matern and the exponential part. Log-uniform priors were imposed for these hyperparameters and the amplitude of the GP varied from $10^{-6}$ to $10^{6}$, and both length scales varied from $10^{-3}$ to $10^{3}$. We employed the nest sampler $MultiNest$ in \textit{Juliet} with 500 live points to explore the parameter space, and obtained optimal systematic error models as well as the transit light curve models of all photometric data sets.

It should be noted that we derived the mid-time of each transit event by modeling each light curve separately with \textit{Juliet} and hence refined the ephemerides. The mid-times of all transit events are listed in Tables \ref{tab2}, \ref{tab3} and \ref{tab4}. In addition, we collected all available mid-times of these three systems from previous works and the ETD website. A linear ephemeris formula 
\begin{displaymath}
		T_{tra}(E) = T_{0} + P \times E 
\end{displaymath}
was used to fit all of the mid-times, where $T_{0}$ is the zero point, $P$ is the orbital period and $E$ is the orbital cycle number. We employed the Markov Chain Monte Carlo (MCMC) algorithm to fit the mid-times and derived the best-fitting linear ephemeris formula and the orbital period values, respectively. The MCMC algorithm was preformed by using the \textit{emcee} package \citep{RN263} and 50,000 MCMC steps with 1,000 burn-in steps was run to ensure the convergence. The final results are
\begin{displaymath} 
T(\mathrm{BJD_{TDB}}-2450000) = 4779.92999(33) + 2.9162420(5) \times E
\end{displaymath}
for the HAT-P-13 system,
\begin{displaymath} 
T(\mathrm{BJD_{TDB}}-2450000) = 5027.59301(19) + 2.7759682(2) \times E
\end{displaymath}
for the HAT-P-16 system and
\begin{displaymath} 
T(\mathrm{BJD_{TDB}}-2450000) = 5779.06707(24) + 2.7186615(3) \times E
\end{displaymath}
for the WASP-32 system. 

\begin{figure*}
  \begin{minipage}[t]{\textwidth}
  \centering
   \includegraphics[width=\hsize]{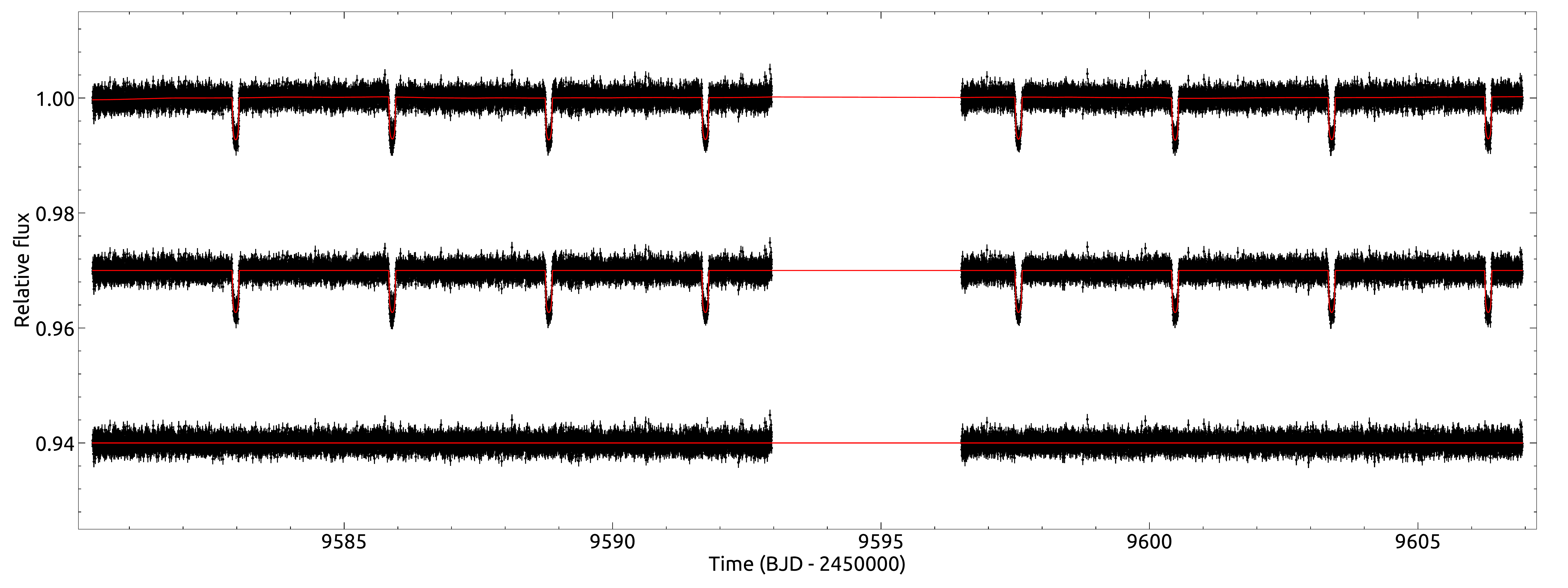}
  \end{minipage}%
  
  \begin{minipage}[t]{\linewidth}
  \centering
   \includegraphics[width=1.0\hsize]{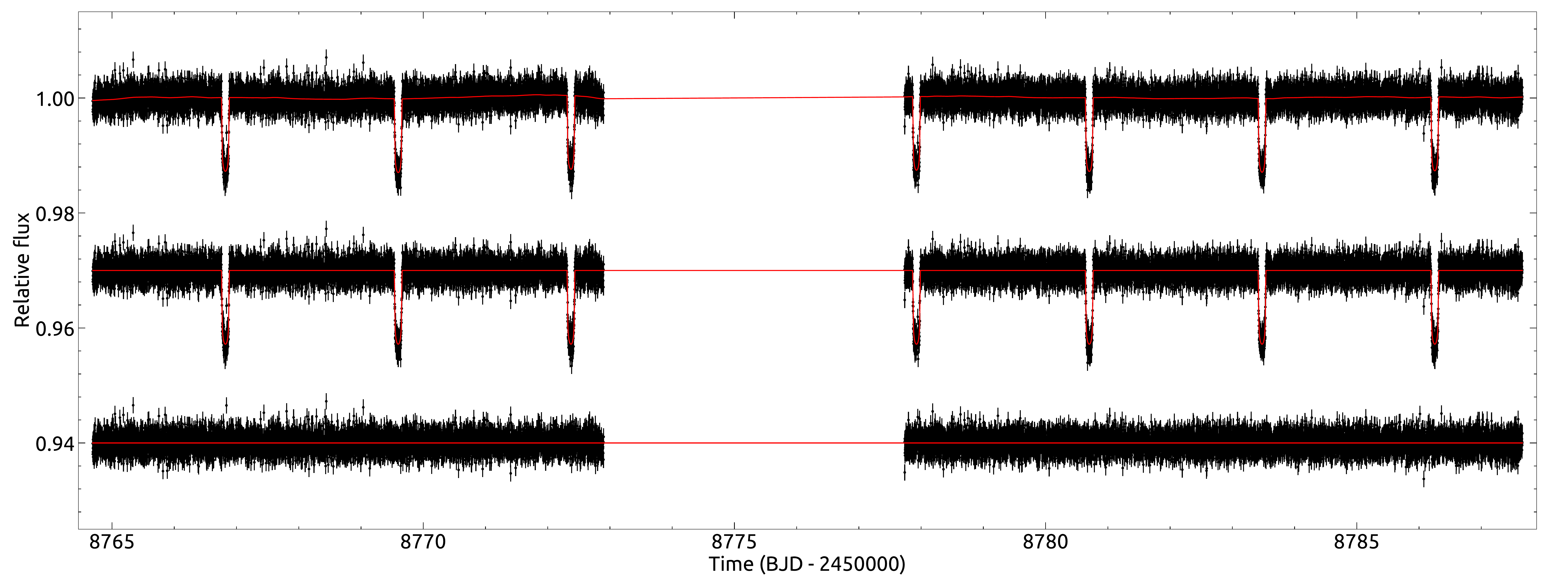}
  \end{minipage}%
  
  \begin{minipage}[t]{\textwidth}
  \centering
   \includegraphics[width=\hsize]{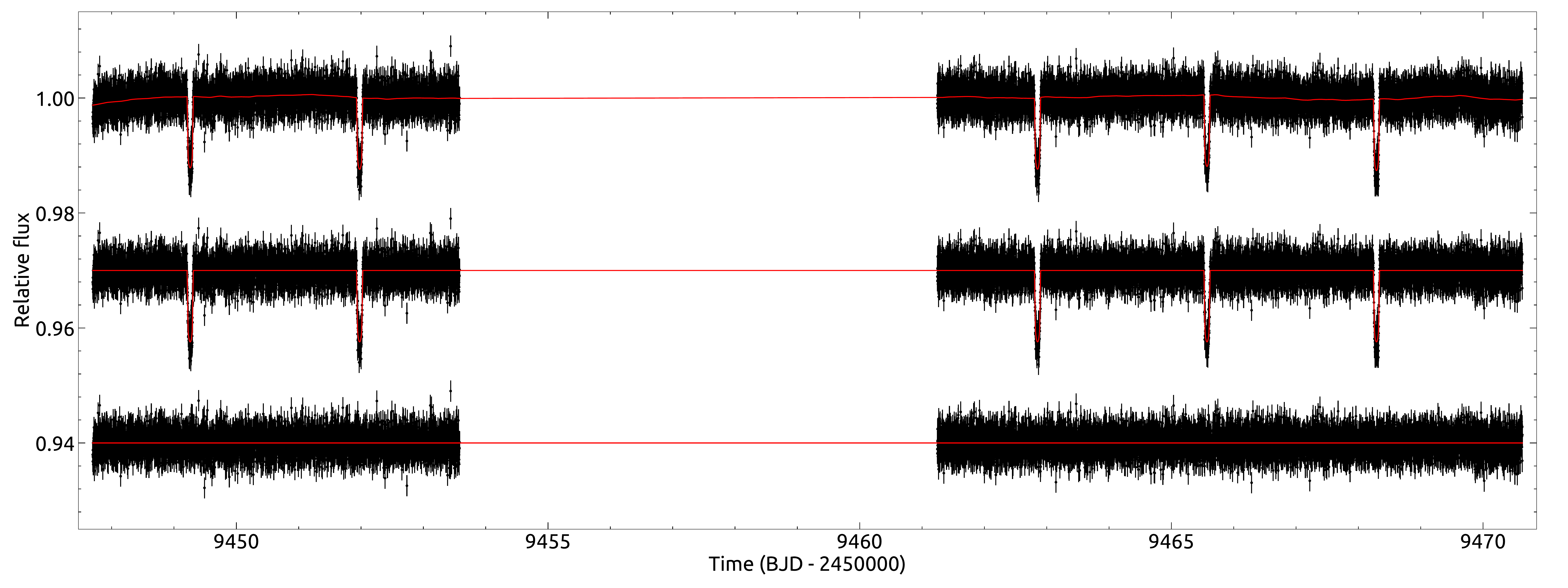}
  \end{minipage}%
      
  \caption{Transit light curves of HAT-P-13 (top panel), HAT-P-16 (middle panel) and WASP-32 (bottom panel) observed by TESS. In each panel, the top is raw light curves with the full median posterior models, the middle is the final light curves with the best-fitting deterministic transit models, and the bottom is the corresponding residuals. Vertical shifts are added for visualization.}
         \label{fig1}
   \end{figure*}

\begin{figure*}
  \begin{minipage}[t]{0.33\hsize}
  \centering
   \includegraphics[width=\hsize]{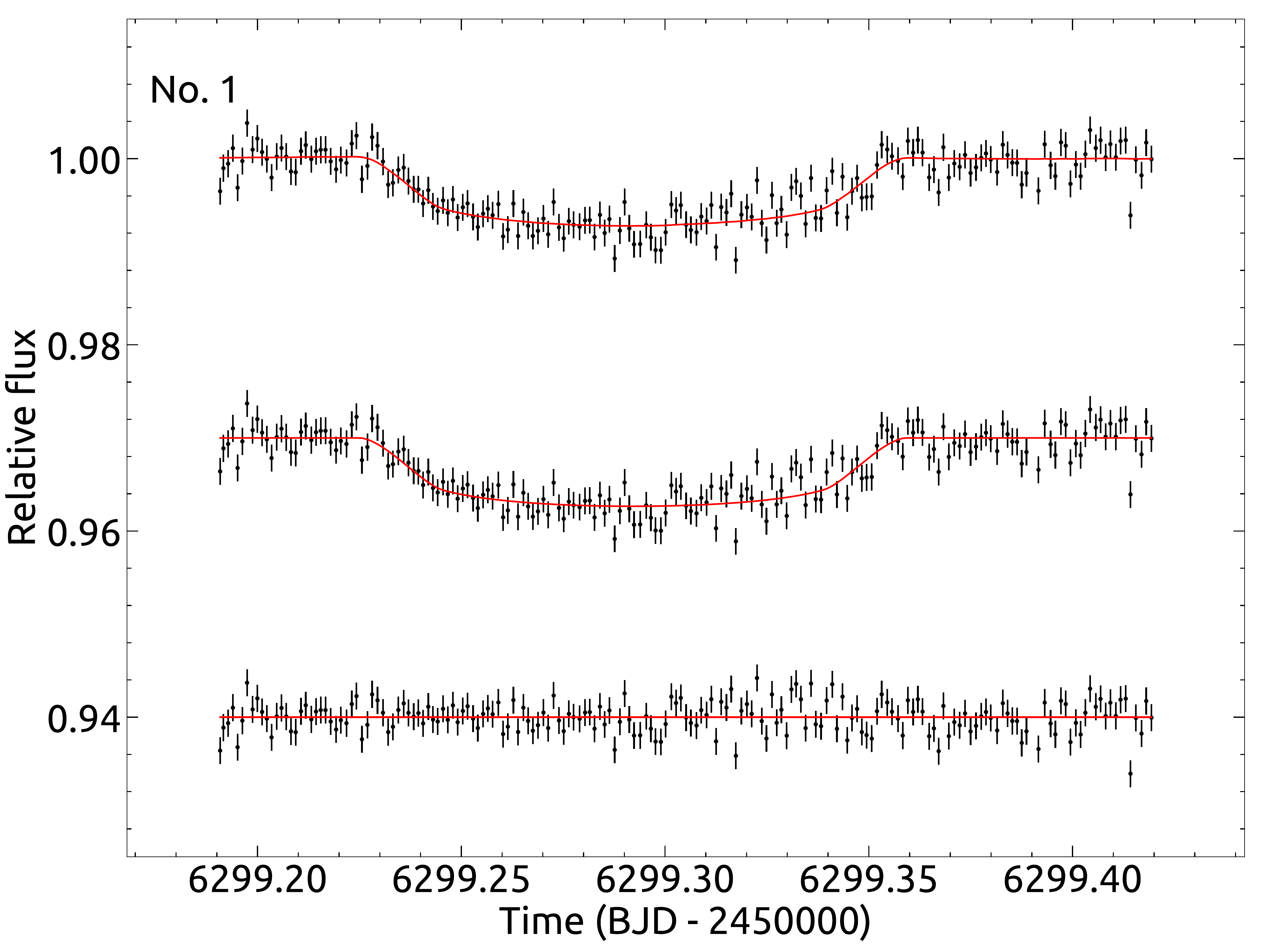}
  \end{minipage}%
  \begin{minipage}[t]{0.33\hsize}
  \centering
   \includegraphics[width=\hsize]{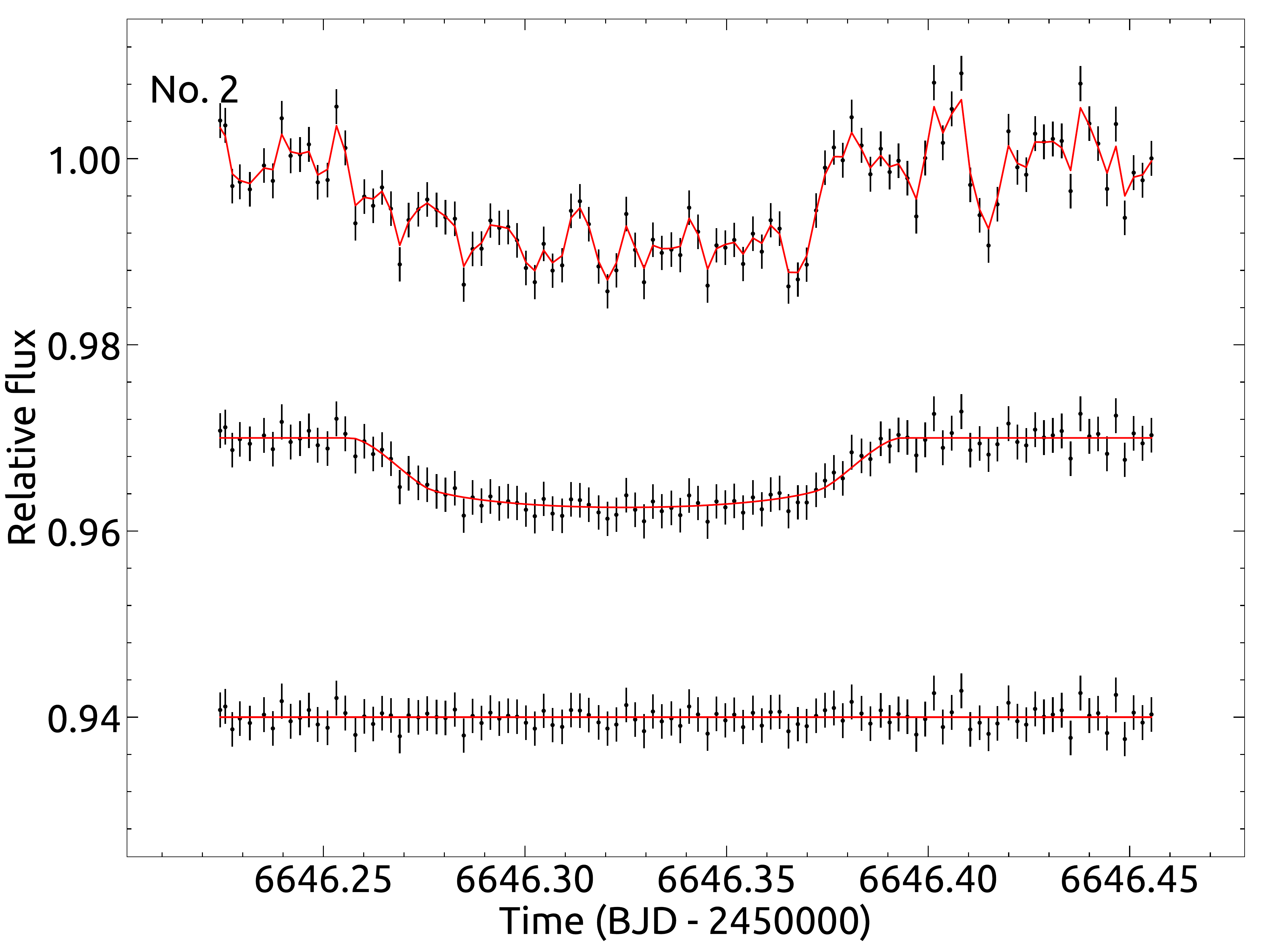}
  \end{minipage}%
  \begin{minipage}[t]{0.33\hsize}
  \centering
   \includegraphics[width=\hsize]{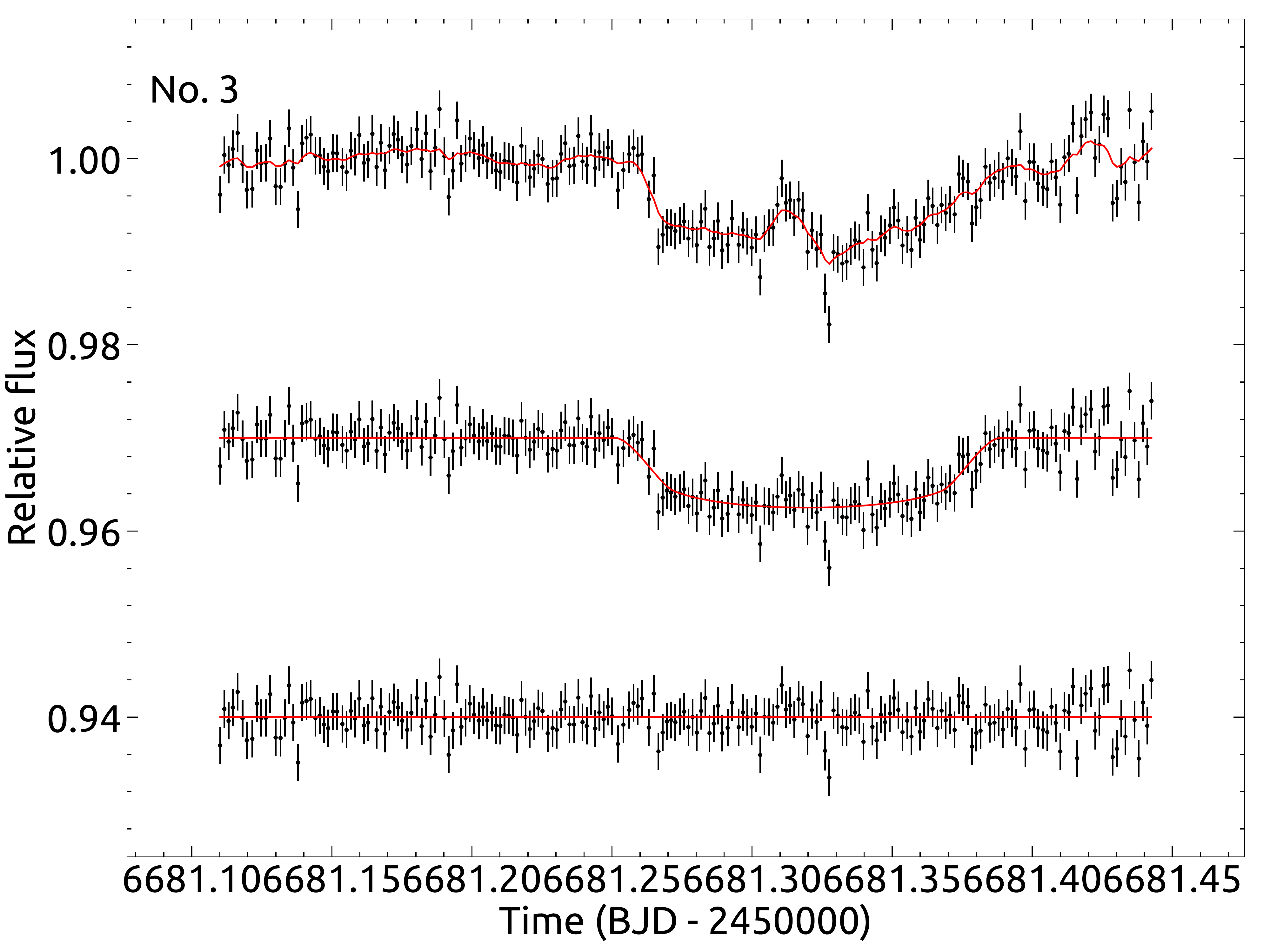}
  \end{minipage}%

  \begin{minipage}[t]{0.33\hsize}
  \centering
   \includegraphics[width=\hsize]{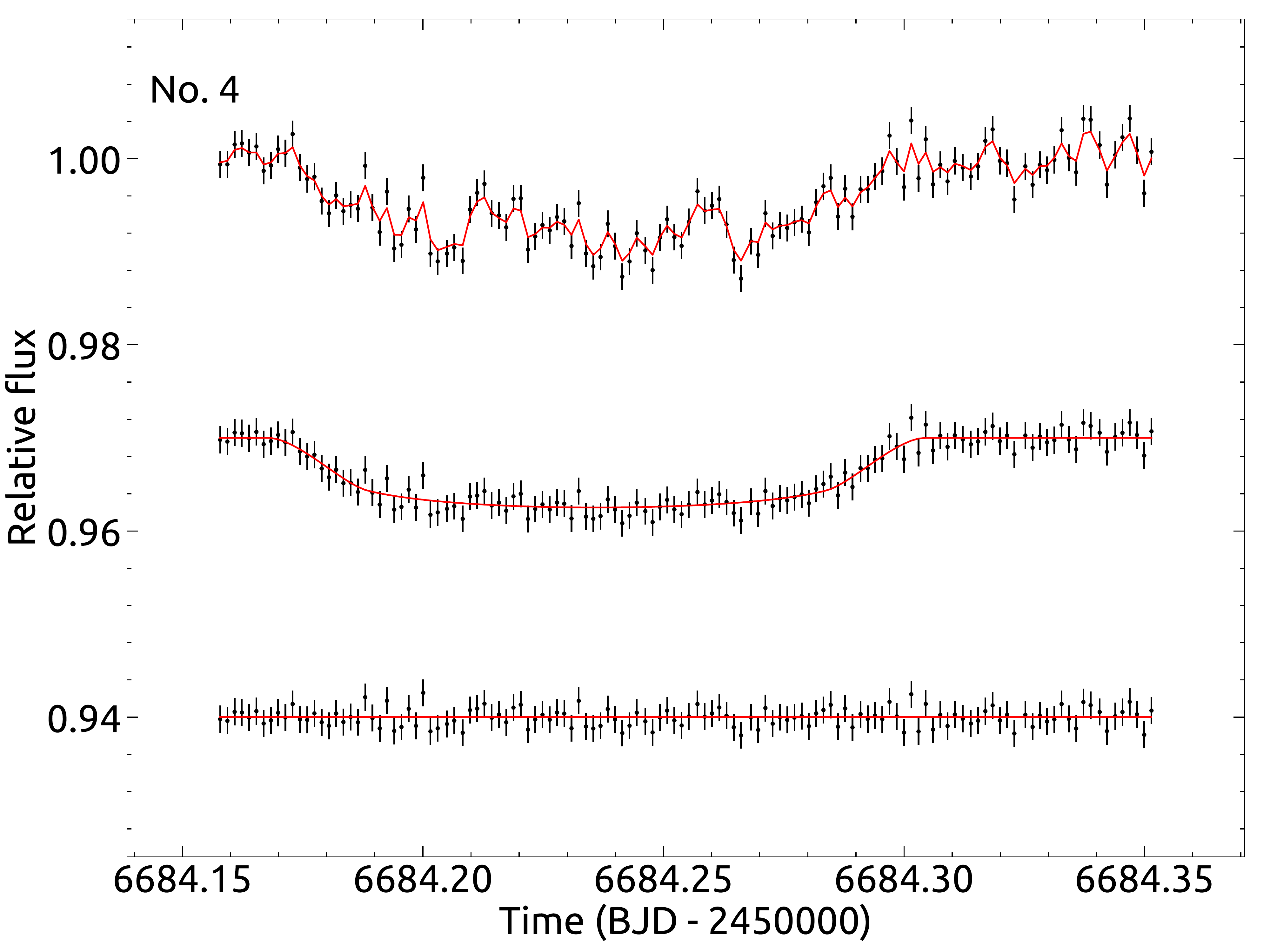}
  \end{minipage}%
  \begin{minipage}[t]{0.33\hsize}
  \centering
   \includegraphics[width=\hsize]{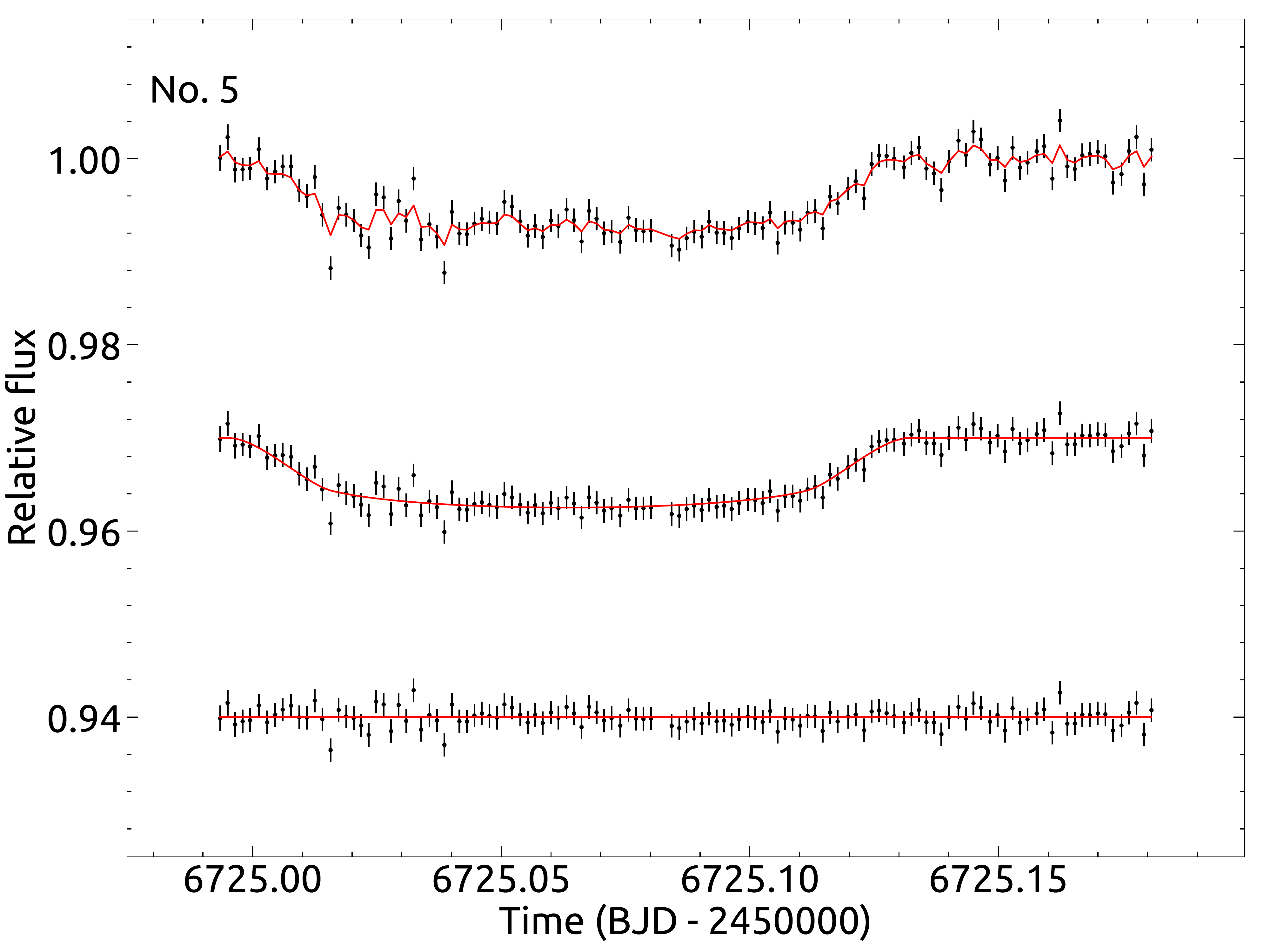}
  \end{minipage}%
  \begin{minipage}[t]{0.33\hsize}
  \centering
   \includegraphics[width=\hsize]{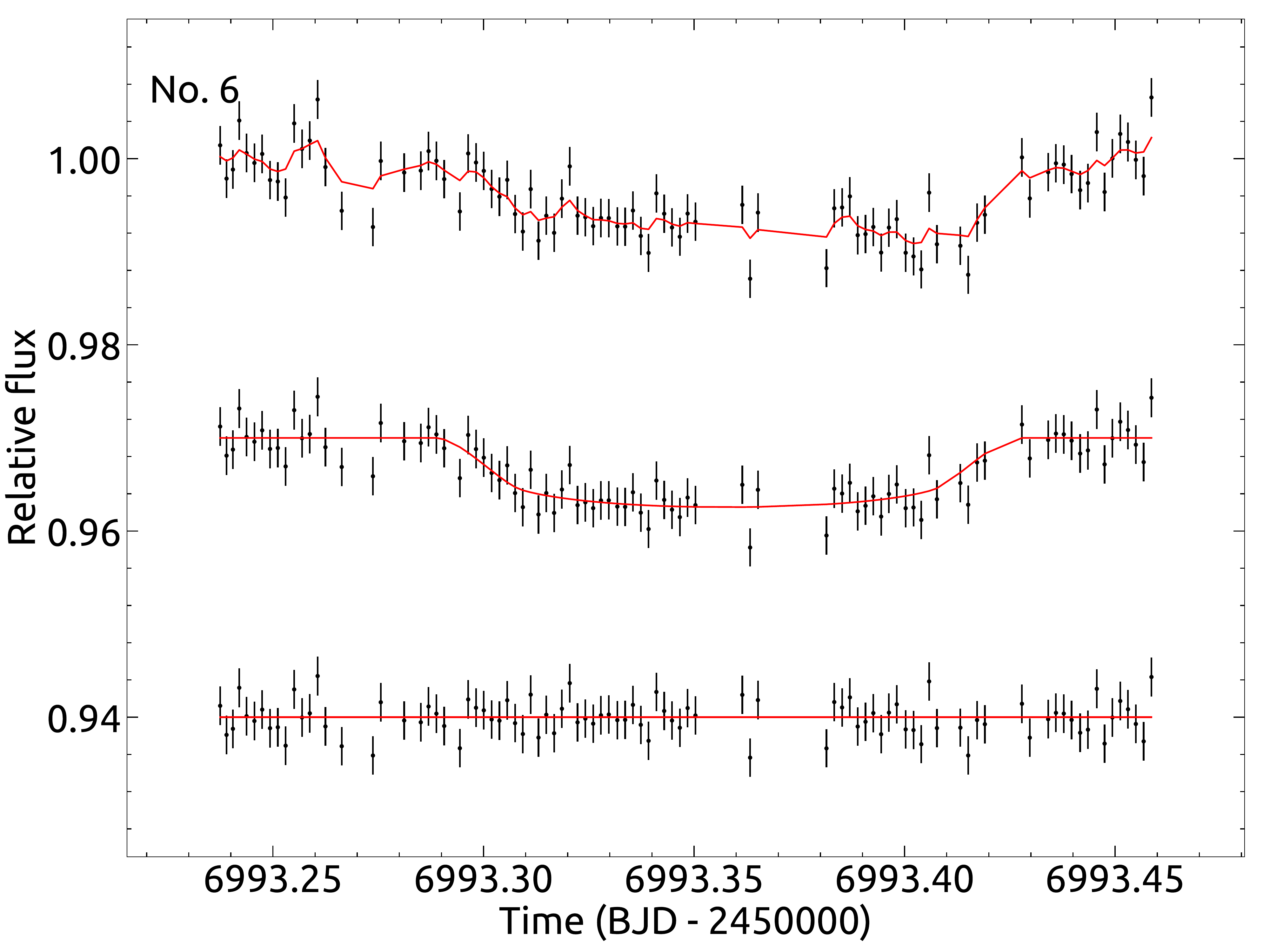}
  \end{minipage}%

   \begin{minipage}[t]{0.33\hsize}
  \centering
   \includegraphics[width=\hsize]{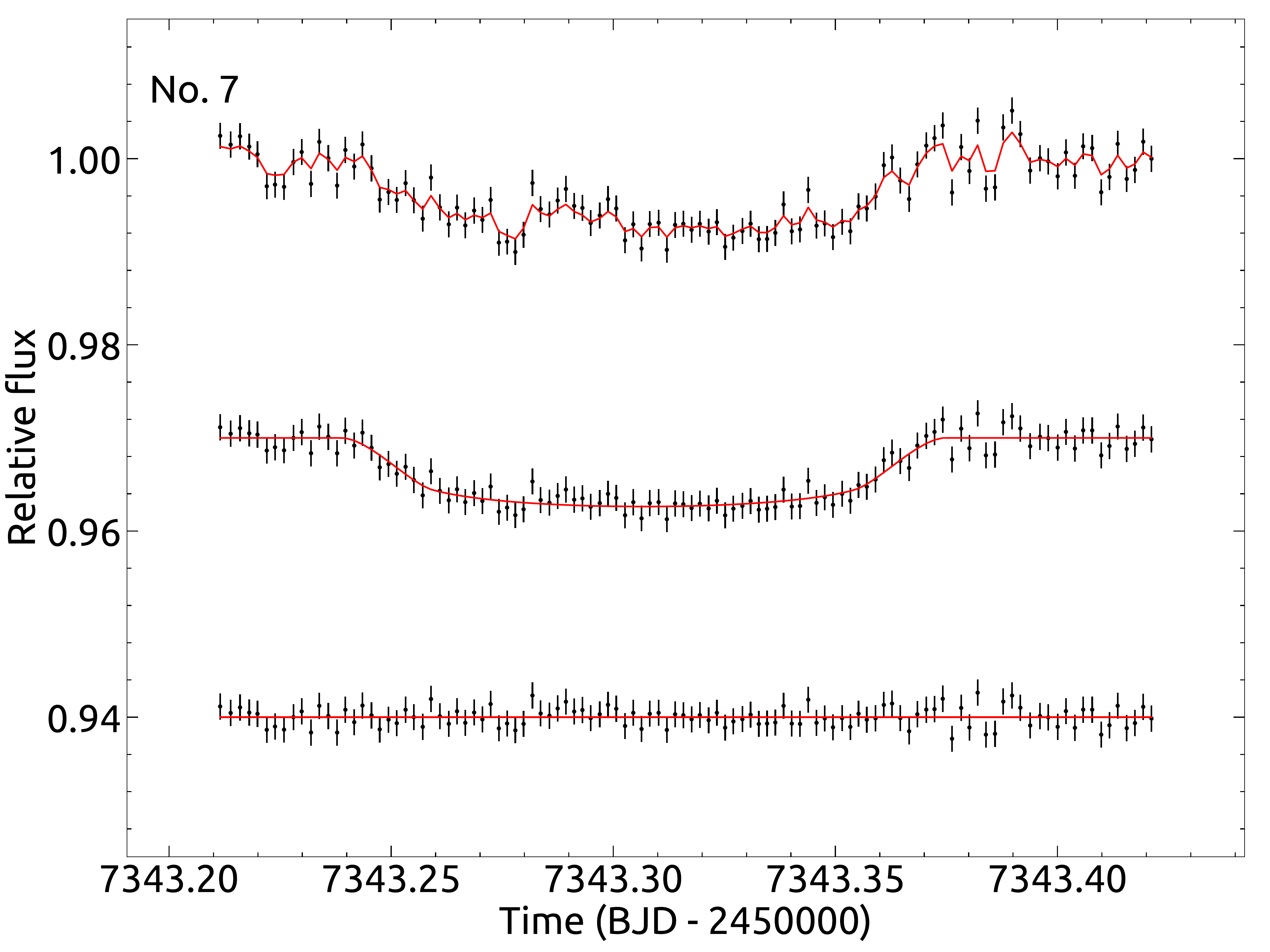}
  \end{minipage}%
  \begin{minipage}[t]{0.33\hsize}
  \centering
   \includegraphics[width=\hsize]{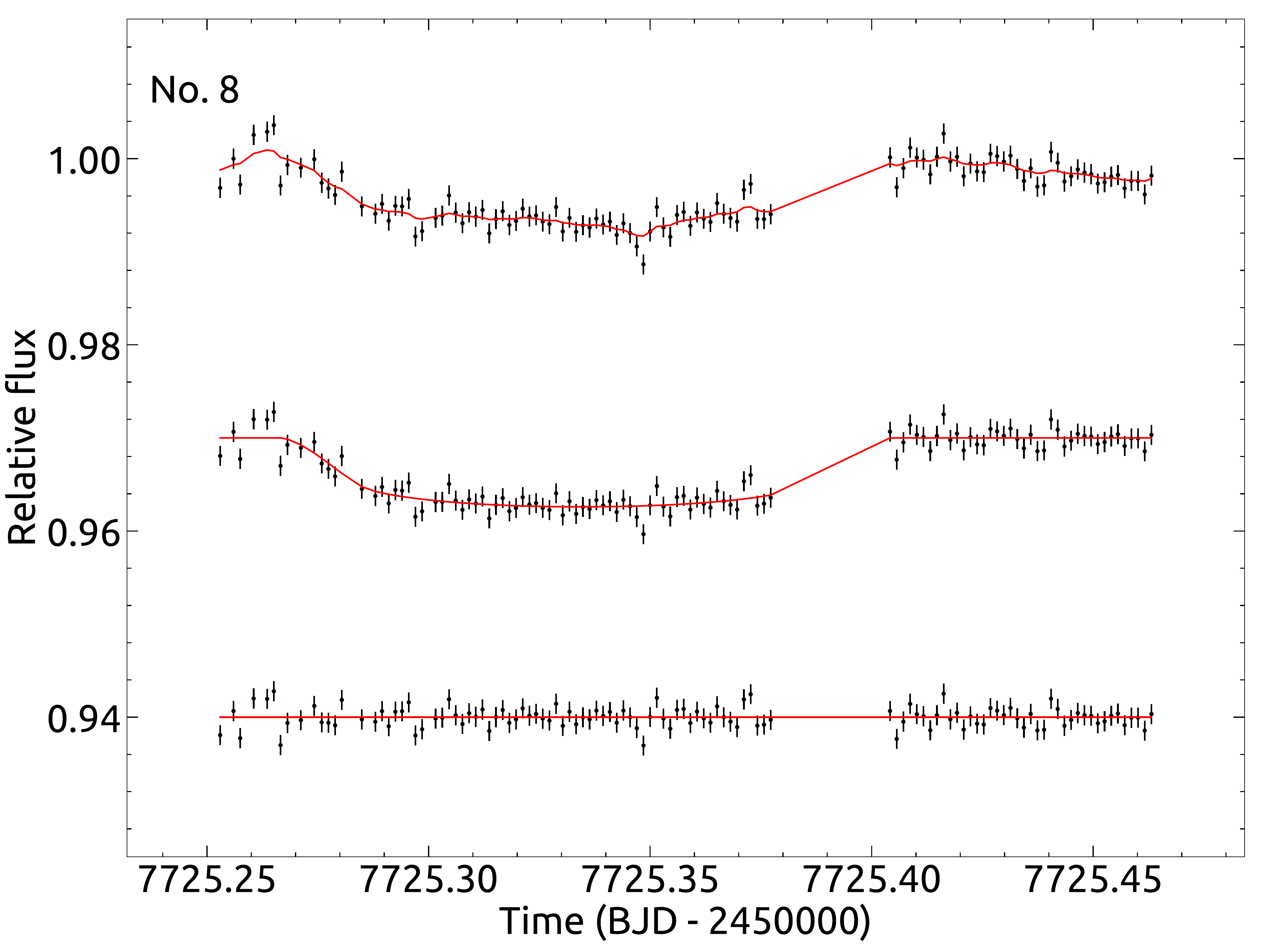}
  \end{minipage}%
  \begin{minipage}[t]{0.33\hsize}
  \centering
   \includegraphics[width=\hsize]{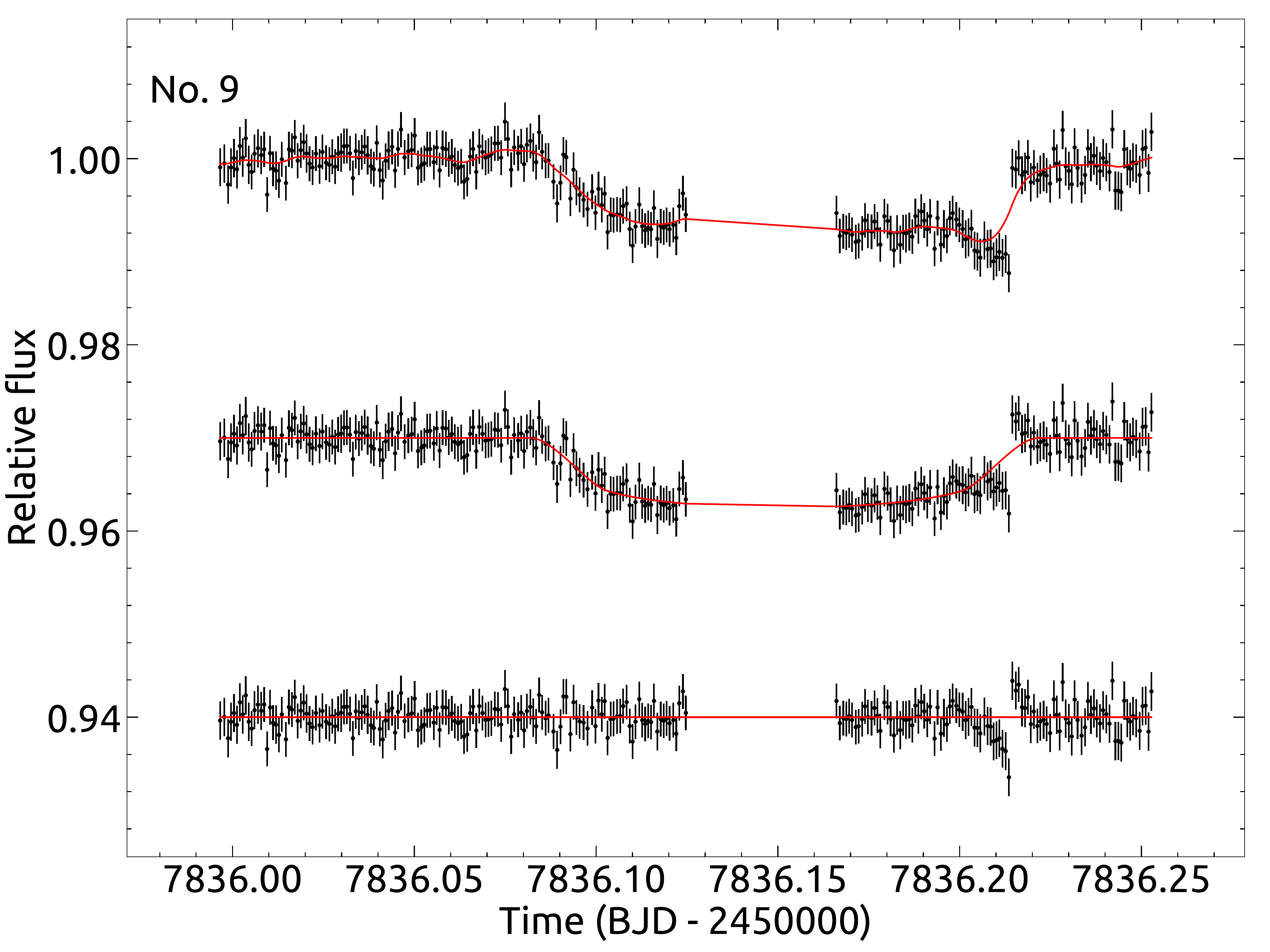}
  \end{minipage}%

  \begin{minipage}[t]{0.33\hsize}
  \centering
   \includegraphics[width=\hsize]{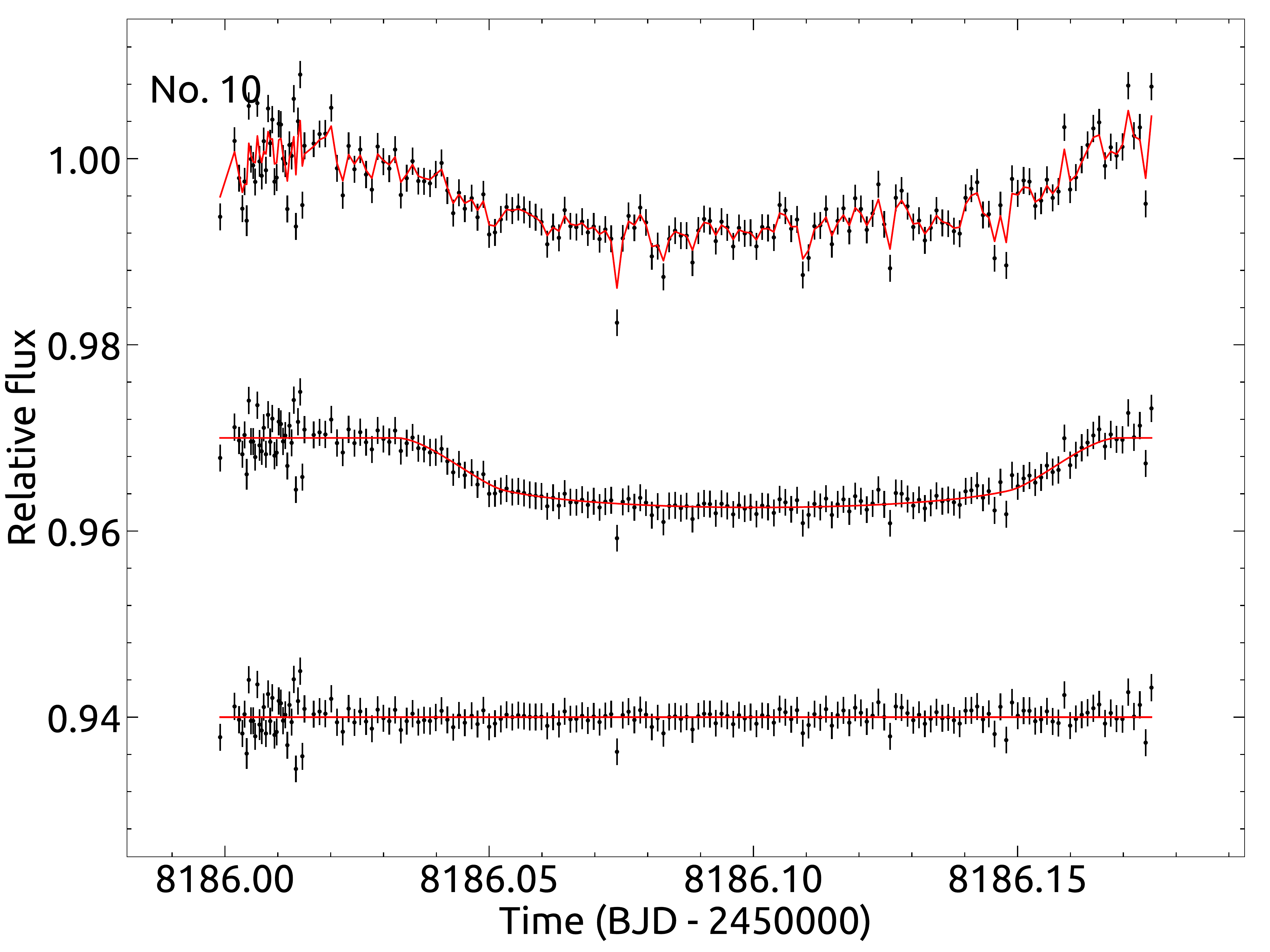}
  \end{minipage}%
  \begin{minipage}[t]{0.33\hsize}
  \centering
   \includegraphics[width=\hsize]{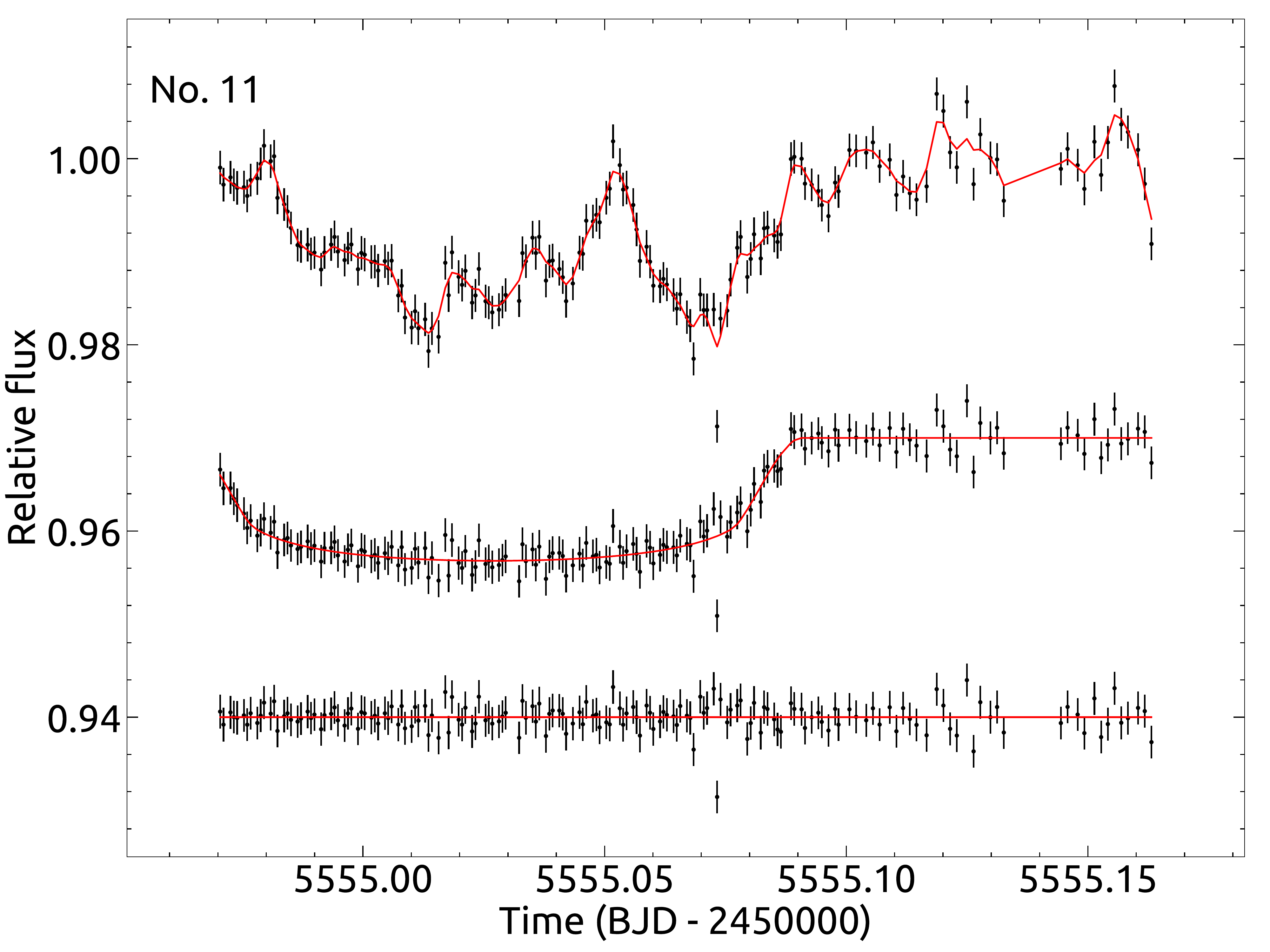}
  \end{minipage}%
  \begin{minipage}[t]{0.33\hsize}
  \centering
   \includegraphics[width=\hsize]{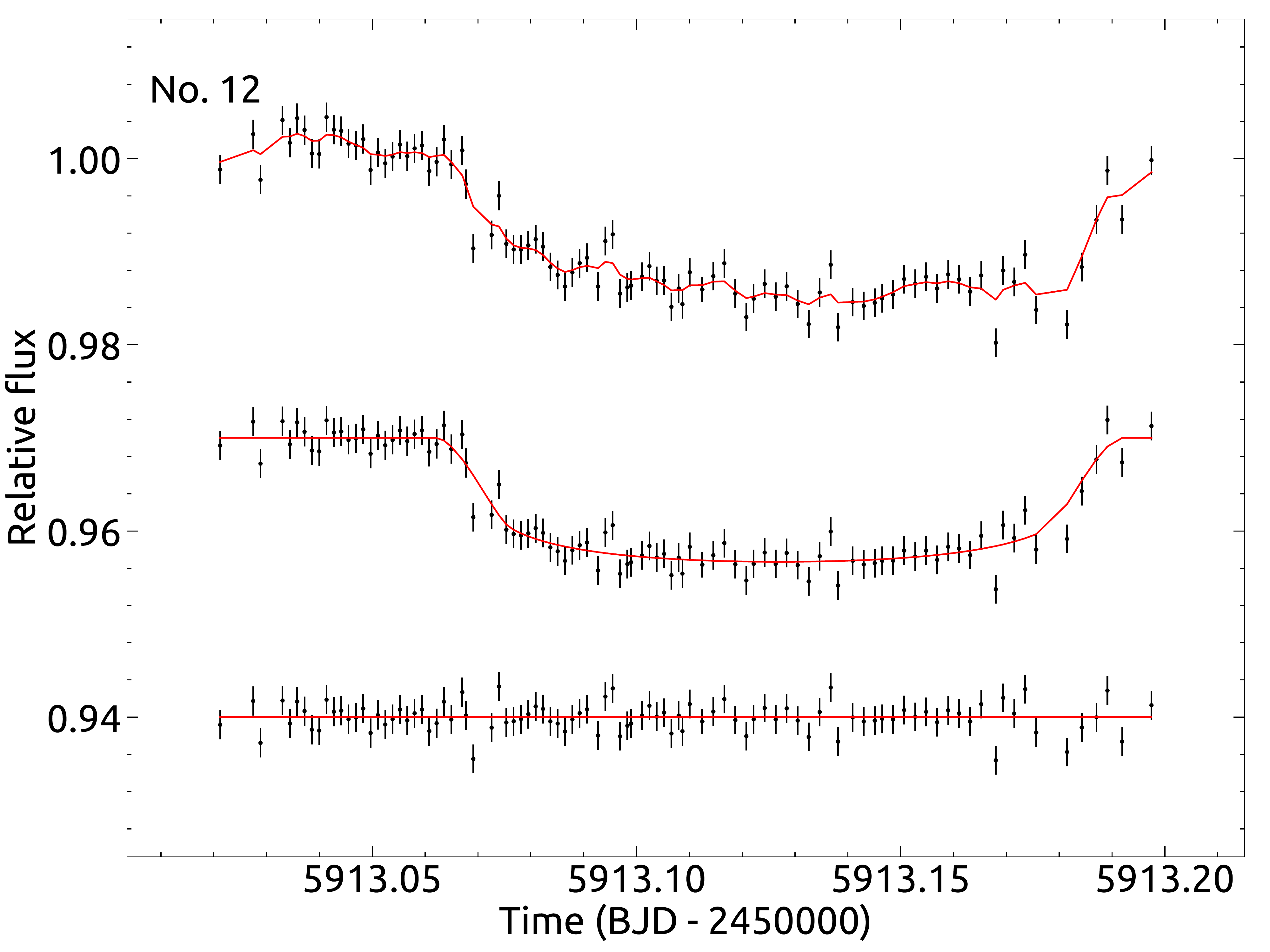}
   \end{minipage}%
   
  \begin{minipage}[t]{0.33\hsize}
  \centering
   \includegraphics[width=\hsize]{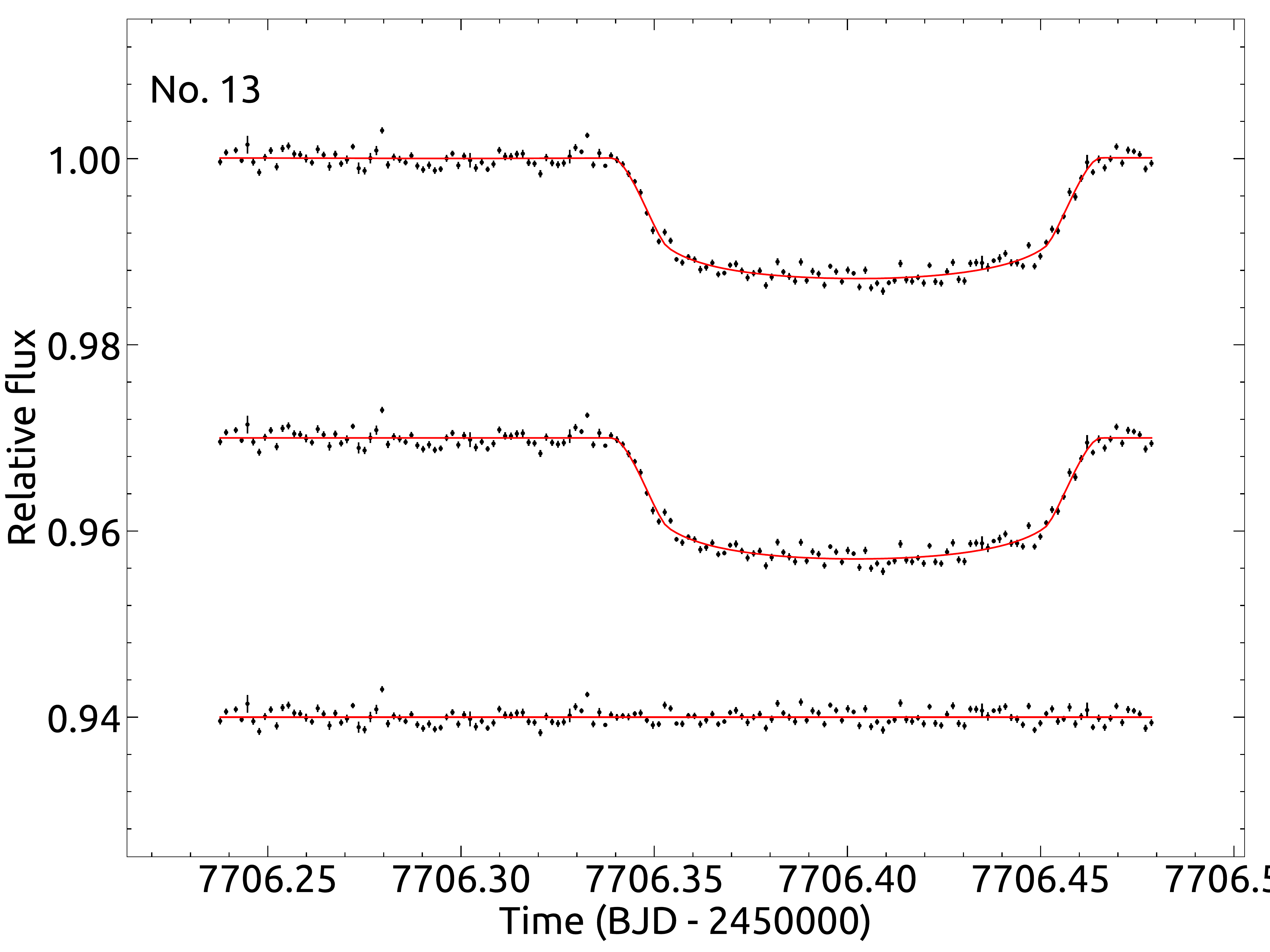}
  \end{minipage}%
   
  \caption{Transit light curves of HAT-P-13 (from No. 1 to 10) and HAT-P-16 (from No. 11 to 13) observed by ground-based telescopes. In each panel, the top is input light curves with the full median posterior models, the middle is the final light curves with the best-fitting deterministic transit models, and the bottom is the corresponding residuals. Vertical shifts are added for visualization.}
  \label{fig2}
\end{figure*}

\begin{figure*}
 \begin{minipage}[t]{0.33\hsize}
  \centering
   \includegraphics[width=\hsize]{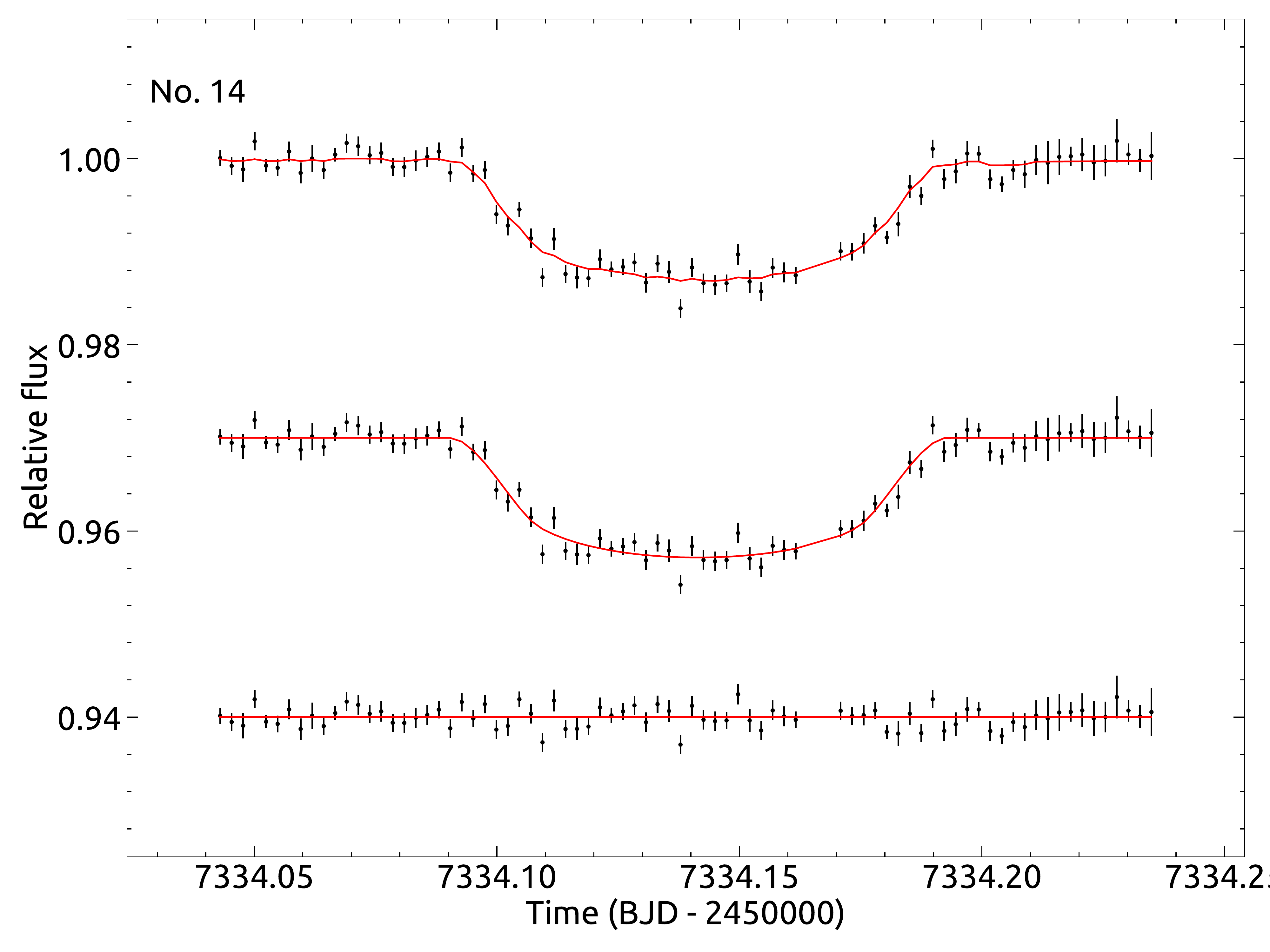}
  \end{minipage}%
  \begin{minipage}[t]{0.33\hsize}
  \centering
   \includegraphics[width=\hsize]{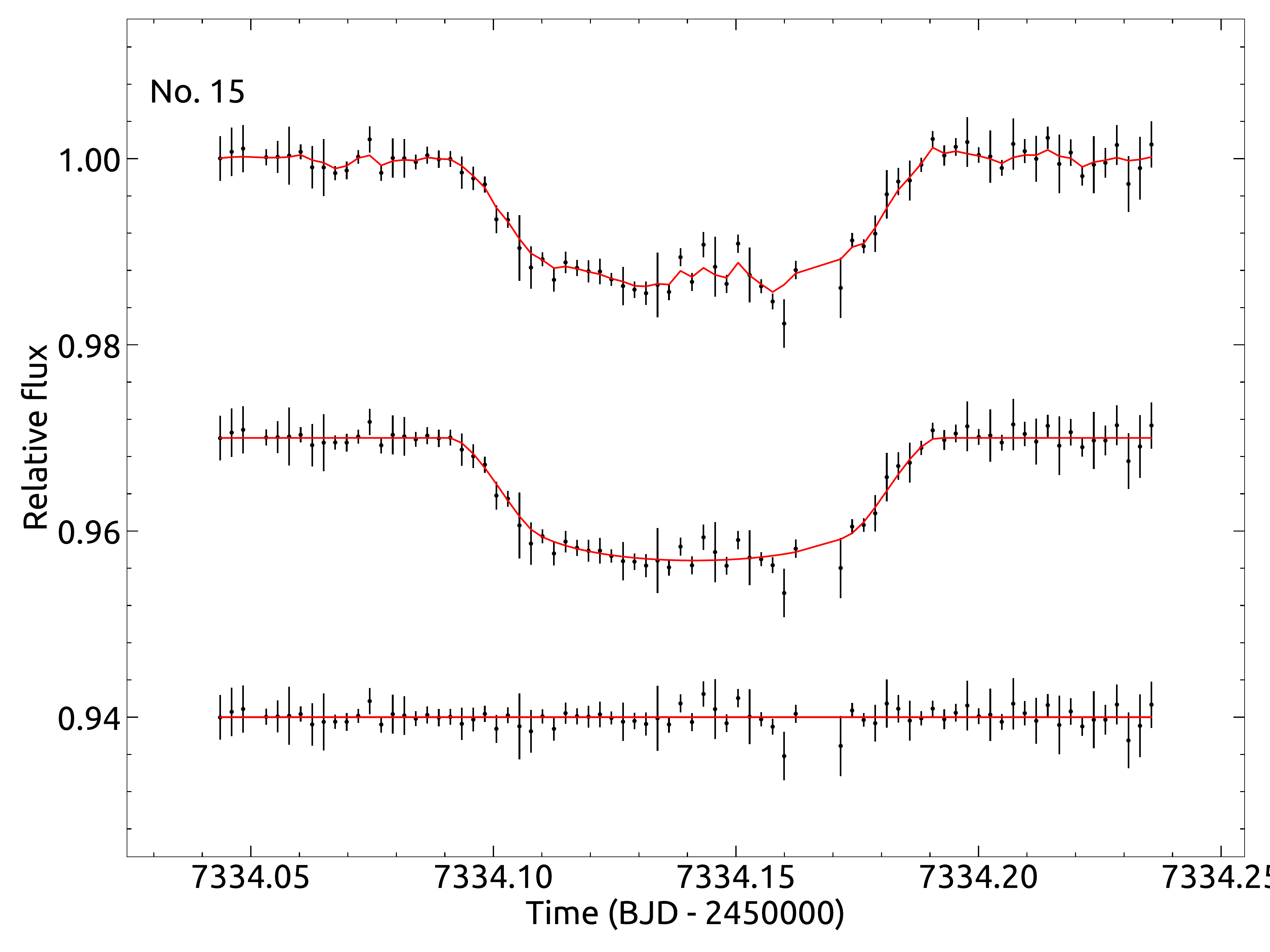}
   \end{minipage}%
   \begin{minipage}[t]{0.33\hsize}
  \centering
   \includegraphics[width=\hsize]{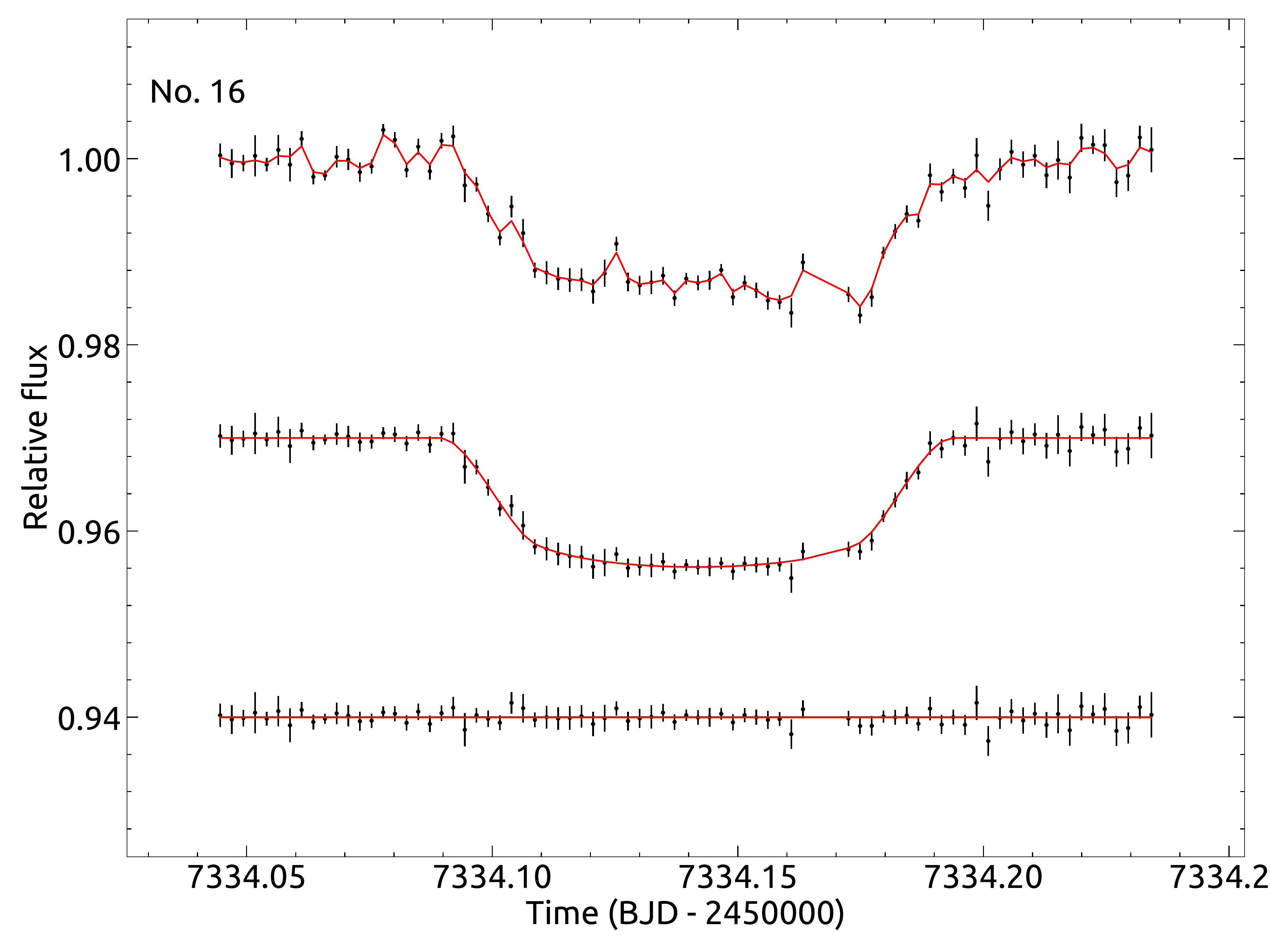}
  \end{minipage}%

  \begin{minipage}[t]{0.33\hsize}
  \centering
   \includegraphics[width=\hsize]{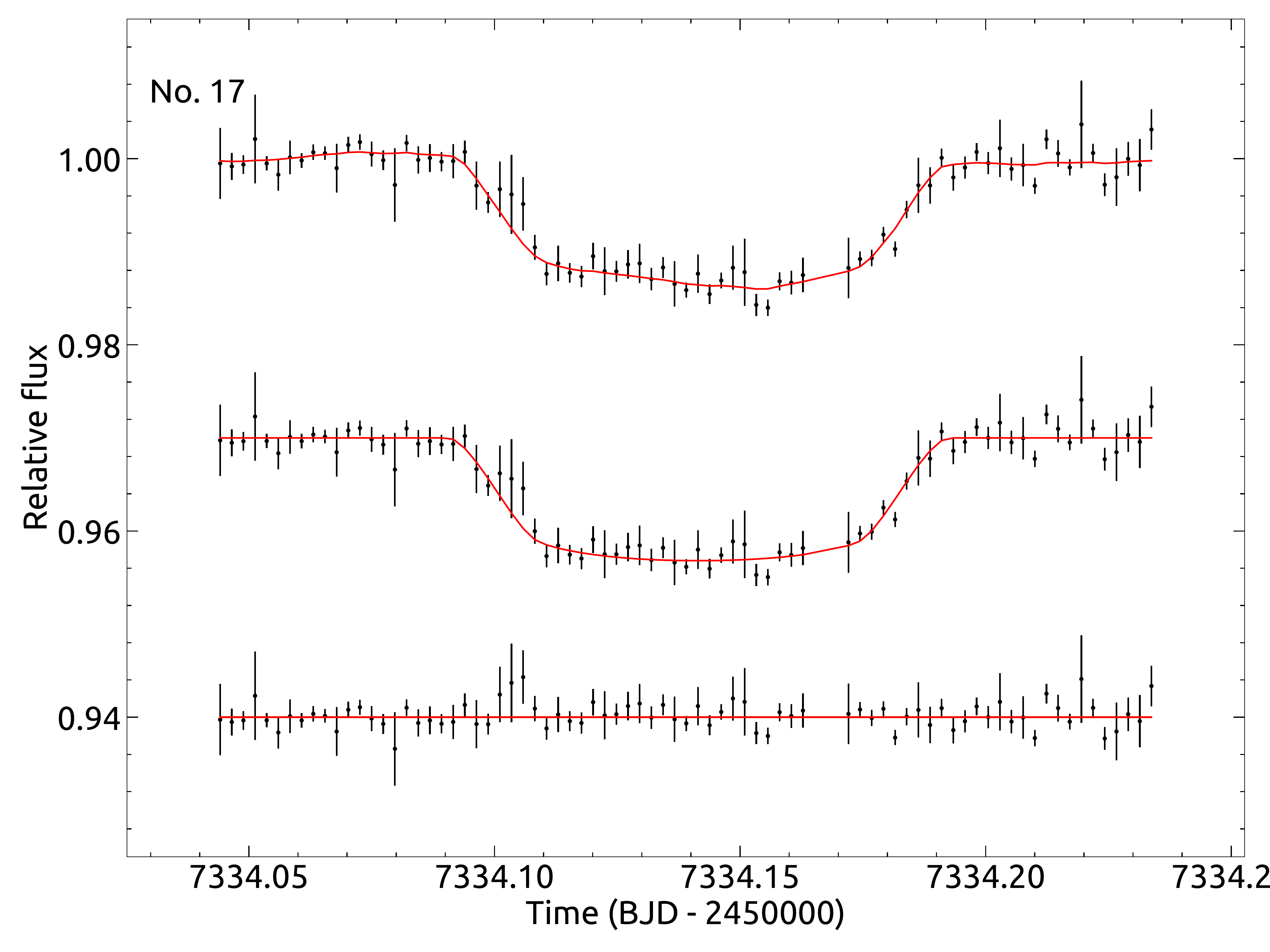}
  \end{minipage}%
  \begin{minipage}[t]{0.33\hsize}
  \centering
   \includegraphics[width=\hsize]{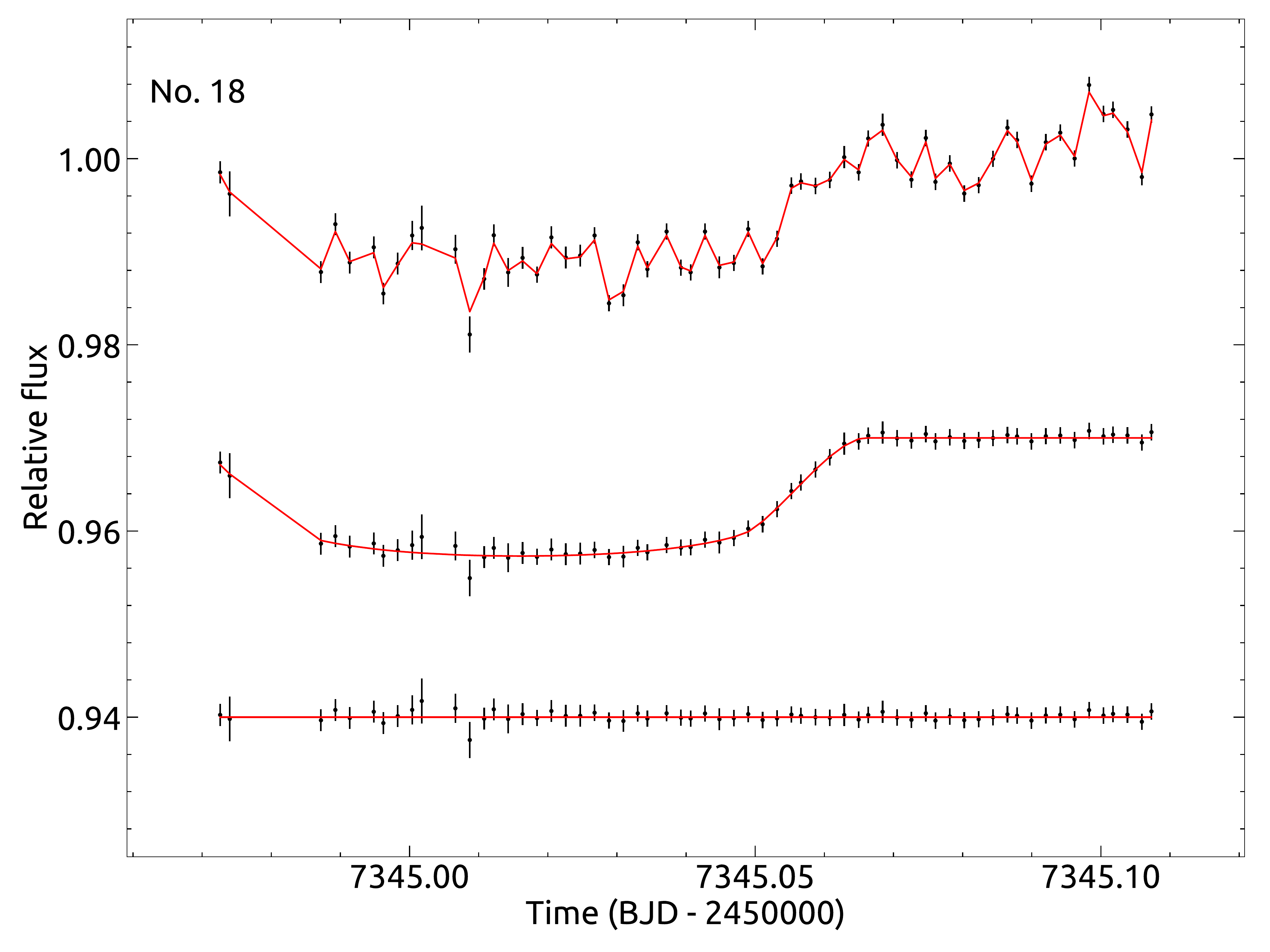}
   \end{minipage}%
   
  \caption{Transit light curves of WASP-32 (from No. 14 to 18) observed by ground-based telescopes. In each panel, the top is input light curves with the full median posterior models, the middle is the final light curves with the best-fitting deterministic transit models, and the bottom is the corresponding residuals. Vertical shifts are added for visualization.}
  \label{fig3a}
\end{figure*}

Finally, we corrected the systematic errors in photometric data through subtracting the optimal results of GP regression from the input light curves. The input light curves with the full median posterior models, the final light curves after the GP correction with the best-fitting deterministic models and the corresponding residuals are presented in Figures \ref{fig1}, \ref{fig2} and  \ref{fig3a}. 

\subsection{Final Transit Modelling }\label{sec:FTM}

We used the MCMC transit modeling code developed by \citet{RN533} to model the above final light curves simultaneously and derived the system parameters. The free parameters of MCMC code of \citet{RN533} are the orbital period of the transiting exoplanet $P$, the mid-time of one reference transit event $T_0$ , the impact parameter $b$, the transit duration $T_{14}$ , the transit depth $\Delta F$, the semi-amplitude of the radial velocity curve $K$, the orbital eccentricity $e$ and the argument of periastron $\omega$, respectively. We refer interest reader to \citet{RN533} for further details. The empirical calibration proposed by \citet{Enoch2010} were used to calculate the masses and radii of host stars, which is based on the effective temperature, metallicity and mean density of the host star. We obtained the four-coefficient limb-darkening coefficients through interpolating the coefficient tables of \citet{RN535} and \citet{RN536} based on stellar atmospheric parameters obtained by \citet{Bakos2009}, \citet{Buchhave2010} and \citet{Maxted2010}.  Circular orbits ($e=0$) were adopted in following analyses, because no significant non-zero eccentricities was found through fitting RV data in the literature.  

\begin{figure*}
   \begin{minipage}[t]{0.45\hsize}
  \centering
   \includegraphics[width=\hsize]{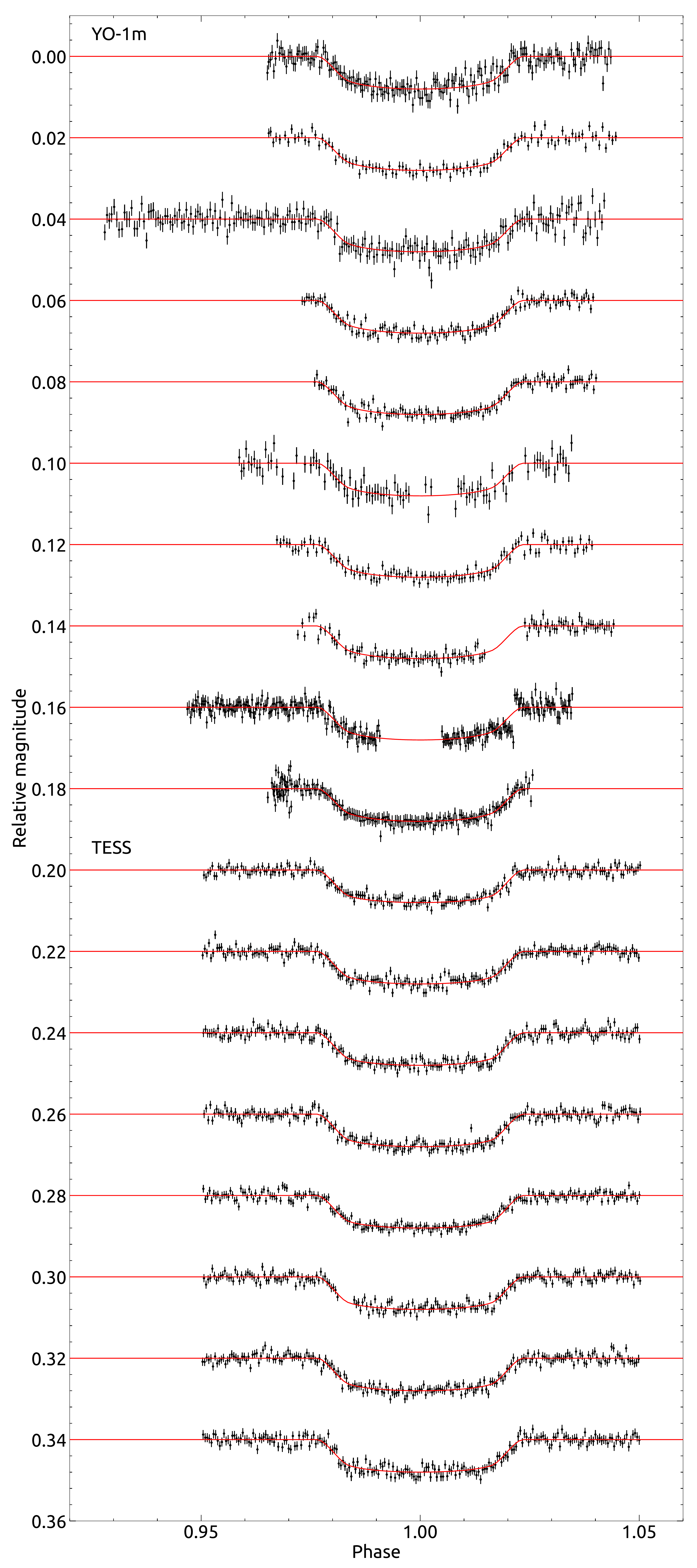}
  \end{minipage}%
  \begin{minipage}[t]{0.45\hsize}
  \centering
   \includegraphics[width=\hsize]{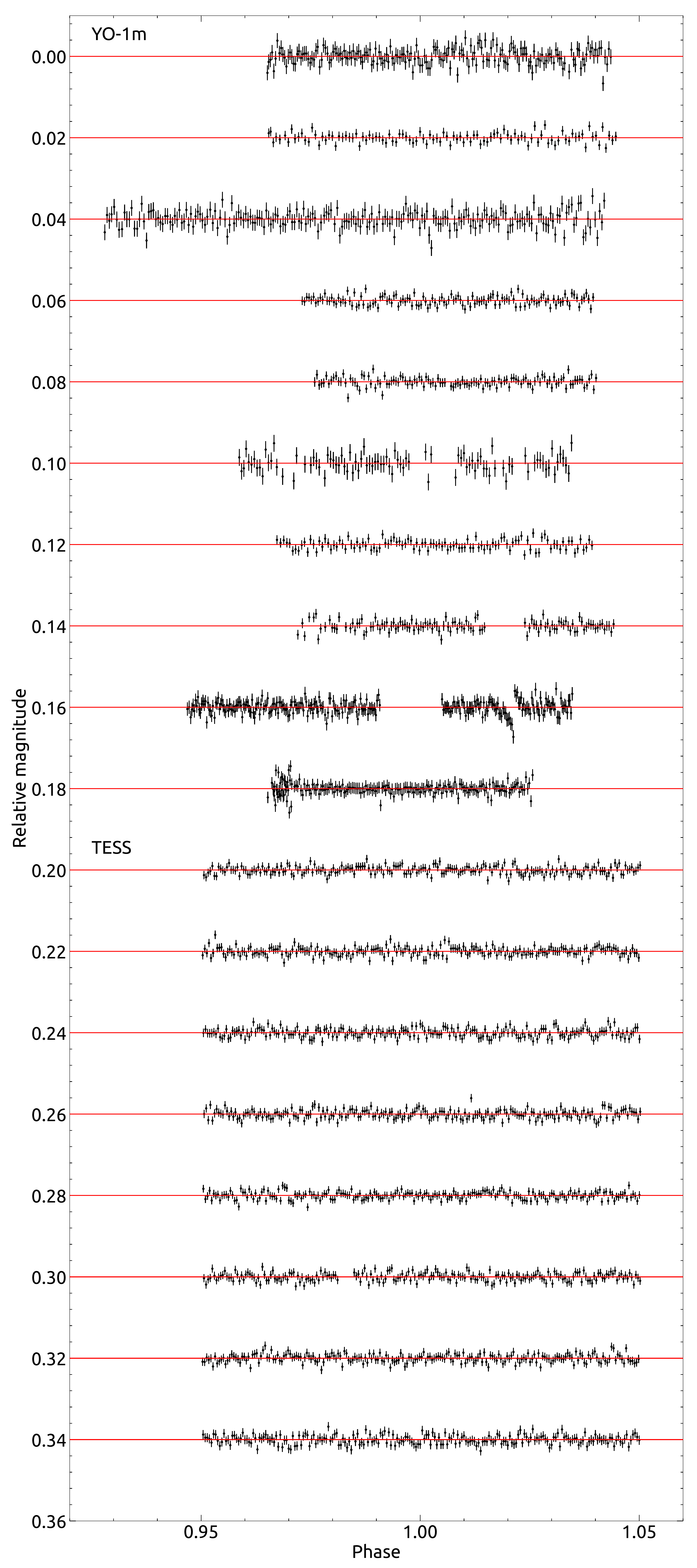}
  \end{minipage}%
      \caption{The final transit light curves of HAT-P-13 with the best-fitting transit models and the residuals.}
         \label{fig3}
   \end{figure*}

\begin{figure*}
    \begin{minipage}[t]{0.45\hsize}
  \centering
   \includegraphics[width=\hsize]{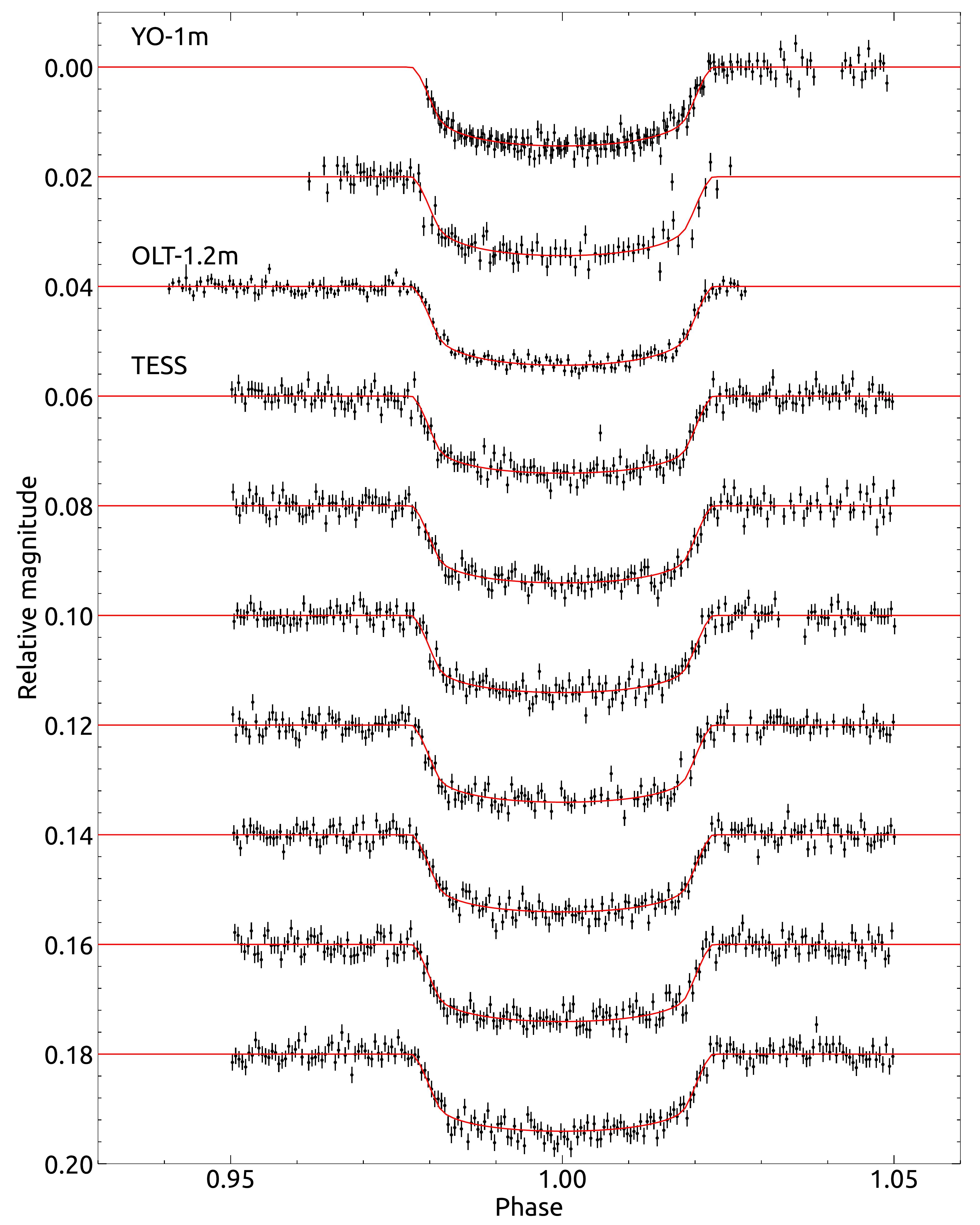}
  \end{minipage}%
  \begin{minipage}[t]{0.45\hsize}
  \centering
   \includegraphics[width=\hsize]{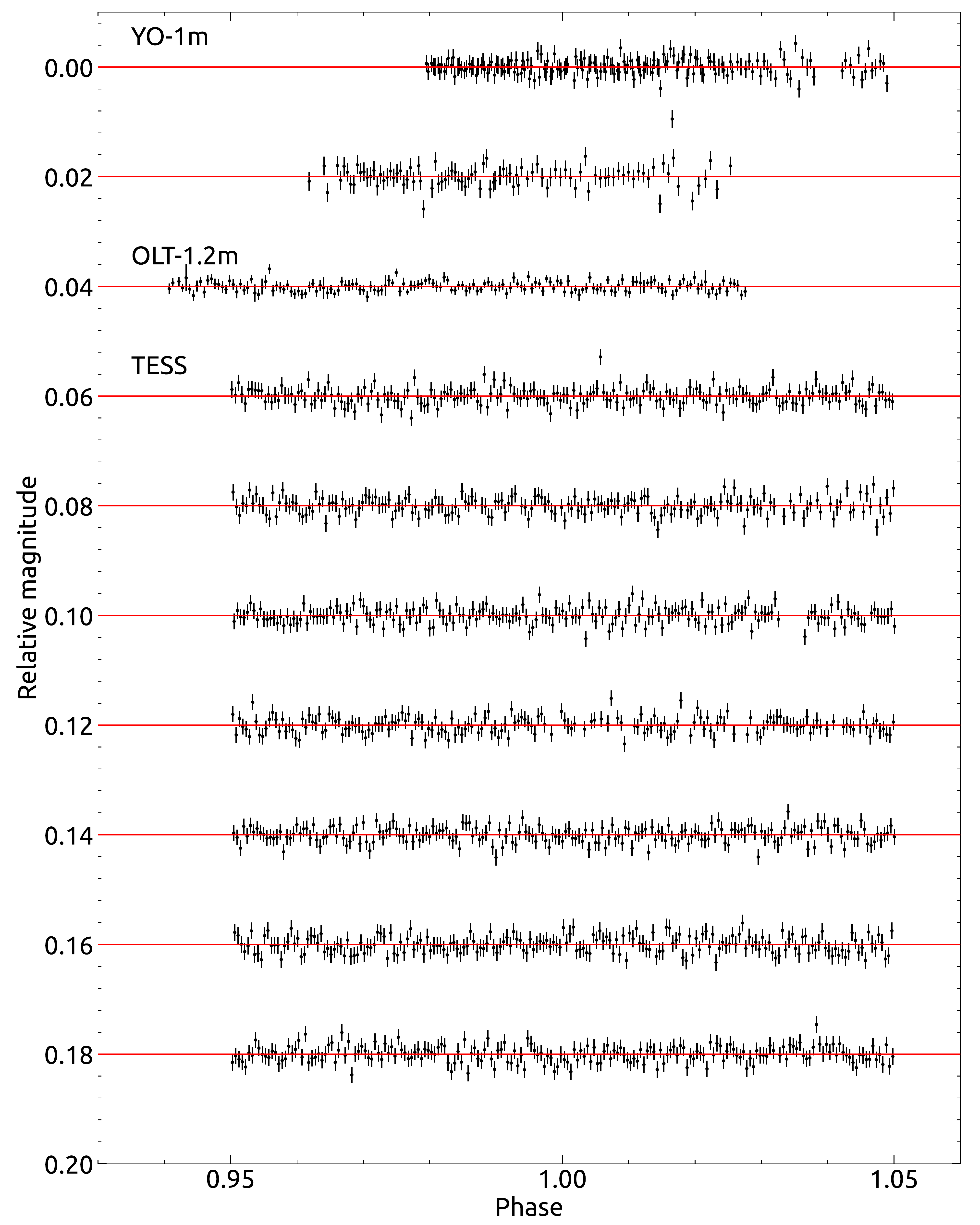}
  \end{minipage}%
      \caption{The final transit light curves of HAT-P-16 with the best-fitting transit models and the residuals.}
         \label{fig4}
   \end{figure*}

\begin{figure*}
    \begin{minipage}[t]{0.45\hsize}
  \centering
   \includegraphics[width=\hsize]{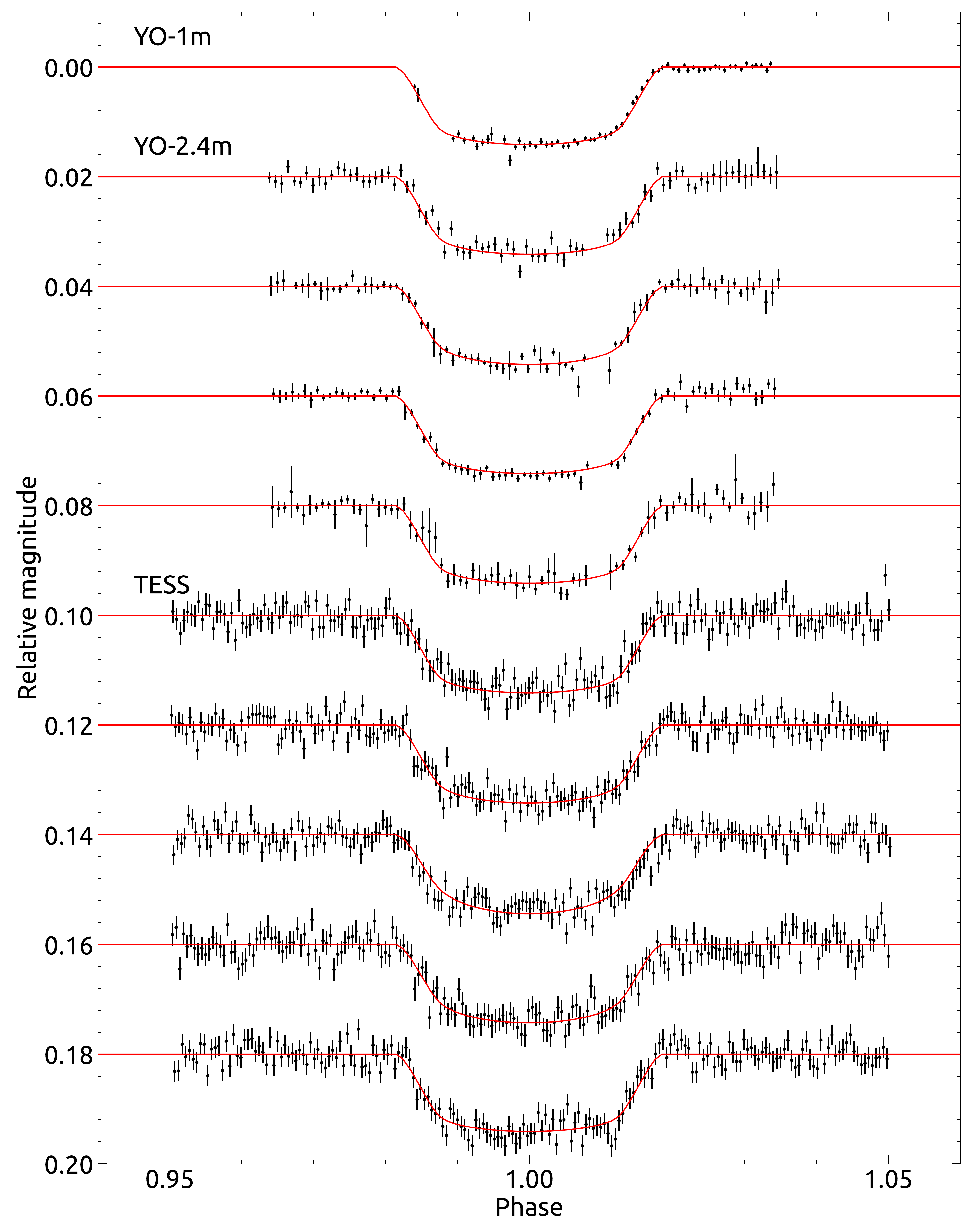}
  \end{minipage}%
  \begin{minipage}[t]{0.45\hsize}
  \centering
   \includegraphics[width=\hsize]{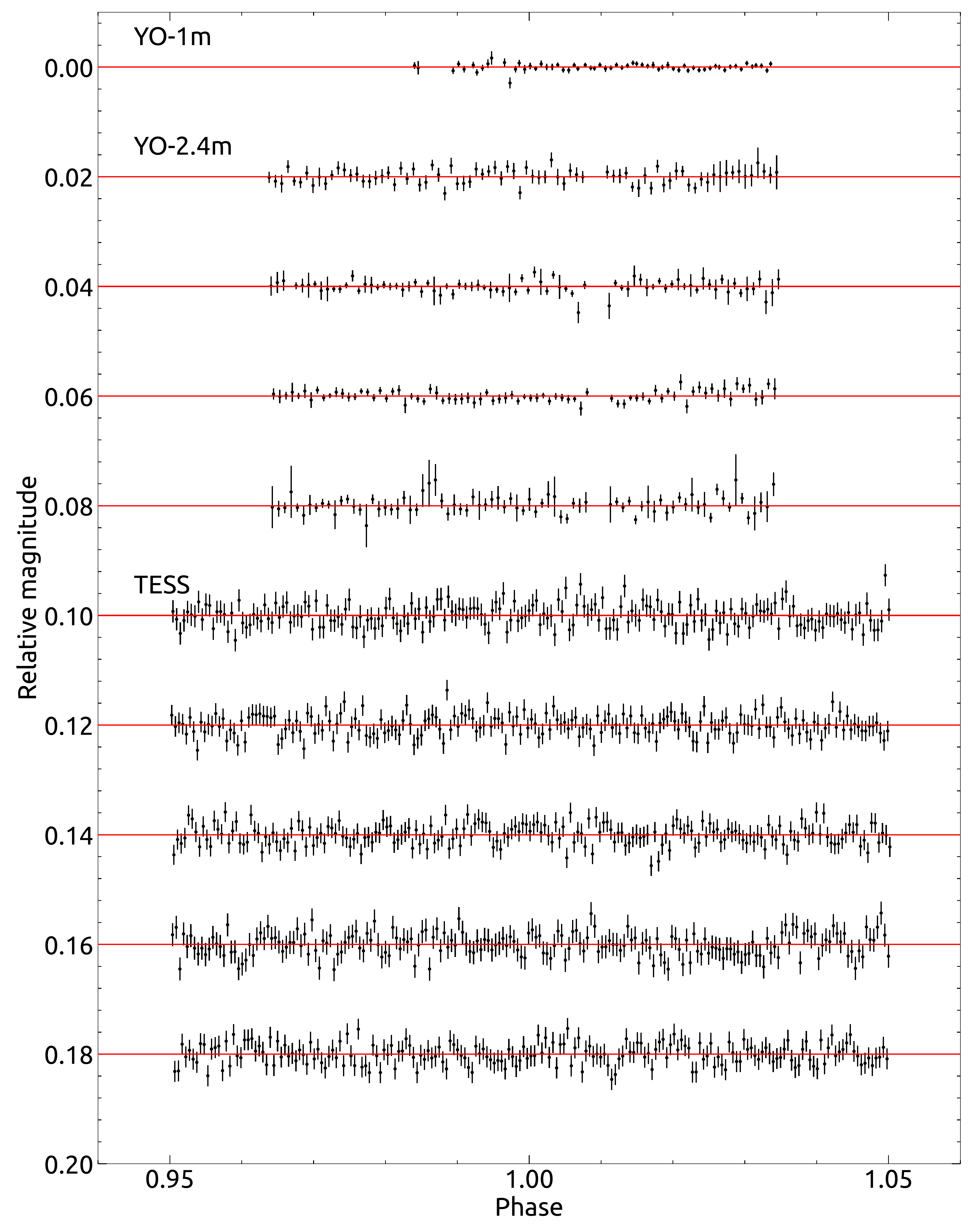}
  \end{minipage}%
      \caption{The final transit light curves of WASP-32 with the best-fitting transit models and the residuals.}
         \label{fig5}
   \end{figure*}

We calculated the final system parameters of each system through modeling all of the light curves simultaneously for each system by employing the code of \citet{RN533}. The refined orbital periods derived in Section \ref{sec:Juliet}were fixed in the global modeling. We ran 5 chains of 17,000 MCMC steps with 2,000 burn-in steps as in \citet{RN338}. We repeated this process for extra ten times to check the convergence of the MCMC sampling, then we obtained consistent results. The final results of the system parameters for each system are listed in Table \ref{tab5}, and the results of discovery papers are also listed there for comparison. Our results are consistent with theirs and have significant improvements. All light curves with best-fitting models and the residuals are shown in Figures \ref{fig3}, \ref{fig4} and \ref{fig5}.

 \begin{table*}
\caption{The transit and occultation times of HAT-P-13b.}
\label{tab2}
\centering
\begin{threeparttable}
\begin{tabular}{c c c c c c c ccccccccccccc }
\hline
\hline
Event & Mid-time & Error & Cycle & Source         &Event & Mid-time & Error & Cycle & Source\\
& ($\mathrm{BJD_{TDB}}$ & (days) & & & & ($\mathrm{BJD_{TDB}}$) & (days) & &\\
&    $\mathrm{-2450000}$) &          & & & & $\mathrm{-2450000}$) & & &\\
\hline
tra&4581.62443 &0.00122 &-68&\citet{Bakos2009}&	tra&5602.31068 &0.00167 &282&\citet{Nascimbeni2011}\\	
tra&4777.01324 &0.00100 &-1&\citet{Bakos2009}&	tra&5613.97390 &0.00225 &286&\citet{RN767}\\	
tra&4779.92990 &0.00063 &0&\citet{Bakos2009}&	tra&5616.89290 &0.00152 &287&\citet{RN767}\\	
tra&4782.84394 &0.00155 &1&\citet{Bakos2009}&	tra&5619.80786 &0.00134 &288&\citet{RN767}\\	
tra&4849.92099 &0.00075 &24&\citet{Bakos2009}&	tra&5622.72289 &0.00120 &289&\citet{Sada2016}\\	
tra&4882.00078 &0.00150 &35&\citet{Bakos2009}&	tra&5622.72351 &0.00166 &289&\citet{RN767}\\	
tra&4960.74005 &0.00178 &62&\citet{Bakos2009}&	tra&5669.38140 &0.00126 &305&\citet{Southworth2012}\\	
tra&5167.79647 &0.00280 &133&\citet{Southworth2012}&	tra&5934.76202 &0.00155 &396&\citet{Sada2016}\\	
tra&5194.03566 &0.00229 &142&\citet{RN767}&	tra&5978.50418 &0.00097 &411&ETD\\	
tra&5196.95450 &0.00127 &143&\citet{RN767}&	tra&6299.29216 &0.00113 &521&This work\\	
tra&5199.86837 &0.00123 &144&\citet{Southworth2012}&	tra&6316.79247 &0.00117 &527&\citet{Sada2016}\\	
tra&5199.86867 &0.00131 &144&\citet{RN767}&	tra&6351.78519 &0.00211 &539&\citet{Sada2016}\\	
tra&5231.94542 &0.00091 &155&\citet{RN767}&	tra&6354.69740 &0.00140 &540&\citet{Turner2016}\\	
tra&5240.69554 &0.00197 &158&\citet{Sada2016}&	tra&6354.70220 &0.00112 &540&\citet{Sada2016}\\	
tra&5249.45117 &0.00200 &161&\citet{Szabo2010}&	tra&6646.32391 &0.00105 &640&This work\\	
tra&5269.86567 &0.00180 &168&\citet{Southworth2012}&	tra&6681.32041 &0.00127 &652&This work\\	
tra&5272.77577 &0.00120 &169&\citet{Southworth2012}&	tra&6684.23580 &0.00074 &653&This work\\	
tra&5272.77627 &0.00250 &169&\citet{Southworth2012}&	tra&6701.73623 &0.00139 &659&\citet{Sada2016}\\	
tra&5275.69207 &0.00180 &170&\citet{Southworth2012}&	tra&6725.06439 &0.00101 &667&This work\\	
tra&5275.69312 &0.00266 &170&\citet{RN767}&	tra&6993.36034 &0.00182 &759&This work\\	
tra&5307.77077 &0.00370 &181&\citet{Southworth2012}&	tra&7063.35170 &0.00125 &783&ETD\\	
tra&5310.69197 &0.00250 &182&\citet{Southworth2012}&	tra&7334.55789 &0.00085 &876&ETD\\	
occ&5326.70818 &0.00406 &187&\citet{Hardy2017}&tra&7343.30634 &0.00088 &879&This work\\	
occ&5355.87672 &0.00226 &197&\citet{Hardy2017}&	tra&7725.33461 &0.00292 &1010&This work\\	
tra&5511.90854 &0.00141 &251&\citet{RN767}&	tra&7754.49545 &0.00084 &1020&ETD\\	
tra&5558.56302 &0.00098 &267&\citet{pal2011}&	tra&7827.40114 &0.00168 &1045&ETD\\	
tra&5561.48416 &0.00400 &268&\citet{pal2011}&	tra&7836.15281 &0.00066 &1048&This work\\	
tra&5564.39876 &0.00180 &269&\citet{Nascimbeni2011}&	tra&8186.10079 &0.00101 &1168&This work\\	
tra&5584.81245 &0.00118 &276&\citet{Sada2016}&	tra&8865.58696 &0.00076 &1401&ETD\\	
tra&5584.81455 &0.00153 &276&\citet{Southworth2012}&	tra&9582.97959	&0.00068 &1647&This work\\
tra&5587.73154 &0.00151 &277&\citet{Sada2016}&	tra&9585.89570	&0.00064 &1648&This work\\
tra&5590.64523 &0.00179 &278&\citet{pal2011}&tra&9588.81230	&0.00063 &1649&This work\\
tra&5593.55879 &0.00185 &279&\citet{Southworth2012}&	tra&9591.72847	&0.00064 &1650&This work\\
tra&5593.56147 &0.00115 &279&\citet{Nascimbeni2011}&	tra&9597.56058	&0.00067 &1652&This work\\
tra&5596.47291 &0.00140 &280&\citet{Southworth2012}&	tra&9600.47816	&0.00067 &1653&This work\\
tra&5596.47327 &0.00202 &280&\citet{Southworth2012}&	tra&9603.39419	&0.00064 &1654&This work\\
tra&5596.47662 &0.00305 &280&\citet{Nascimbeni2011}&	tra&9606.31091	&0.00061 &1655&This work\\
tra&5599.39267 &0.00075 &281&\citet{Nascimbeni2011}&	tra&9253.44747  &0.00118 &1534&ETD\\	
tra&5599.39446 &0.00100 &281&\citet{Southworth2012}&      &                 &            &       &  \\
\hline  
\hline 
\end{tabular}
\end{threeparttable}
\end{table*}

 \begin{table*}
\caption{The transit times of HAT-P-16b.}
\label{tab3}
\centering
\begin{threeparttable}
\begin{tabular}{c c c c c c c ccccccccccccc }
\hline
\hline
Event & Mid-time & Error & Cycle & Source         &Event & Mid-time & Error & Cycle & Source\\
& ($\mathrm{BJD_{TDB}}$ & (days) & & & & ($\mathrm{BJD_{TDB}}$ & (days) & &\\
&    $\mathrm{-2450000}$) &          & & & & $\mathrm{-2450000}$) &  & &\\
\hline
tra&5027.59293&0.00031&0&\citet{Ciceri2013}&	tra&7642.55557&0.00062&942&ETD\\											
tra&5085.88780&0.00049&21&\citet{Ciceri2013}&	tra&7692.52421&0.00092&960&ETD\\											
tra&5085.88864&0.00006&21&\citet{Wang2021}&	tra&7706.40229&0.00032&965&This work\\											
tra&5096.99125&0.00009&25&\citet{Wang2021}& tra&7706.40419&0.00087&965&ETD\\											
tra&5124.75086&0.00009&35&\citet{Wang2021}&	tra&7773.02393&0.00014&989&\citet{Wang2021}\\											
tra&5135.85362&0.00050&39&\citet{Ciceri2013}&	tra&8036.74082&0.00068&1084&ETD\\											
tra&5135.85449&0.00006&39&\citet{Wang2021}&tra&8350.42812&0.00062&1197&ETD\\											
tra&5463.41931&0.00080&157&ETD&	tra&8375.41142&0.00063&1206&ETD\\											
tra&5463.42067&0.00049&157&ETD&	tra&8475.34489&0.00082&1242&ETD\\											
tra&5471.74748&0.00047&160&\citet{Sada2016}&tra&8733.51150&0.00073&1335&ETD\\											
tra&5482.85087&0.00066&164&ETD&	tra&8747.38887&0.00074&1340&ETD\\											
tra&5485.62913&0.00050&165&ETD& tra&8766.82141&0.00041&1347&This work\\											
tra&5499.50837&0.00019&170&\citet{Ciceri2013}&	tra&8769.59792&0.00046&1348&This work\\										
tra&5555.02710&0.00063&190&This work&	tra&8772.37361&0.00043&1349&This work\\											
tra&5796.53707&0.00034&277&\citet{Ciceri2013}&	tra&8777.92488&0.00046&1351&This work\\										
tra&5829.84931&0.00059&289&ETD&	tra&8780.70199&0.00040&1352&This work\\											
tra&5835.40206&0.00091&291&ETD&	tra&8780.70211&0.00068&1352&ETD\\											
tra&5843.72852&0.00081&294&ETD&	tra&8783.47734&0.00041&1353&This work\\											
tra&5893.69673&0.00065&312&\citet{Sada2016}&tra&8786.25400&0.00042&1354&This work\\											
tra&5913.12765&0.00087&319&This work&	tra&8797.35862&0.00103&1358&\citet{Aladag2021}\\										
tra&6190.72516&0.00059&419&\citet{Sada2016}&	tra&8811.24399&0.00147&1363&\citet{Aladag2021}\\								
tra&6204.60421&0.00032&424&\citet{Ciceri2013}&	tra&8822.34086&0.00055&1367&ETD\\											
tra&6204.60451&0.00030&424&\citet{Ciceri2013}&	tra&8836.22609&0.00123&1372&\citet{Aladag2021}\\							
tra&6226.81172&0.00057&432&ETD&	tra&9055.52400&0.00069&1451&ETD\\											
tra&6540.49484&0.00054&545&ETD&	tra&9105.49026&0.00049&1469&ETD\\											
tra&6573.80636&0.00057&557&\citet{Sada2016}&	tra&9130.47387&0.00066&1478&ETD\\											
tra&6598.79110&0.00060&566&\citet{Turner2016}&	tra&9130.47461&0.00052&1478&\citet{Aladag2021}\\							
tra&6598.79150&0.00079&566&ETD&	tra&9144.34689&0.00076&1483&\citet{Aladag2021}\\											
tra&6601.56852&0.00066&567&ETD&	tra&9155.45802&0.00062&1487&ETD\\											
tra&6604.34400&0.00084&568&ETD&	tra&9491.34883&0.00073&1608&ETD\\											
tra&7259.47015&0.00099&804&ETD&	tra&9541.31793&0.00044&1626&ETD\\											
tra&7334.42244&0.00078&831&ETD&	tra&9552.42074&0.00060&1630&ETD\\											
tra&7345.52642&0.00095&835&ETD&	tra&9566.29983&0.00072&1635&ETD\\
\hline  
\hline 
\end{tabular}
\end{threeparttable}
\end{table*}

\begin{table}
\caption{The transit times of WASP-32.}             
\label{tab4}      
\centering          
\begin{tabular}{c c c c c}  
\hline\hline       
Event & Mid-time & Error & Cycle & Source \\
& ($\mathrm{BJD_{TDB}-2450000}$) & (days) & &  \\
\hline
tra&5496.32648&0.00101&-104&ETD\\
tra&5507.19973&0.00074&-100&\citet{Sun2015}\\
tra&5803.53497&0.00072&9&ETD\\
tra&5803.53506&0.00076&9&ETD\\
tra&5849.75088&0.00069&26&\citet{Sada2012}\\
tra&6252.11417&0.00139&174&\citet{Sun2015}\\
tra&7331.42313&0.00033&571&ETD\\
tra&7334.14167&0.00019&572&This work\\
tra&7345.01634&0.00069&576&This work\\
tra&7619.60135&0.00097&677&ETD\\
tra&7668.53409&0.00134&695&ETD\\
tra&8731.53383&0.00074&1086&ETD\\
tra&9449.25924&0.00061&1350&This work\\
tra&9451.97865&0.00059&1351&This work\\
tra&9462.85368&0.00061&1355&This work\\
tra&9465.57222&0.00071&1356&This work\\
tra&9468.28993&0.00058&1357&This work\\
\hline  
\hline 
\end{tabular}
\end{table}

\begin{figure*}
   \centering
   \includegraphics[width=0.8\hsize]{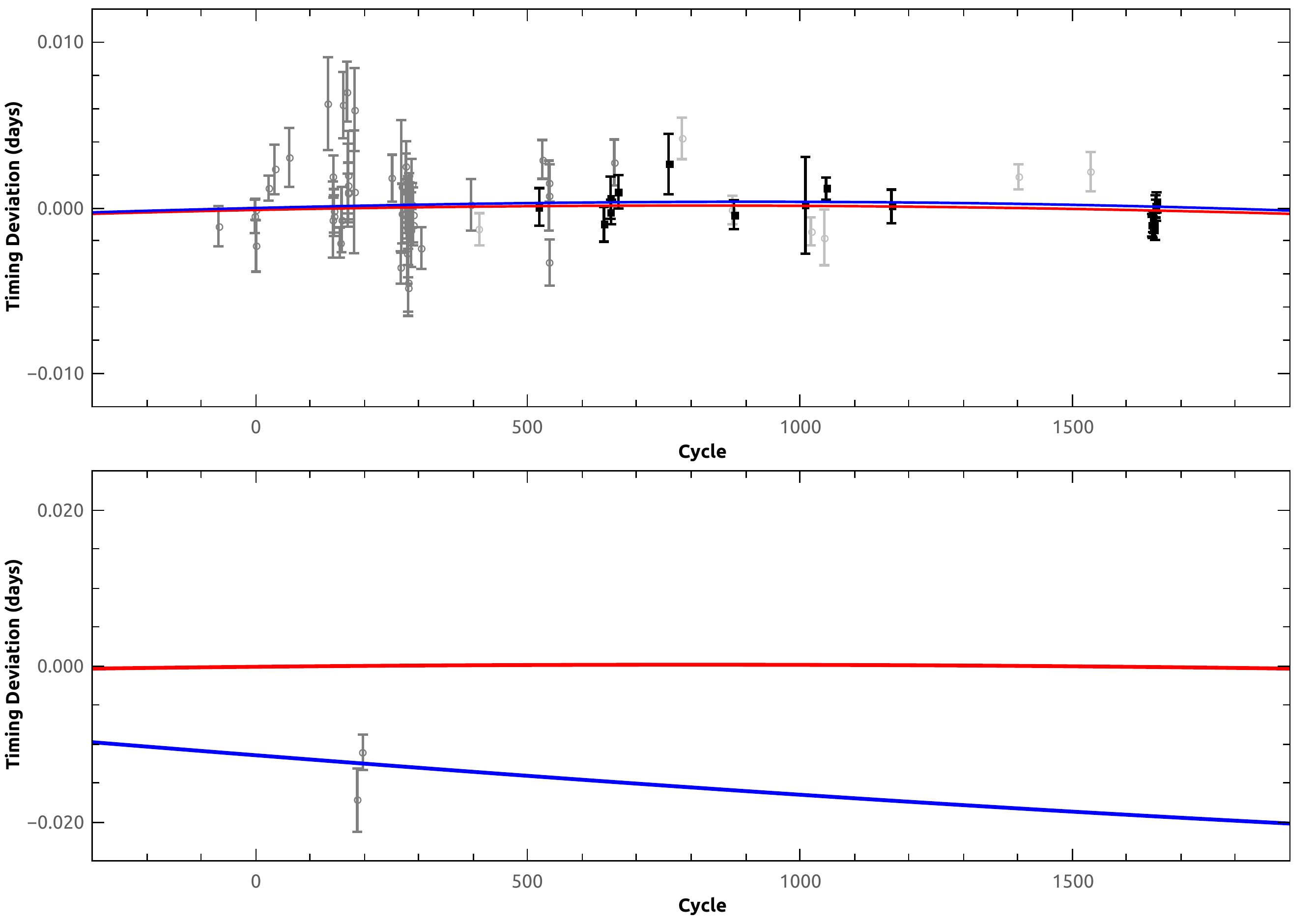}
      \caption{Upper panel: transit timing variation after subtracting the constant-period model of HAT-P-13. The black squares are the new transit times from TESS and our ground-based photometry, the dark gray dots are from the data of previous works, and the light gray ones denote those from ETD; Lower panel: secondary eclipse times variation after subtracting the constant-period model. The red curve shows the expected orbital decay model and the blue curve shows the apsidal precession model.}
         \label{fig6}
\end{figure*}

\begin{figure*}
   \centering
   \includegraphics[width=0.8\hsize]{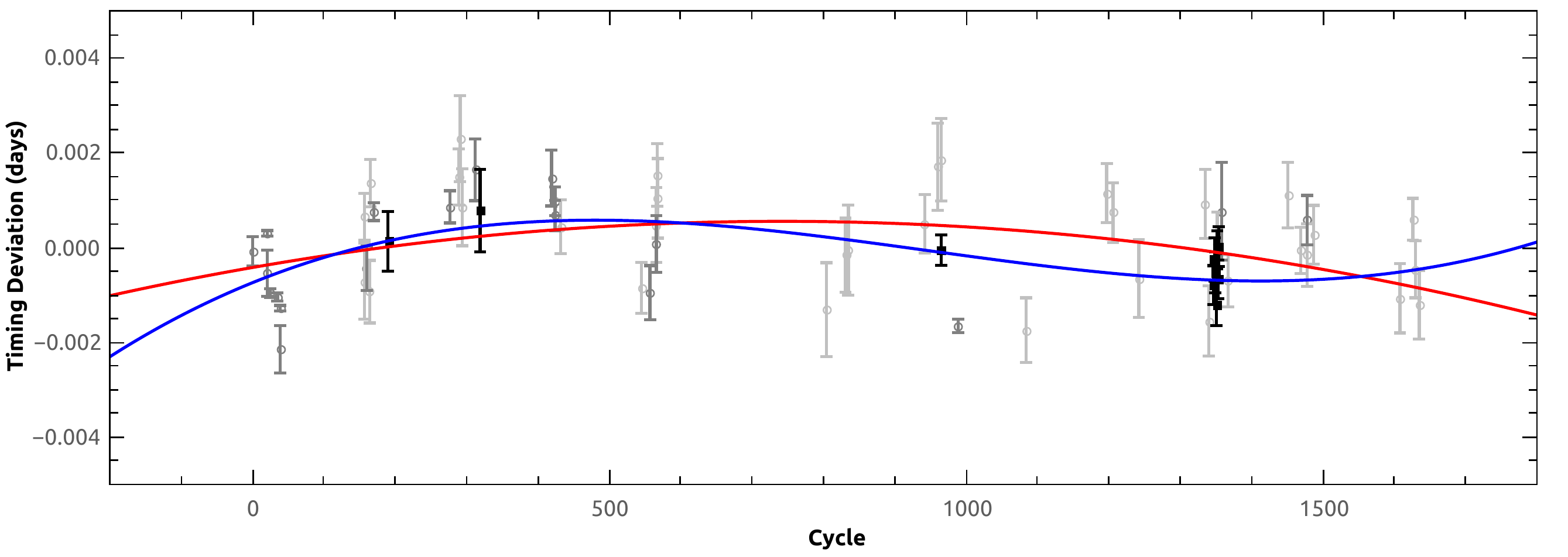}
      \caption{Transit timing variation after subtracting the constant-period model of HAT-P-16. The black squares are the new transit times from TESS and our ground-based photometry, the dark gray dots are from the data of previous works, and the light gray ones denote those from ETD. The red curve shows the expected orbital decay model and the blue curve shows the apsidal precession model.}
         \label{fig7}
\end{figure*}

\begin{figure*}
   \centering
   \includegraphics[width=0.8\hsize]{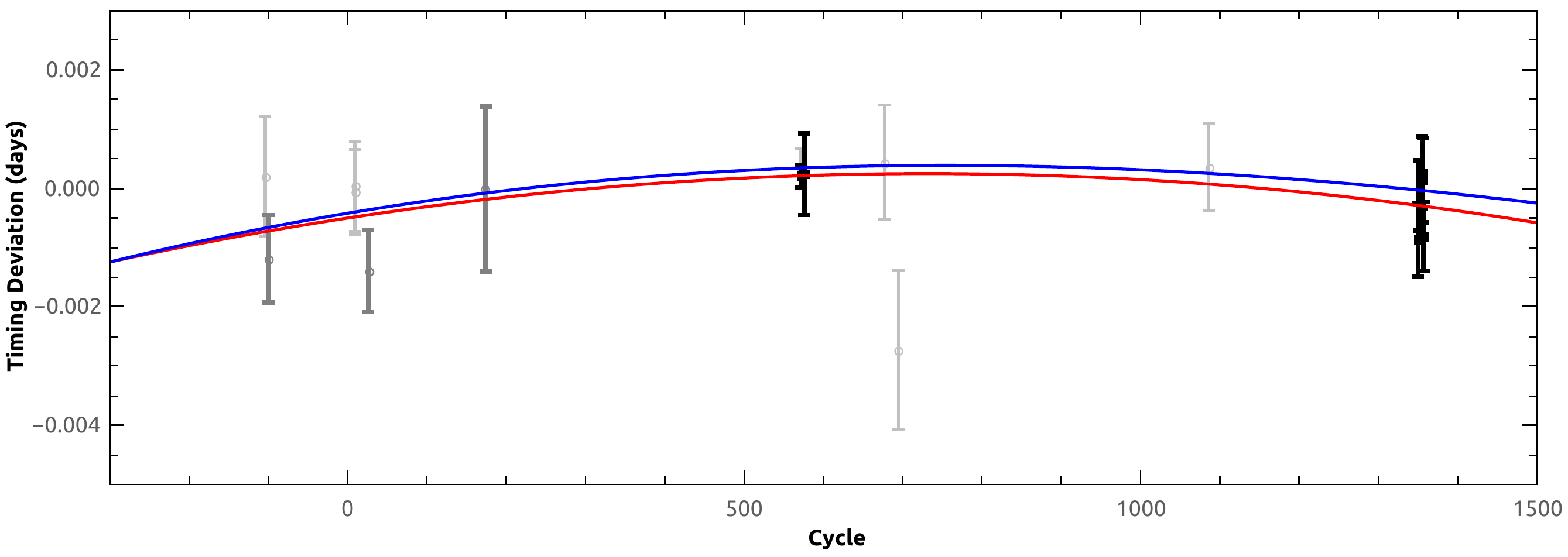}
      \caption{Transit timing variation after subtracting the constant-period model of WASP-32. The black squares are the new transit times from TESS and our ground-based photometry, the dark gray dots are from the data of previous works, and the light gray ones denote those from ETD. The red curve shows the expected orbital decay model and the blue curve shows the apsidal precession model.}
         \label{fig8}
\end{figure*}

\begin{figure*}
   \centering
   \includegraphics[width=1.0\hsize]{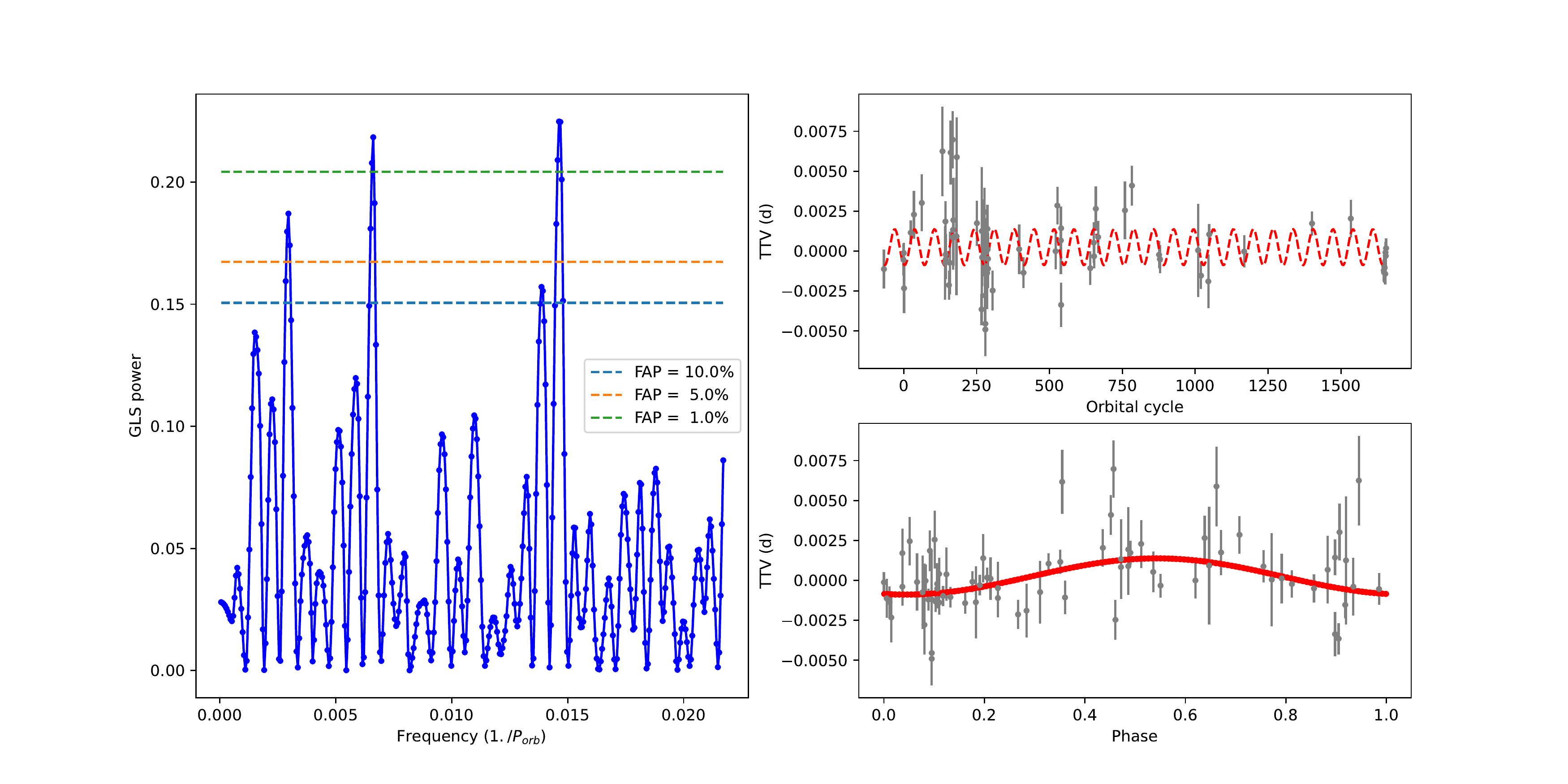}
      \caption{The GLS periodogram of HAT-P-13b' TTV and the optimal sine curve fitting associated with the largest GLS power.} 
         \label{fig_gls1}
\end{figure*}

\begin{figure*}
   \centering
   \includegraphics[width=1.0\hsize]{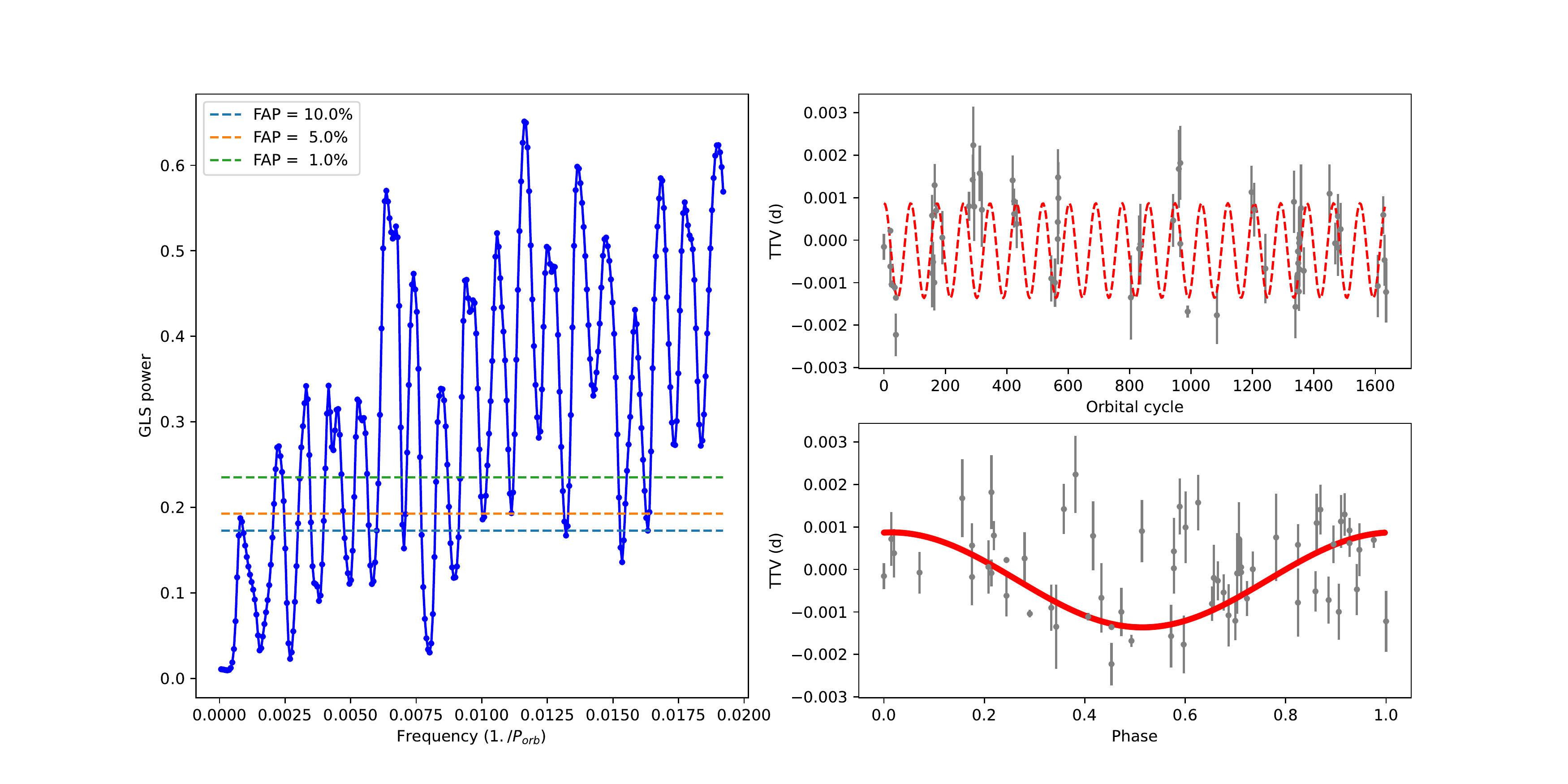}
      \caption{The GLS periodogram of HAT-P-16b' TTV and the optimal sine curve fitting associated with the largest GLS power.}
         \label{fig_gls2}
\end{figure*}

\begin{figure*}
  \begin{minipage}[t]{0.34\linewidth}
  \centering
   \includegraphics[width=\linewidth]{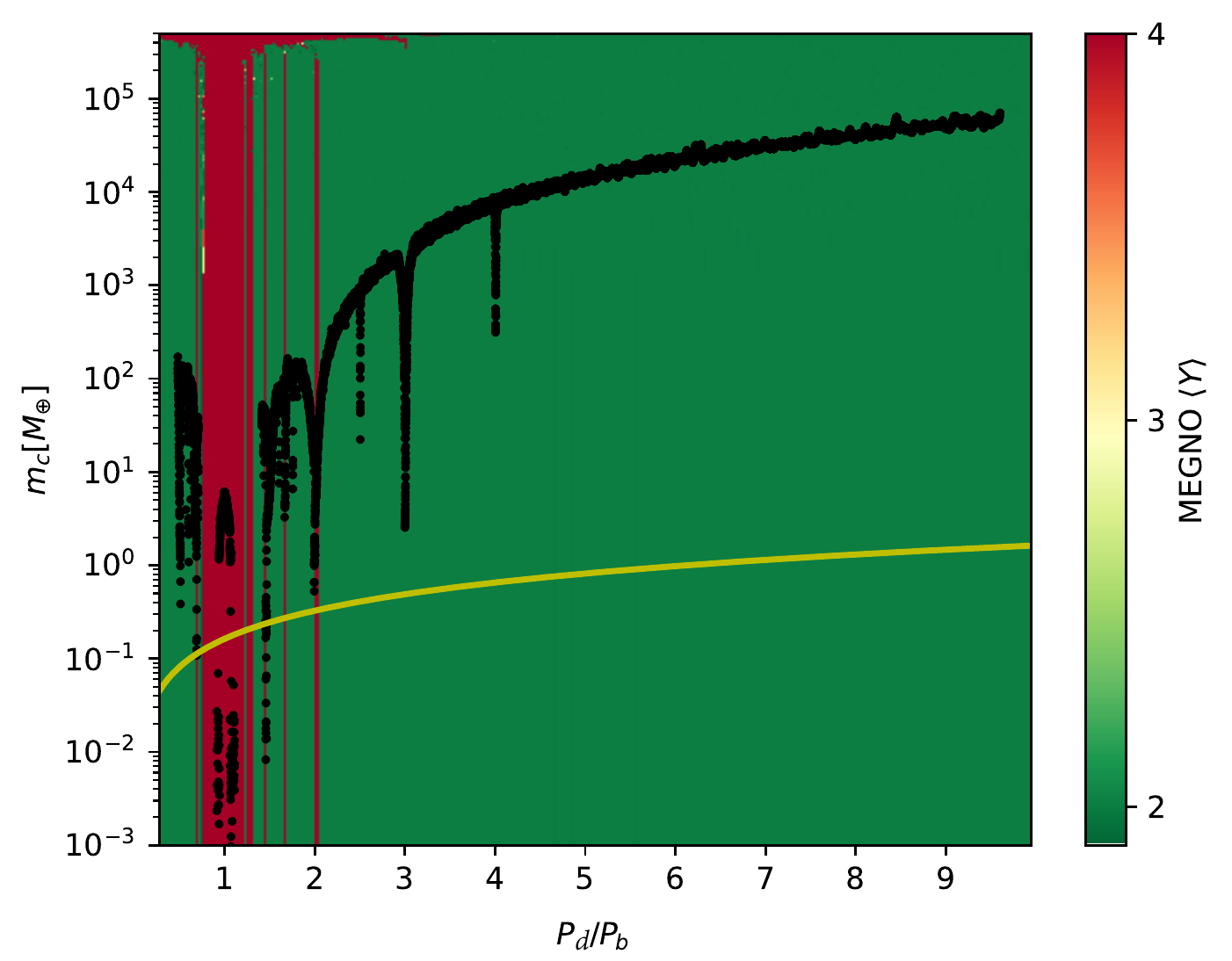}
  \end{minipage}%
  \begin{minipage}[t]{0.34\textwidth}
  \centering
   \includegraphics[width=\linewidth]{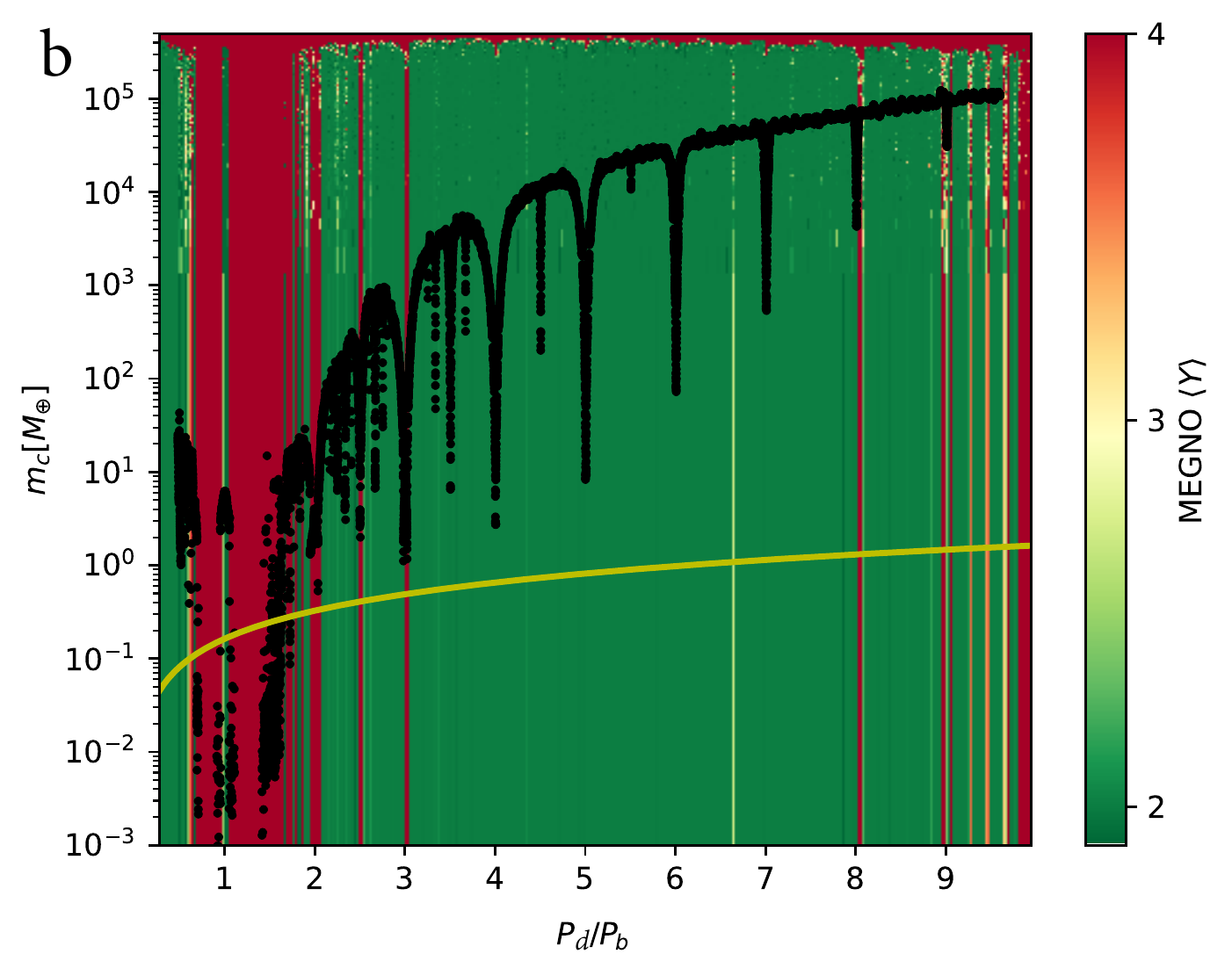}
  \end{minipage}%
  \begin{minipage}[t]{0.34\textwidth}
  \centering
   \includegraphics[width=\linewidth]{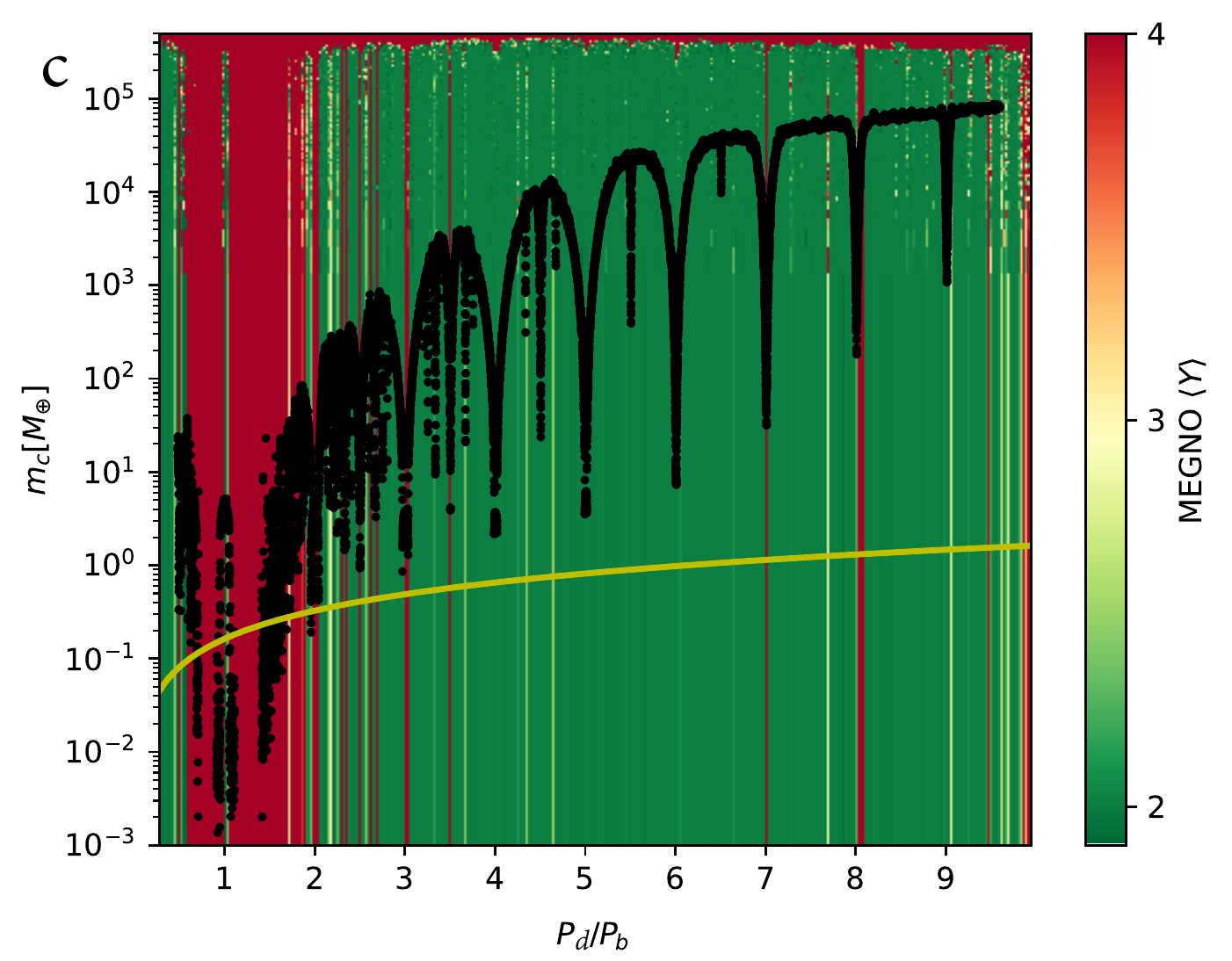}
  \end{minipage}%
    \caption{The MEGNO maps of HAT-P-13 based on three different initial conditions. Panel a: the perturber initially on coplanar , circular orbit. Panel b: the perturber initially on coplanar, slightly eccentric orbit (e = 0.1). Panel c: the perturber initially on inclined, slightly eccentric orbit (e = 0.1). The black scatters are the upper mass limit derived from the RMS of HAT-P-13b's TTVs, while the yellow line is from the constraints of RMS of HAT-P-13's RV residuals after eliminating the components of HAT-P-13 b \& c. The parameter spaces for regular orbital architectures color-coded with green are separated from the chaotic ones on the MEGNO maps.}
    \label{fig9}
\end{figure*}

\begin{figure*}
  \begin{minipage}[t]{0.34\linewidth}
  \centering
   \includegraphics[width=\linewidth]{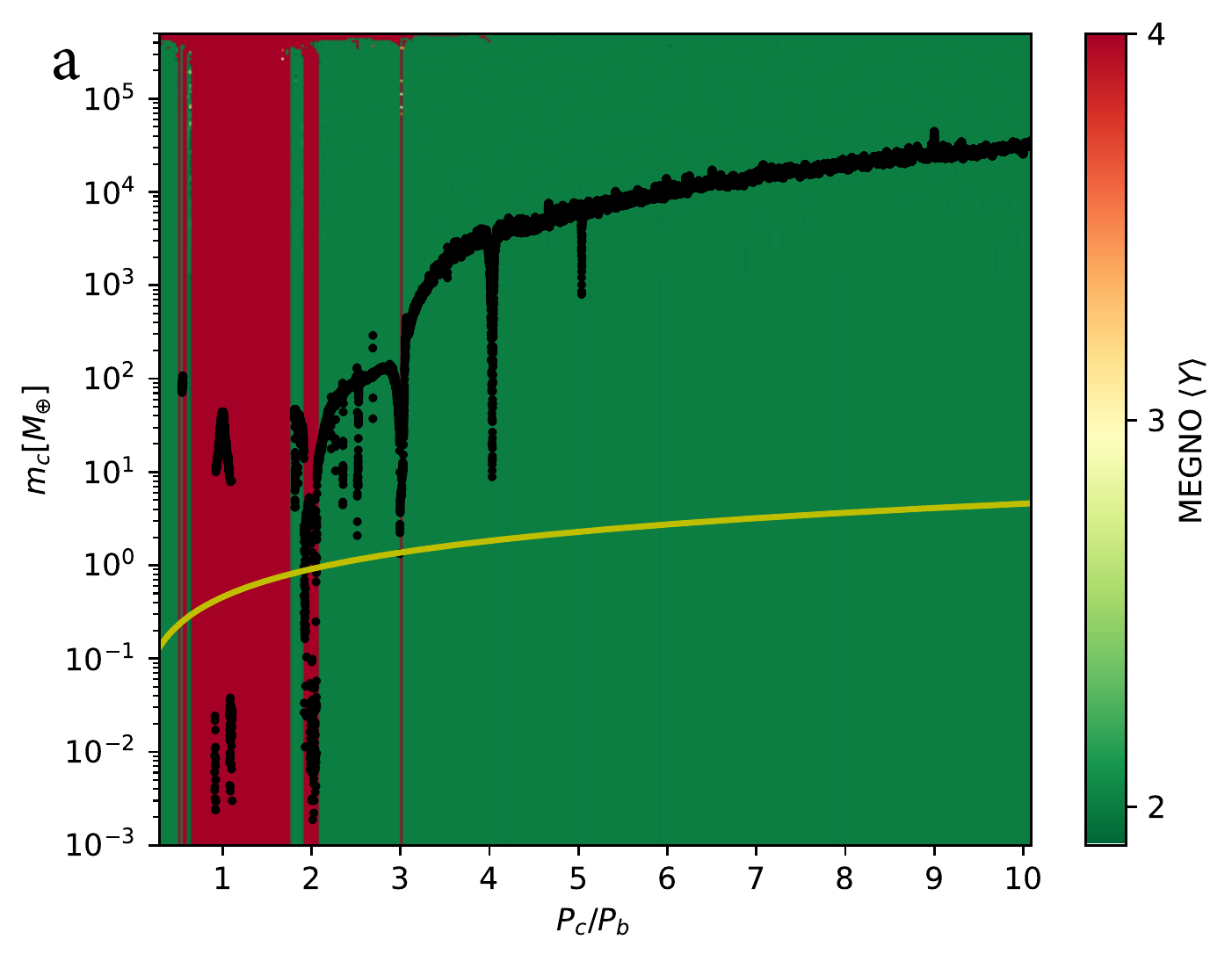}
  \end{minipage}%
  \begin{minipage}[t]{0.34\textwidth}
  \centering
   \includegraphics[width=\linewidth]{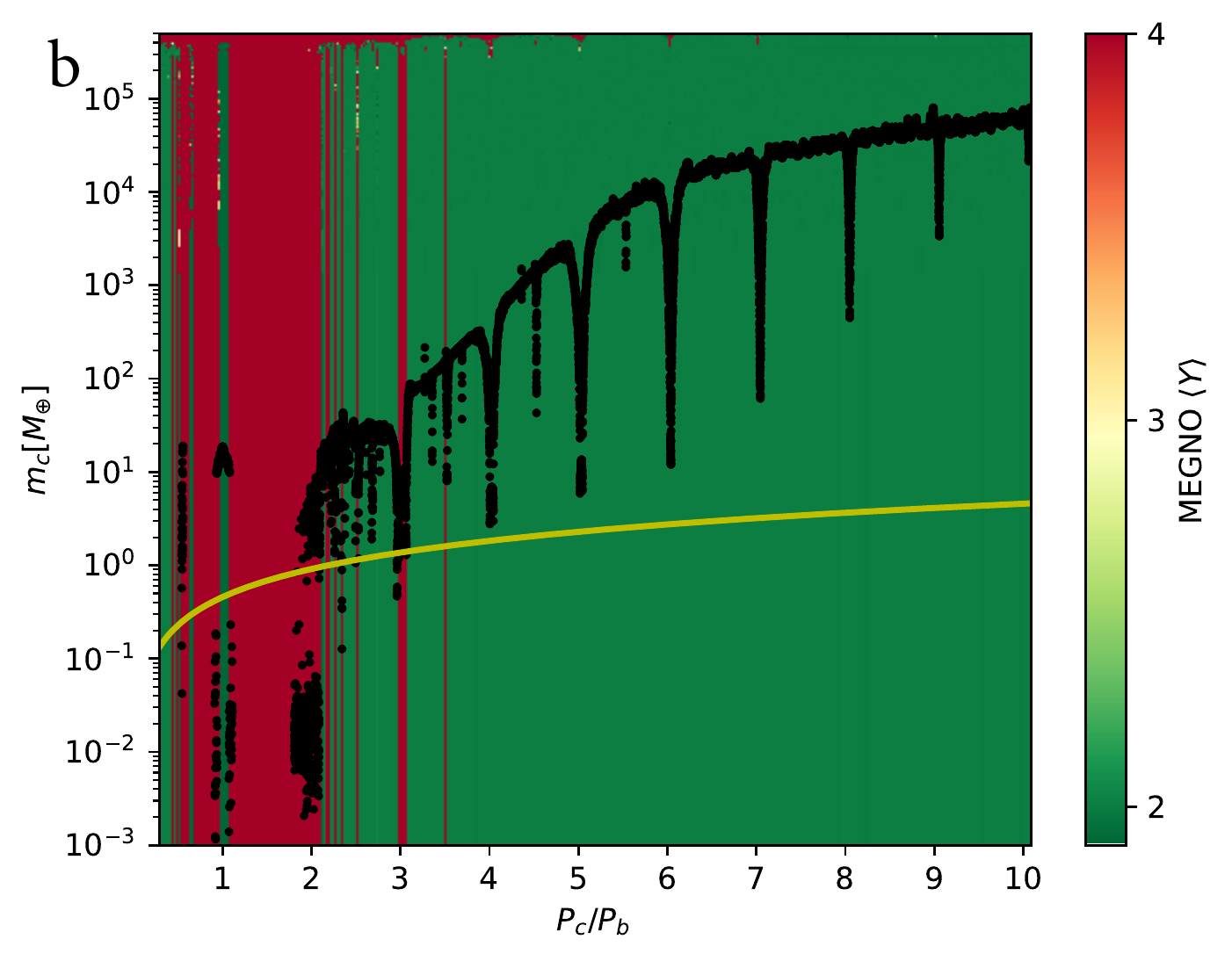}
  \end{minipage}%
  \begin{minipage}[t]{0.34\textwidth}
  \centering
   \includegraphics[width=\linewidth]{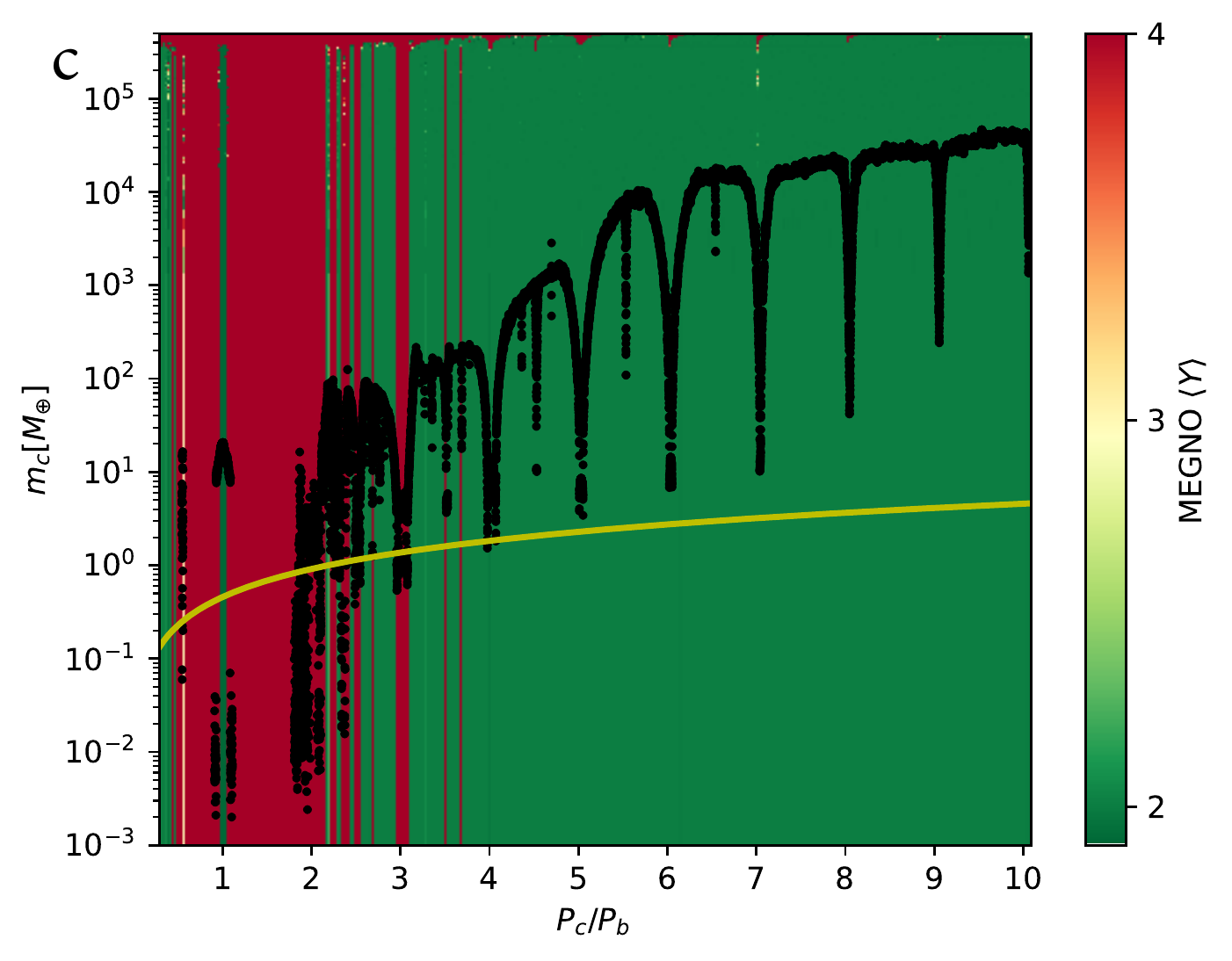}
  \end{minipage}%
    \caption{The MEGNO maps of HAT-P-16 based on three different initial conditions. All labels are identical to those in Figure. \ref{fig9}.}
    \label{fig10}
\end{figure*}

\begin{figure*}
  \begin{minipage}[t]{0.34\linewidth}
  \centering
   \includegraphics[width=\linewidth]{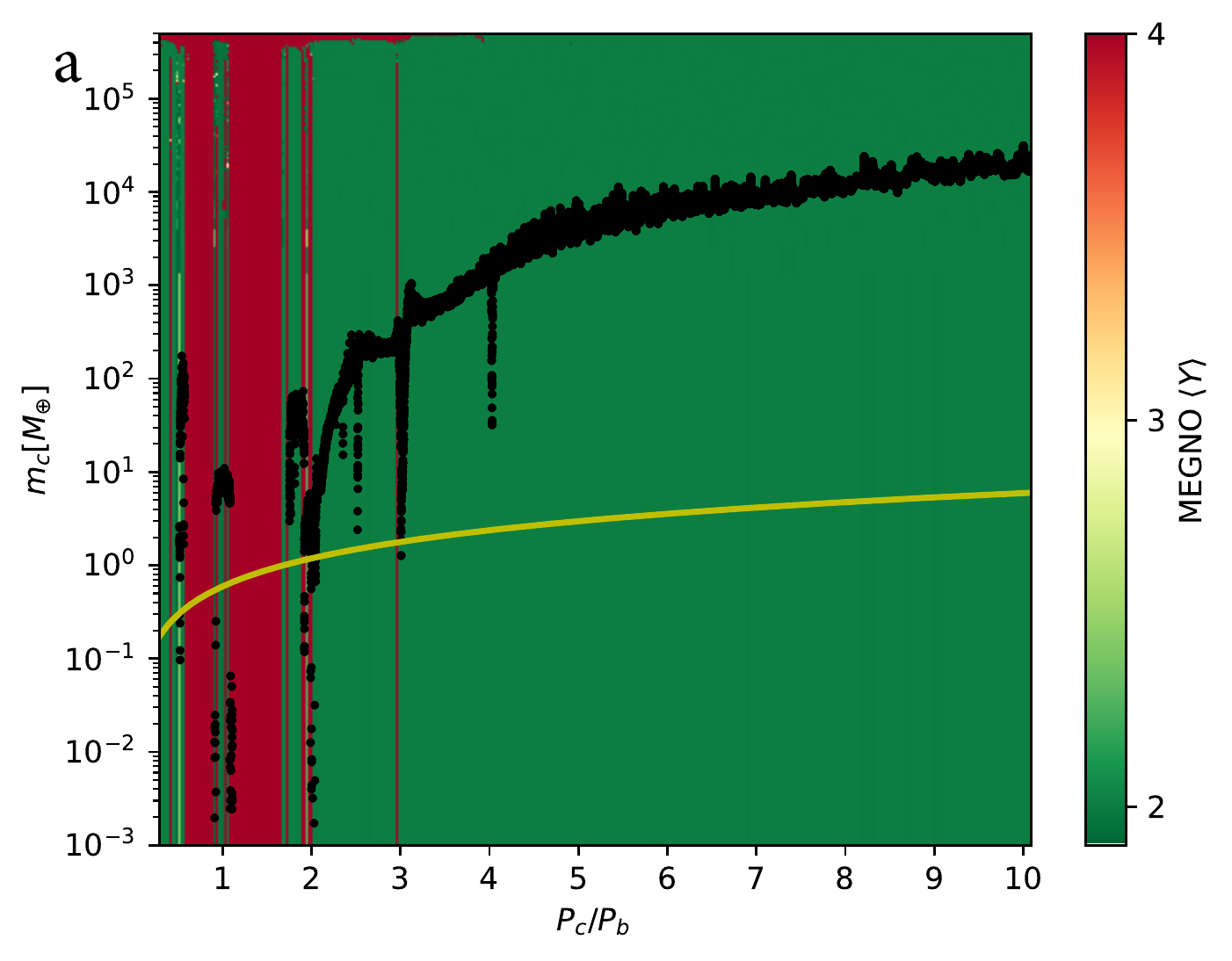}
  \end{minipage}%
  \begin{minipage}[t]{0.34\textwidth}
  \centering
   \includegraphics[width=\linewidth]{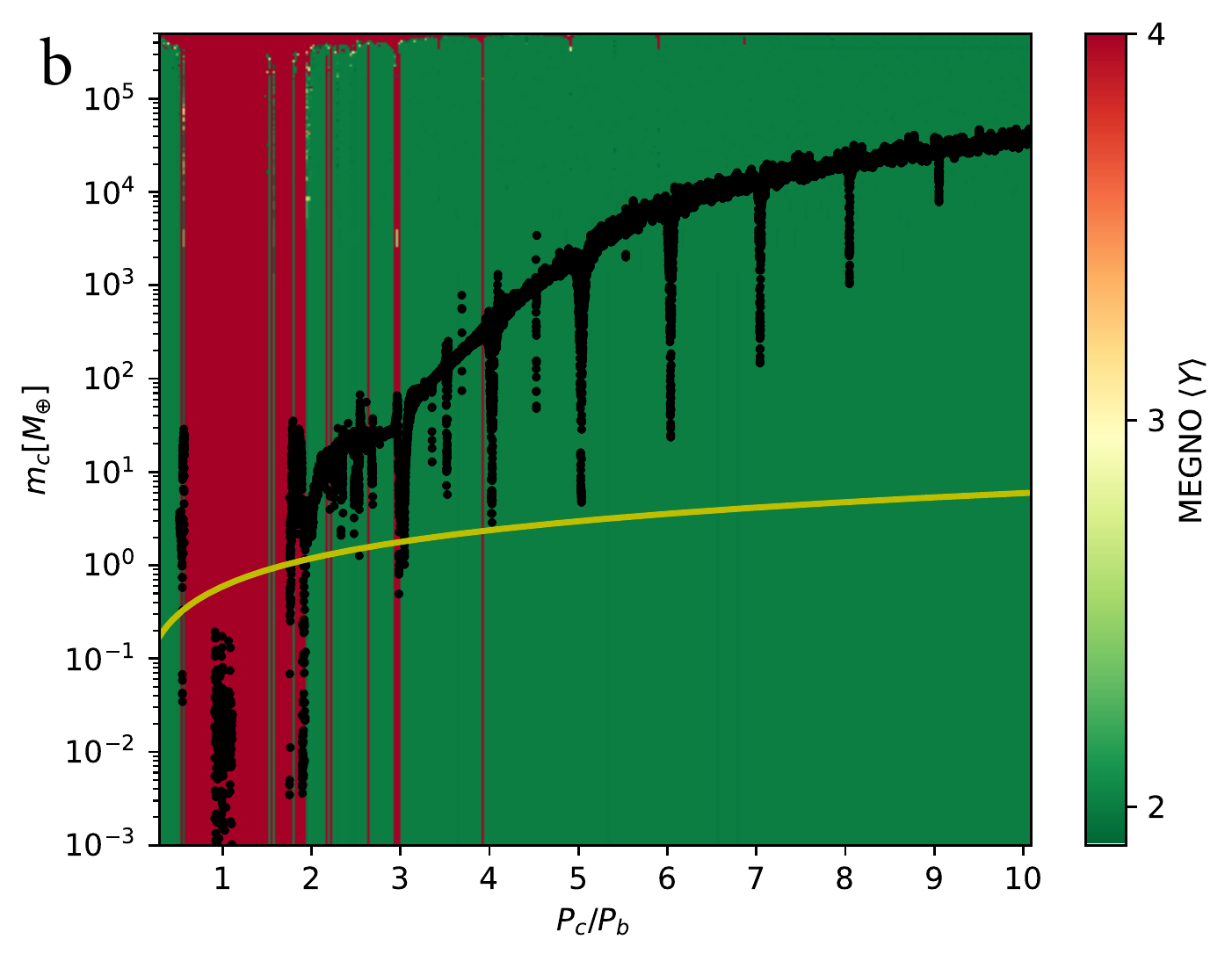}
  \end{minipage}%
  \begin{minipage}[t]{0.34\textwidth}
  \centering
   \includegraphics[width=\linewidth]{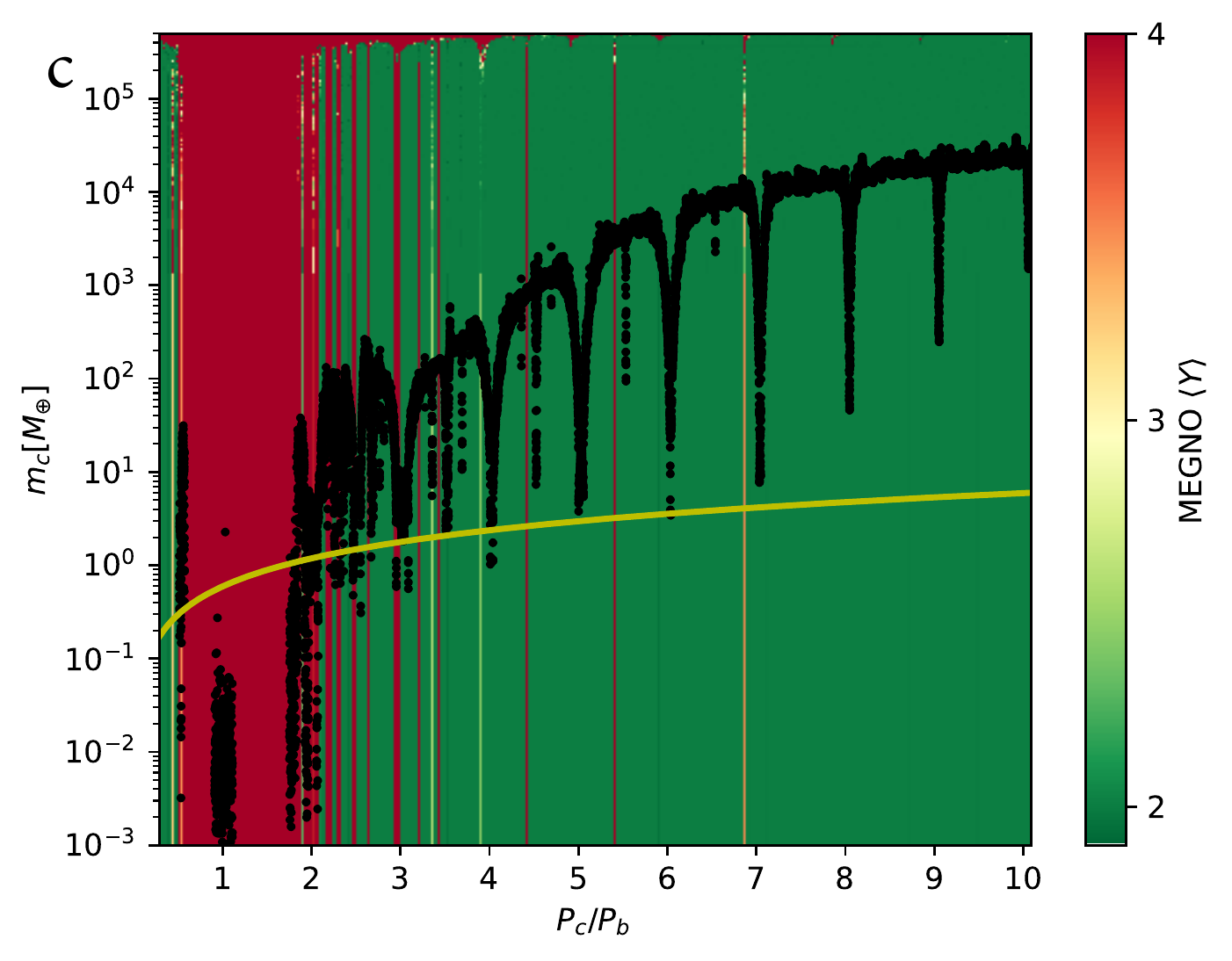}
  \end{minipage}%
    \caption{The MEGNO maps of WASP-32 based on three different initial conditions. All labels are identical to those in Figure.\ref{fig9}.}
    \label{fig11}
\end{figure*}

 \begin{table*}
\caption{System parameters of HAT-P-13, HAT-P-16 and WASP-32.}
\label{tab5}
\centering
\begin{threeparttable}
\begin{tabular}{c c c c c c c c }
\hline
\hline
Target&Parameter & \citet{Bakos2009} & This work \\ 
\hline
HAT-P-13&Orbital period (days) & 2.916260$\pm$0.000010 & 2.9162420$\pm$0.0000005\\
&Transit epoch (BJD-2450000) & 4779.92979$\pm$0.00038 & 4779.92999$\pm$0.00033\\
&Transit Duration (days) & 0.1345$\pm$0.0017 & 0.1344$\pm$0.0009\\
&Planet/star area ratio & 0.0071$\pm$0.0002 &0.0071$\pm$0.0001\\
&Impact parameter & 0.668$^{+0.032}_{-0.045}$ &0.711$\pm$0.002 \\
&Orbital separation (AU)&0.0427$^{+0.0006}_{-0.0012}$&0.0418$\pm$0.0003 \\
&Orbital inclination (deg)&83.4$\pm$0.6&82.77$\pm$0.33 \\
&Orbital eccentricity& 0.021$\pm$0.009 & 0 (fixed)\\
&Stellar radius ($R_{\odot}$) &1.56$\pm$0.08 & 1.590$\pm$0.042 &\\
&Planet radius ($R_{Jup}$) &1.281$\pm$0.079&1.306$\pm$0.042 \\
&Stellar effective temperature (K) &5653$\pm$90 & 5651$\pm$93 &\\
&Planet temperature (K) &1653$\pm$45&1681$\pm$34 \\
\hline\hline       
Target&Parameter & \citet{Buchhave2010} & This work \\ 
\hline
HAT-P-16&Orbital period (days) & 2.775960$\pm$0.000003 & 2.7759682$\pm$0.0000002\\
&Transit epoch (BJD-2450000) & 5027.59293$\pm$0.00031 & 5027.59301$\pm$0.00019\\
&Transit Duration (days) & 0.1276$\pm$0.0013 & 0.1257$\pm$0.0006\\
&Planet/star area ratio & 0.0115$\pm$0.0002 &0.0112$\pm$0.0001\\
&Impact parameter & 0.439$^{+0.065}_{-0.098}$ &0.272$\pm$0.008 \\
&Orbital separation (AU)&0.0413$\pm$0.0004&0.0405$\pm$0.0002 \\
&Orbital inclination (deg)&86.6$\pm$0.7&87.93$\pm$0.62 \\
&Orbital eccentricity& 0.036$\pm$0.004 & 0 (fixed)\\
&Stellar radius ($R_{\odot}$) &1.237$\pm$0.054 & 1.157$\pm$0.030 &\\
&Planet radius ($R_{Jup}$) &1.289$\pm$0.066& 1.194$\pm$0.037 \\
&Stellar effective temperature (K) &6158$\pm$80 & 6158$\pm$79 &\\
&Planet temperature (K) &1626$\pm$40&1587$\pm$26 \\
\hline\hline       
Target&Parameter & \citet{Maxted2010} & This work \\ 
\hline
WASP-32&Orbital period (days) & 2.718659$\pm$0.000008 & 2.7186615 $\pm$0.0000003\\
&Transit epoch (BJD-2450000) & 5151.0546$\pm$0.0005 & 5779.06707 $\pm$0.00024\\
&Transit Duration (days) & 0.101$\pm$0.002 & 0.0999$\pm$0.0006\\
&Planet/star area ratio & 0.0124$\pm$0.0004 &0.0124$\pm$0.0001\\
&Impact parameter & 0.64$\pm$0.04 &0.682$\pm$0.014 \\
&Orbital separation (AU)&0.0394$\pm$0.0003&0.0394$\pm$0.0003 \\
&Orbital inclination (deg)&85.3$\pm$0.5&84.85$\pm$0.21 \\
&Orbital eccentricity& 0.0180$\pm$0.0065 & 0 (fixed)\\
&Stellar radius ($R_{\odot}$) &1.11$\pm$0.05 & 1.114$\pm$0.023 &\\
&Planet radius ($R_{Jup}$) &1.18$\pm$0.07& 1.209$\pm$0.031 \\
&Stellar effective temperature (K) &6100$\pm$100 & 6104$\pm$103&\\
&Planet temperature (K) &1560$\pm$50&1565$\pm$30 \\
\hline  
\hline 
\end{tabular}
\end{threeparttable}
\end{table*}

\section{TTV modeling}\label{sec:dis}

\subsection{TTV signals} \label{ttvsection}

As introduced in Section \ref{sect:intro}, a few physical processes could stimulate perturbation on hot Jupiter's orbits and/or distort observed transit light curves, for example, spot crossing events, and hence generate measurable TTVs. We considered four scenarios to interpret TTVs of these systems, namely linear ephemeris, orbital decay, apsidal precession and planetary gravitational perturbation. The spot crossing event is not discussed here because of the absence of apparent patterns of spot modulation in TESS light curves. In this subsection, we firstly modeled observed transit times based on the former three models.

The first model is based on the assumption that the orbital period is a constant as mentioned before,
\begin{displaymath}
		T_{tra}(E) = T_{0} + P \times E 
\end{displaymath}
\begin{displaymath}
		T_{occ}(E) = T_{0} + \frac{P}{2} + P \times E
\end{displaymath}
, where $T_{0}$ is zero point and $E$ is the cycle number, respectively.

 The second model assumes that the orbital period is changing with a steady rate \citep[e.g,][]{RN212,RN209,RN214},
\begin{displaymath}
		T_{tra}(E) = T_{0} + P \times E + \frac{1}{2}\frac{dP}{dE}E^2
\end{displaymath}
\begin{displaymath}
		T_{occ}(E) = T_{0} + \frac{P}{2} + P \times E + \frac{1}{2}\frac{dP}{dE}E^2
\end{displaymath}
, where $dP/dE$ denotes the decay rate. 

And the third model is based on the assumption that the planetary orbit is 
slightly eccentric and its argument of pericenter is precessing uniformly over time \citep[e.g,][]{RN212,RN209,RN214},
\begin{displaymath}
		T_{tra}(E) = T_{0} + P_{s} \times E - \frac{eP_{a}}{\pi}\mathrm{cos} \omega(E)
\end{displaymath}
\begin{displaymath}
		T_{occ}(E) = T_{0} + \frac{P_{a}}{2} + P_{s} \times E + \frac{eP_{a}}{\pi}\mathrm{cos} \omega(E)
\end{displaymath}
\begin{displaymath}
		\omega(E) = \omega_{0} +  \frac{d\omega}{dE}E
\end{displaymath}
\begin{displaymath}
		P_{s} = P_{a}\bigg ( 1-\frac{1}{2\pi}\frac{d\omega}{dE}\bigg )
\end{displaymath}
where $P_{s}$ is the sidereal period, $P_{a}$ is the anomalistic period, $e$ is the orbital eccentricity, $\omega$ is the argument of pericenter and $d\omega/dE$ is the precession rate \citep{RN265}, respectively. We employed the MCMC sampler \textit{emcee} to fit TTV signals based on these three assumptions and obtained the optimal parameters and uncertainties \citep{RN263}. In the MCMC sampling, 50000 MCMC samples with 1000 burn-in ones was taken to ensure the convergence. The results of fitted parameters with relative uncertainties are listed in Table \ref{tab8}, and TTVs of these systems with the optimal orbital decay and apsidal precession models are shown in Figures \ref{fig6}, \ref{fig7} and \ref{fig8}, respectively.

To make a comparsion between the alternative models, the Bayesian Information Criterion \citep[BIC;][]{RN267} is often adopted. BIC is parameterized as
\begin{displaymath}
		\mathrm{BIC} = \chi^{2} + k \mathrm{log} n
\end{displaymath}
, where $k$ denotes the number of free parameters and $n$ represents the number of data points.

\begin{table*}
\caption{Timing model parameters of HAT-P-13, HAT-P-16 and WASP-32.}             
\label{tab8}      
\centering    
\begin{threeparttable}                     
\begin{tabular}{c c c c}        
\hline\hline                 
Parameter & HAT-P-13 & HAT-P-16 & WASP-32 \\   
\hline                        
\hline
\multicolumn{4}{l}{\textit{Constant period model}} \\
$P$ (days) & $2.9162420\pm0.0000005$ & $2.7759682\pm0.0000002$ & $2.7186615\pm0.0000003$\\
$T_{0}$ (BJD-2450000) & $4779.92999\pm0.00033$ & $5027.59301\pm0.00019$ & $5779.06707\pm0.00023$ \\
$N_{dof}$ & 75 & 61 & 18 \\
$\chi^{2}_{min}$ & 210.947 & 1033.220 & 17.981 \\
BIC & 219.635 & 1041.506 & 23.647 \\
\hline      
\multicolumn{4}{l}{\textit{Orbital decay model}} \\    
$P$ (days) & $2.9162427\pm0.0000017$ & $2.7759708\pm0.0000009$ & $2.7186636\pm0.0000009$ \\
$T_{0}$ (BJD-2450000) & $4779.92987\pm0.00046$ & $5027.59259\pm0.00023$ & $5779.06657\pm0.00031$  \\
$dP/dE$ (days/orbit) & $(-0.823\pm2.109)\times10^{-9}$ & $(-3.536\pm1.212)\times10^{-9}$ & $(-2.806\pm1.156)\times10^{-9}$ \\
$N_{dof}$ & 74 & 60 & 17\\
$\chi^{2}_{min}$ & 208.062 & 863.871 & 12.217\\
BIC & 221.093 & 876.301 & 20.716\\
\hline       
\multicolumn{4}{l}{\textit{Apsidal precession model}} \\    
$P_{s}$ (days) & $2.9162397\pm0.0000011$ & $2.7759926\pm0.0000091$ & $2.7186701\pm0.0000031$ \\
$T_{0}$ (BJD-2450000) & $4779.92426\pm0.00106$ & $5027.56989\pm0.00788$ & $5779.04116\pm0.01074$\\
e & $0.01145\pm0.00280$ & $0.03503\pm0.01658$ & $0.03577\pm0.01493$\\
$\omega_{0}$ (rad) & $2.139\pm0.107$ & $3.904\pm0.126$ & $3.744\pm0.147$ \\
$d\omega/dE$ (rad/orbit) & $0.00035\pm0.00019$ & $0.00085\pm0.00015$ & $0.00036\pm0.00012$ \\
$N_{dof}$ & 72 & 58 & 15\\
$\chi^{2}_{min}$ & 171.827 & 703.054 & 13.949\\
BIC & 193.546 & 723.769 & 28.115\\
\hline
\hline       
\end{tabular}
\end{threeparttable}
\end{table*}

\subsubsection{HAT-P-13}

As shown in Figure \ref{fig6}, the transit and occultation times were modelled simultaneously using the procedure mentioned in \ref{ttvsection}. The apsidal precession model has a lower chi-square (i.e., $\chi^{2}_{min}=171.827$) and its BIC value is also the smallest, compared with the constant period model (i.e., $\chi^{2}_{min}=210.947$) and the orbital decay model (i.e., $\chi^{2}_{min}=208.062$). The apsidal precession model is more favored than other two models with $\Delta(\mathrm{BIC_{3,1}}) = 26.089$ and $\Delta(\mathrm{BIC_{3,2}}) = 27.547$. Bayes factors are $B_{3,1} = 4.63\times10^5$ and $B_{3,2} = 9.59\times10^5$, respectively, which suggests that observed TTV signal significantly supports the apsidal precession model. However, as shown in Figure \ref{fig6}, the apsidal precession model could not well reproduce the observed TTV, in particular the spike signals on some epochs, which possibly suggests that TTV of HAT-P-13 is originated from the orbital perturbations of close companion(s).

\subsubsection{HAT-P-16}

As shown in Figure \ref{fig7}, the transit times of HAT-P-16b show significant TTV signal, which deviates from the constant period model with $\chi^{2}_{min}=1033.220$ for the degree of freedom of 61. Comparing with the constant period model, we found that the transit times of HAT-P-16 are favored the orbital decay model ($\chi^{2}_{min}=863.871$) and the apsidal precession model ($\chi^{2}_{min}=703.054$). Because several transit times obtained by \citet{Aladag2021} significantly deviates from other TTV measurements, they are eliminated in the TTV modelling process. The apsidal precession model has the lowest BIC value with $\Delta(\mathrm{BIC_{3,1}}) = 317.737$ and $\Delta(\mathrm{BIC_{3,2}}) = 152.532$. This means a predominant interpretation of the apsidal precession model to the observed TTV signal of HAT-P-16. Even if the apsidal model is preferred by BIC, a reduced $\chi^{2}$ of much large than 10 may imply that the model is possible wrong and/ or the errors of some transit times are underestimated.

\subsubsection{WASP-32}

As shown in Figure \ref{fig8}, compared with the constant period model (i.e., $\chi^{2}_{min}=17.981$) and the apsidal precession model (i.e., $\chi^{2}_{min}=13.949$), the orbital decay model fits the timing data of WASP-32 better (i.e., $\chi^{2}_{min}=12.217$) and also has smaller BIC value. The orbital decay model is the suprior one for interpreting the TTV of WASP-32 with $\Delta(\mathrm{BIC_{2,1}}) = 2.931$ and $\Delta(\mathrm{BIC_{2,3}}) = 7.399$, which means that the observations of WASP-32 slightly favour the orbital decay model.

\subsection{Upper Mass Limit of a Hypothetical Perturber}\label{upp}

TTV signals of transiting exoplanetary systems provide an opportunity to constrain masses of additional perturbing planets, given that these TTVs are triggered by the gravitational perturbation from other planets. While inverting TTV signals has succeeded in detecting new exoplanets \citep[e.g.,][]{Sun2019}, this technique cannot be utilised directly to these three systems, because only TTV measurements with enough high cadence, high SNR and even long observation baseline, like some TTV measurements collected by {\it Kepler}, may well constrain the orbital period and orbital architecture of the perturber. However, we could utilise the RMS of the TTVs for the sparse measurements to approximate estimate the property and orbital parameter of extra planets. Before conducting the {\it N}-body simulation, we computed the generalised Lomb-Scargle (GLS) periodograms of these three systems' TTVs to search for potential periodic signals \citep{Zechmeister2009}. As a result, we found that the measured TTVs of HAT-P-13b and HAT-P-16b exhibited significant periodic signals. The GLS periodograms and the optimal sine curve fittings for the TTVs of those two system are showed in \ref{fig_gls1} and \ref{fig_gls2}, respectively.

In order to constrain the property and orbital parameter of additional planets, we employed a {\it N}-body code to perform direct orbit integration, which was also used in the literature \citep[e.g.,][]{RN734,RN733,Wang2021,RN700,RN749,Bai2022}. In principle, we numerically integrated the orbits of each system with a hypothetic perturbing planet in the system and compared the RMS of their TTVs with the observed. We employed similar methodology as described in \citet{Bai2022} to acquire the upper mass limits. 

During integrating the orbits of each system, a large number of trial orbital periods for the perturbing planet was sampled while the other orbital parameters were fixed to specific values. Because TTV signals are dominated by the perturber's mass, the orbital period, the eccentricity and the mutual inclination of the orbit \citep{Agol2005,Holman2005,Nesvorny2008}, we considered three different orbital architectures for hypothetical perturbers as a brief demonstration in the following : (a) all planets on coplanar and circular orbits initially; (b) the perturbing planet on a coplanar and slightly eccentric orbit initially (i.e., $e_{c}=0.1$); and (c) the perturbing planet on an inclined and slightly eccentric orbit initially (i.e., $i_{c}=i_{b}-30\degr$, $e_c=0.1$, where $i_{b}$ and $i_{c}$ are the inclinations of the known hot jupiter and the perturber, respectively). Hereafter, these orbital architectures are repectively referred to Case a, Case b and Case c. With the exception of the orbital period of the perturbing planet, the other parameters adopted selected values; the orbital period of the perturber $P_{c}$ was uniformly-spaced sampled from 1 day to 10$P_{b}$ with a step of 0.001 days. 

Furthermore, the radial velocities (RVs) of the host star generated by a hypothetical perturber would supply extra constraints on the property of the purterber. Given that the planets are not locked in orbital resonance, the RVs of the host star could be represented as the sum of the RVs induced by each planetary component's Keplerian motion \citep{Ford2006}. Therefore, the residuals of RVs after eliminating the components from known transiting planets, would in principle stem from additional bodies. High precision RV measurements with a good coverage in its orbital phase will required for providing constraints on the orbital period and mass of the perturbing planet. Here we utilized RMS of residuals of each planet's RVs instead of residuals to statistically place constraints on the mass for the hypothetical perturbers. The amplitude $K$ of RV curves induced by a planet on its host star is represented :
\begin{displaymath}
\left(\frac{M_p\sin i_p}{M_\oplus}\right)=11.19\left(\frac{K}{m/s}\right)\sqrt{1-e^2}\left(\frac{M_*}{M_\odot}\right)^{2/3}\left(\frac{P_{orb}}{1 yr}\right)^{1/3}
\end{displaymath}
where $P_{orb}$, $M_p$ and $M_*$ denote the orbital period of the planet, the mass of the planet and the mass of the host star, respectively. Given that the orbital eccentricities of $e_c\leq 0.1$ for the perturber, the RMS of RV curve statistically equals $\sqrt{2}/2$ times of its amplitude. The mass limits of the hypethetical perturbers are plotted with the yellow curves in Figures \ref{fig9}, \ref{fig10} and \ref{fig11} based on the RMS of HAT-P-13, HAT-P-16 and WASP-32's RV residuals.

Extra constraints could also be imposed on the property of the perturber due to the requirement of the stability for the long-run orbital evolution of planetary system. We compute the Mean Exponential Growth factor of Nearby Orbits \citep[MEGNO;][]{Cincotta2000,Gozdziewski2001,Cincotta2003} to meet our purpose by utilising REBOUND to numerically integrate the orbits and simultaneously calculate associated variational equations of motion over a large number of initial orbital states \citep{Rein2015,Rein2016}. During the simulation, the orbital period and the mass of the purterber were varied; each initial grid based on previously mentioned orbital architectures was integrated  for 500 yrs, which would be helpful for finding the location of weak chaotic high-order mean-motion resonances. MEGNO is widely employed to quantitatively estimate the stochastic behaviour for a non-linear dynamical system and thus capture the chaotic resonances \citep{Gozdziewski2001,Hinse2010}. For an initial orbital state, once the chaotic behavior was seized by MEGNO, there is no doubt about its erratic nature in the future \citep{Hinse2010}.

In the following subsections, the MEGNO results for each system are presented. For each case, we found the common unstable regions were in the vicinity of the transiting planet labeled as yellow and/ or red in MEGNO maps (corresponding to $\langle Y\rangle > 3.5$).

\subsubsection{HAT-P-13b}

We obtained comparable results to those of  \citet{Agol2005} and \citet{Holman2005}, which implies that TTVs are amplified when the orbital architectures are in or near mean motion resonances (hereafter, MMR). In addition, $\rm TTV_{\rm RMS}$ of both Case b and c are far more complex than that in Case a, in which high-order MMRs (e.g., $P_c/P_b\simeq 5:1, 6:1$ and so on.) trigger large TTV signals and thus the relative upper mass limit drops rapidly, in contrast to those of Case a. Surprisingly, the observed $\rm TTV_{\rm RMS}$ for HAT-P-13b could be well reproduced by a $\sim 2.8~M_{\oplus}$ super Earth near 1:1 MMR with HAT-P-13b in Case b , although RVs arosed by this co-orbital purterber are larger than RMS of the RV residuals. Through over-plotting the RMS of RV residuals ($\rm RV_{\rm RMS}$) that removes the contributions from known planets, the constraints placed by $\rm RV_{\rm RMS}$ are more stringent than those from $\rm TTV_{\rm RMS}$ on the upper mass limits; similar simulation results have been obtained for HAT-P-16 and WASP-32 systems. As presented in panel a of Figure~\ref{fig9}, a coplanar perturbing planet with mass of $0.011-0.24~M_{\oplus}$ initially on circular orbit would generate a $\rm TTV_{\rm RMS}$ of $204.4\,{\rm s}$ when it is near 3:2 MMR with HAT-P-13b. For the perturber initially on slightly eccentric and inclined orbit, the upper mass limits near 2:1 MMR from the TTV simulation are under those of $\rm RV_{\rm RMS}$, as shown in panel c of Figure~\ref{fig9}.

\subsubsection{HAT-P-16b}
The measured $\rm TTV_{\rm RMS}$ of HAT-P-16b was $85.5 \rm s$. In Case a, all stable orbital architectures could not reproduce the observed TTV scatter as shown in Figure \ref{fig10}. When hypothetical planets of $0.0037~M_{\oplus}$ and $1.29~M_{\oplus}$ reside near 1:1 and 3:1 MMRs in Case b, they can yield the observed $\rm TTV_{\rm RMS}$; and $0.417~M_{\oplus}$, $0.749~M_{\oplus}$, $1.479~M_{\oplus}$ near 7:3, 5:2 and 4:1 MMRs in Case c, respectively.

\subsubsection{WASP-32b}
The measured $\rm TTV_{\rm RMS}$ of WASP-32b was $70.8 \rm s$. When hypothetical bodies of $1.0\times10^{-3}~M_{\oplus}$ near the 1:1 MMR, $ 0.0029~M_{\oplus}$ near the 2:1 MMR and $0.99~M_{\oplus}$ near the 3:1 MMR in Case a, they could cause the observed TTV scatter. For Case b, extral planets of $0.107~M_{\oplus}$ and $0.759~M_{\oplus}$ will generate that $\rm TTV_{\rm RMS}$, when they are near 2:1 and 3:1 MMRs with WASP-32b, respectively. For Case c, the extral planets with masses of $0.536~M_{\oplus}$, $0.285~M_{\oplus}$, $0.531~M_{\oplus}$, $0.971~M_{\oplus}$ and $3.426~M_{\oplus}$ near 7:3, 5:2, 3:1, 4:1 and 6:1 MMRs, respectively, would well reproduced the observed $\rm TTV_{\rm RMS}$ for WASP-32b. See Figure~\ref{fig11} for further details.

\section{Conclusions}
\label{sec:con}

We have carried out an analysis for the transiting exoplanetary systems HAT-P-13, HAT-P-16 and WASP-32 based on new photometric data observed by TESS, OLT-1.2m, YO-1m and YO-2.4m telescopes and the data from ETD website. We employ the GP to correct the systematic errors hidden in the light curves, use the MCMC technique to model the final light curves and derive the system parameters. The refined system parameters are consistent with the previous results. We find that both HAT-P-13 b and HAT-P-16 b show significant timing variations which can be explained with apsidal precession, the timing variation of WASP-32b may be led by a decaying orbit due to tidal dissipation or apsidal precession. However, these TTVs could also be reproduced by the gravitational perturbations of close planetary companions.


\section*{Acknowledgements}
We acknowledge the support of the staffs of the Hamburg 1.2m, Lijiang 2.4m and Kunming 1m telescopes. Funding for the Lijiang 2.4m telescope
has been provided by Chinese Academy of Sciences and the People's
Government of Yunnan Province. We appreciate the referee for his/her
helpful suggestions and comments, which led to a significant
improvement to the paper. This work is supported by National Natural Science Foundation of China through grants No. U1531121, No. 10873031, No. 11473066 and No. 12003063. We also acknowledge the science research grant from the China Manned Space Project with NO. CMS-CSST-2021-B09. This paper includes data collected by the TESS mission, which is funded by the NASA Explorer Program.

 This work has been conducted under the frame between China Scholarship Council (CSC) and Deutscher Akademischer Austausch Dienst (DAAD). The joint research project between Yunnan Observatories and Hamburg Observatoy is funded by Sino-German Center for Research Promotion (GZ1419).

%

\section*{Data availability}
{\it TESS} data products can be accessed through the official NASA website \url{https://heasarc.gsfc.nasa.gov/docs/tess/data-access.html}.\\

The data that support the plots within this paper and other findings of this study are available from the corresponding authors upon reasonable request.

%

%

\bibliographystyle{mnras} 
\bibliography{HAT13}

\bsp	
\label{lastpage}
\end{document}